%% file: main.tex
 \documentclass[final,1p,times,twocolumn]{elsarticle}

\usepackage{graphicx,color}

\usepackage{amssymb}

\usepackage{multirow}

\journal{Nucl. Instr. and Meth. A}

\begin{document}

\begin{frontmatter}



\title{Calibration of the Super-Kamiokande Detector}


\input {authors}

\begin{abstract}
Procedures and results on hardware-level detector calibration
in Super-Kamiokande (SK) are presented in this paper.
In particular, we report improvements made in our calibration methods
for the experimental phase IV in which new readout electronics have
been operating since 2008.

The topics are separated into two parts.
The first part describes the determination of constants needed to interpret
the digitized output of our electronics so that we can obtain physical numbers
such as photon counts and their arrival times for each photomultiplier tube
(PMT).
In this context, we developed an in-situ procedure to determine high-voltage
settings for PMTs in large detectors like SK, as well as a new method for
measuring PMT quantum efficiency and gain in such a detector.

The second part describes modeling of the detector in Monte Carlo
simulations, including, in particular, the optical properties of the water
target and their variability over time.
Detailed studies on water quality are also presented.

As a result of this work, we have achieved a precision sufficient for physics
analyses over a wide energy range (from a few MeV to above 1 TeV).
For example, charge determination was at the level of 1\%,
and the timing resolution was 2.1 ns at the one-photoelectron charge level
and 0.5 ns at the 100-photoelectron charge level.
\end{abstract}


\end{frontmatter}


\section{Overview}
 The Super-Kamiokande (SK) is an imaging water Cherenkov detector used to study
neutrinos from various sources (the Sun, the atmosphere, past supernovae,
and an accelerator) 
and to search for proton decay.
Since the beginning of the experiment, various calibration procedures have been
employed to determine the properties of the detector. 
Calibrations carried out in the initial phases of SK were
reported in~\cite{Fukuda:2002uc}.  This paper discusses recent 
improvements and adaptions to our calibration methods.

In Section~\ref{sec:skdet}, we provide an overview of the SK detector with
emphasis on recent electronics upgrade.
Calibration of the inner detector is described in Section~\ref{sec:inner}.
Section~\ref{sec:pmt} describes basic calibration of photomultiplier tubes
(PMTs). Section~\ref{sec:track} examines optical photon propagation in the
detector. Section~\ref{outer} describes calibration of the outer detector.

 Some specialized calibration procedures are reported elsewhere;
use of an electron accelerator ~\cite{Nakahata:1999} and
${}^{16}$N source~\cite{Blaufuss:2001}
for solar and supernova neutrino analyses~\cite{Hosaka:2006}.
Use of cosmic ray muons, $\pi^0$ generated in the detector,
and muon decay electrons for atmospheric and accelerator neutrino oscillation,
and proton decay analyses are described in~\cite{Ashie:2005,Regis:2012sn}.


\section{Super-Kamiokande detector}\label{sec:skdet}
\subsection{Introduction}
 The SK detector is a cylindrical tank (39.3\,m diameter and 41.4\,m height)
filled with 50\,kilotons of ultra-pure water.
It is located 1000\,m underground (2,700\,m water equivalent)
in the Kamioka mine in Gifu Prefecture, Japan.
Inside its stainless steel tank, a stainless steel structure supports
the PMTs and optically separates the tank into inner (ID) and outer detector
(OD)~\cite{Fukuda:2002uc}.
The optical isolation of ID and OD is effected by
black polyethylene terephthalate sheets (``black sheets'')
on the inside of the barrier and highly reflective Tyvek sheets on its outside.
Tank walls are also lined with Tyvek sheets.
The experiment began data-taking in April 1996 and was shut down for maintenance in July
2001, this initial phase is called ``SK-I''. Because of an accident during 
the ensuing upgrade work, the experiment resumed in October 2002
with only about half of its original number of ID-PMTs.
To prevent further accidents from that time on, all ID-PMTs were encased in
fiber-reinforced plastic (FRP) cases with acrylic front windows.
The phase from October 2002 until another shutdown for full
reconstruction, which started in October 2005, is called ``SK-II''.
In July 2006, the experiment resumed with full number of PMTs,
it stopped briefly for an electronics upgrade in September of 2008.
That phase is called ``SK-III''.
The period of data-taking after the electronics upgrade,
which started after September 2008, is called ``SK-IV'',
that period is ongoing as of the time of writing this paper.
The photodetector coverage differed in most phases.
The SK phases and their main characteristics are summarized in
Table~\ref{tab:sk}. A detailed description of the SK-I detector was published
~\cite{Fukuda:2002uc}, where detector calibrations were also discussed.
Since then, there have been major improvements to the calibrations, especially
those related to the optimization of high-voltage settings for each
PMT, in understanding of the PMT-to-PMT variations in quantum efficiency and
gain, and in understanding of optical properties of water in the detector.
These improvements are described in this paper.
\begin{table}[htbp]
\begin{center}
%
%
%
\begin{tabular}{|c|c||c|c|c|c|}
 \hline
\multicolumn{2}{|c||}{Phase} & SK-I & SK-II & SK-III & SK-IV \\
 \hline
\multirow{2}{*}{Period} & start & 1996 Apr. & 2002 Oct. & 2006 Jul. & 2008 Sep. \\
                        & end   & 2001 Jul. & 2005 Oct. & 2008 Sep. & (running) \\
 \hline
Number & \multirow{2}{*}{ID} & 11146 & 5182 & 11129 & 11129 \\
of                           &       & (40\%) & (19\%) & (40\%) & (40\%) \\
 \cline{2-6}
PMTs                         & OD & \multicolumn{4}{c|}{1885} \\
 \hline
\multicolumn{2}{|c||}{Anti-implosion} & \multirow{2}{*}{no} & \multirow{2}{*}{yes} & \multirow{2}{*}{yes} & \multirow{2}{*}{yes} \\
\multicolumn{2}{|c||}{container} & & & & \\
 \hline
\multicolumn{2}{|c||}{OD segmentation} & \multirow{2}{*}{no} & \multirow{2}{*}{no} & \multirow{2}{*}{yes} & \multirow{2}{*}{yes} \\
\multicolumn{2}{|c||}{(Figure~\ref{fig:odseg})} & & & & \\
 \hline
\multicolumn{2}{|c||}{Front-end}   & \multicolumn{3}{c|}{ATM (ID)} & \multirow{2}{*}{QBEE}\\
\multicolumn{2}{|c||}{electronics} & \multicolumn{3}{c|}{OD QTC\cite{Fukuda:2002uc} (OD)} & \\
 \hline
\end{tabular}
\end{center}
\caption{Detector configuration for the SK experiment phases up to the current time. The values in parentheses below the number of PMTs in the ID show percent photo-coverage of the surface.}
\label{tab:sk}
\end{table}

\subsection{SK-IV electronics upgrade}\label{sec:elec}
 In phases SK-I, SK-II, and SK-III, signals from the ID-PMTs were processed
by custom electronics modules called analog timing modules
(ATMs)~\cite{Tanimori:1989}. These modules contained both charge-to-analog
converters (QAC) and time-to-analog converters (TAC). The dynamic range was
from 0 $\sim$ 450\,pico Coulomb (pC) with 0.2\,pC resolution for charge
and from -300$\sim$ 1000\,ns with 0.4\,ns resolution for time.
There were two pairs of QAC/TAC for each PMT input signal, this prevented
dead time and allowed the readout of multiple sequential hits that may arise,
e.g. from electrons that are decay products of stopping muons.

 To ensure stable data-taking over the next decade and to improve
the throughput of data acquisition systems, the SK electronics system was
upgraded in September 2008. For SK-IV, new front-end electronics modules, called
QTC-based electronics with Ethernet (QBEE)~\cite{qbee}, were developed and
installed for measuring arrival times and integrated charges of signals
for ID-PMTs and the OD-PMTs.
The QBEE provides high-speed signal processing by combining pipelined
components, these included a newly developed custom charge-to-time converter
(QTC) in the form of an application-specific integrated circuit (ASIC),
a multi-hit time-to-digital converter (TDC),
and field-programmable gate array (FPGA)~\cite{Nishino:2009zu}.
Each QTC input has three gain ranges - ``Small'', ``Medium''
and ``Large'' - the resolutions for each are shown in Table~\ref{tab:range}.
For each range, analog to digital conversion is conducted separately,
but the only range used is that with the highest resolution that is not being
saturated. The overall charge dynamic range of the QTC is $0.2\sim2500$\,pC,
which is about five times larger than that of the old ATM.
The charge and timing resolution of the QBEE at the single photoelectron
level is 0.1\,photoelectrons and 0.3\,ns respectively, both are better
than the intrinsic resolution of the 20-inch PMTs used in SK.
The QBEE achieves good charge linearity over a wide dynamic range.
The integrated charge linearity of the electronics is better than 1\%.
The thresholds of the discriminators in the QTC are set to $-0.69$\,mV
(equivalent to 0.25\,photoelectron, which is the same as for SK-III).
This threshold was chosen to replicate the behavior of the detector during
its previous ATM-based phases.

\begin{table}[htbp]
\begin{center}
\begin{tabular}{|c|c|c|}
 \hline
  Range & Measuring Region & Resolution  \\
 \hline
Small & 0 - 51\,pC & 0.1\,pC/count (0.04\,pe/count)  \\
 \hline
Medium & 0 - 357\,pC & 0.7\,pC/count (0.26\,pe/count)  \\
 \hline
Large & 0 - 2500\,pC & 4.9\,pC/count (1.8\,pe/count)  \\
 \hline
\end{tabular}
\end{center}
\caption{
Summary of QTC ranges for charge acquisition. 
}
\label{tab:range}
\end{table}
\subsection{Water tank}\label{sec:tank}
\subsubsection{Tank geometry}
During the reconstruction period between SK-II and SK-III,
we surveyed the tank to verify the blueprint-based MC implementation
of the geometry.
The detector surface is subdivided into a barrel part and two round parts,
top and bottom part.
%
%
The positions of 14 PMTs (ten around the center of the barrel at various
azimuthal positions, three PMTs at top and one at bottom)
were measured using a transit theolodite.
All were consistent with their respective design values up to a few cm.
This difference was small enough that we did not need to alter fiducial volume
estimates.
%
%
%

\subsubsection{Geomagnetic field inside the tank}
 As described in \cite{Fukuda:2002uc}, 26 sets of horizontal and vertical
coils are arranged around the inner surface of the tank
to neutralize the geomagnetic field that would otherwise affect
photoelectron trajectories in the PMTs. A 100\,mG field applied parallel to the
dynode plate of a 20-inch PMT reduces the hit collection efficiency by 10\%,
and it does so in an asymmetric manner~\cite{Suzuki:1993zu}.

\begin{figure}[htbp]
\begin{center}
\includegraphics[scale=0.28]{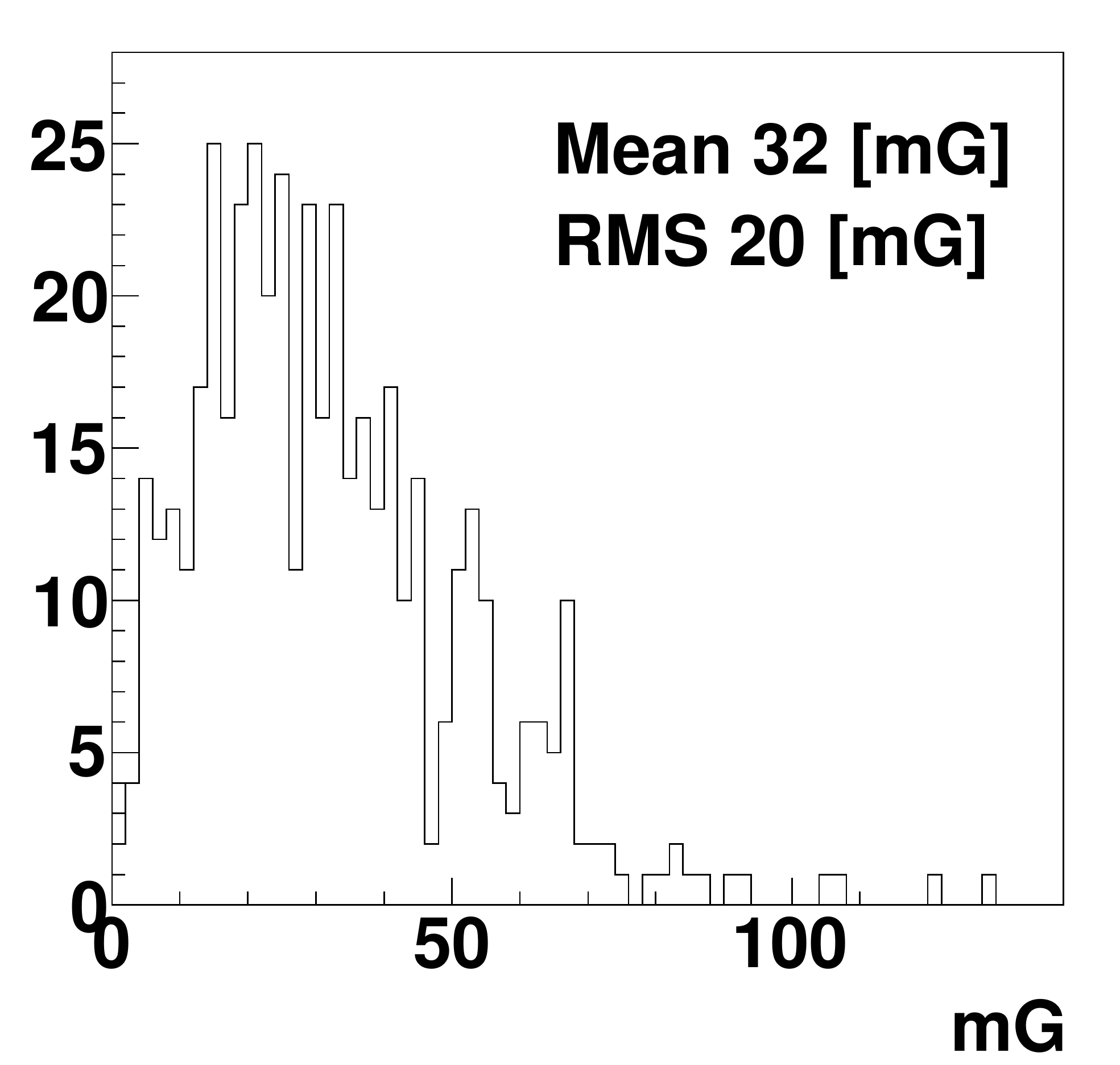}
\includegraphics[scale=0.28]{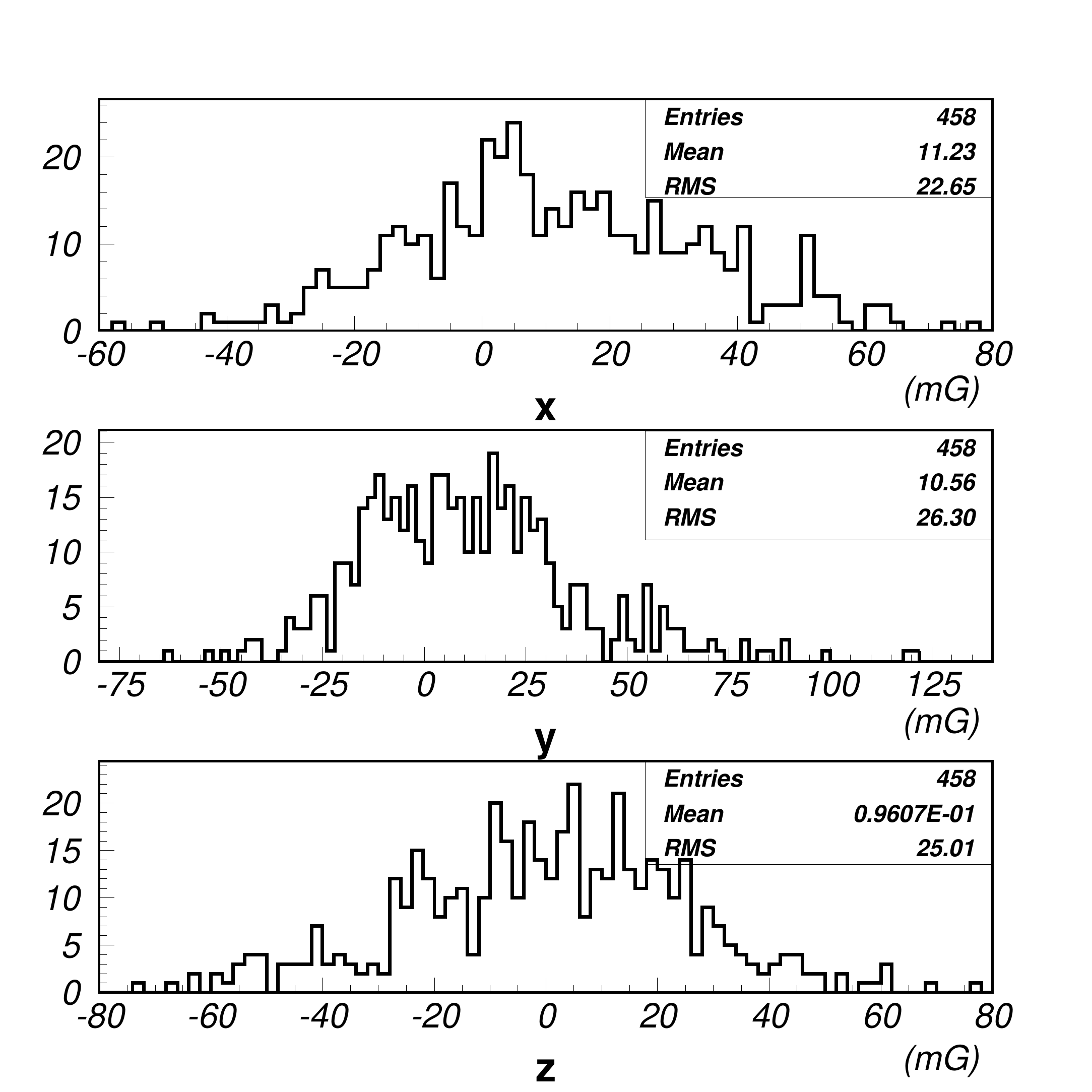}
\caption{Distribution of magnitude of the residual magnetic field at different locations in the detector. The left figure shows the magnitude; the right figures show the value along the usual SK coordinate system axes.}
\label{fig:mag01}
\end{center}
\end{figure}

\begin{figure}[htbp]
\begin{center}
\includegraphics[scale=0.18]{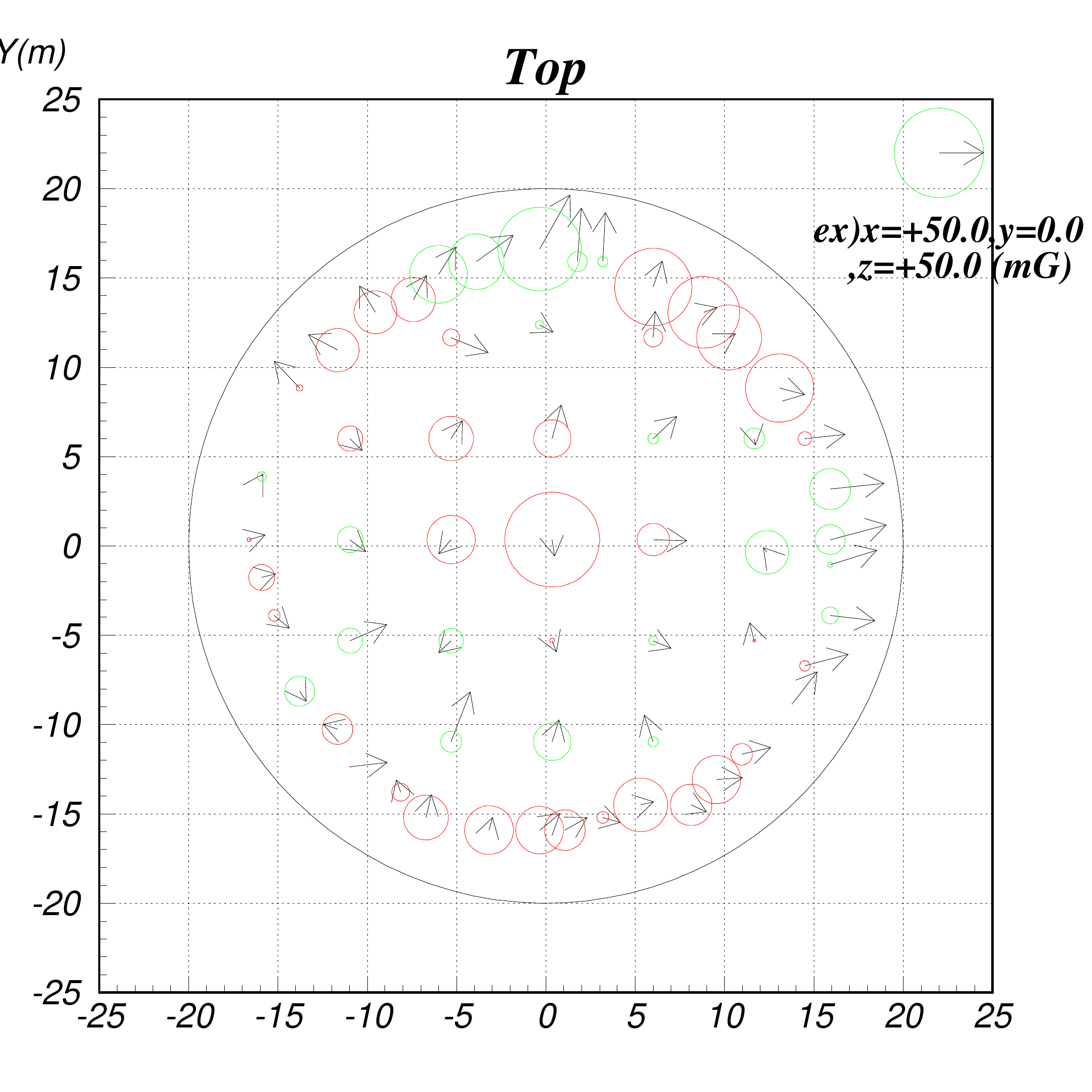}
\includegraphics[scale=0.18]{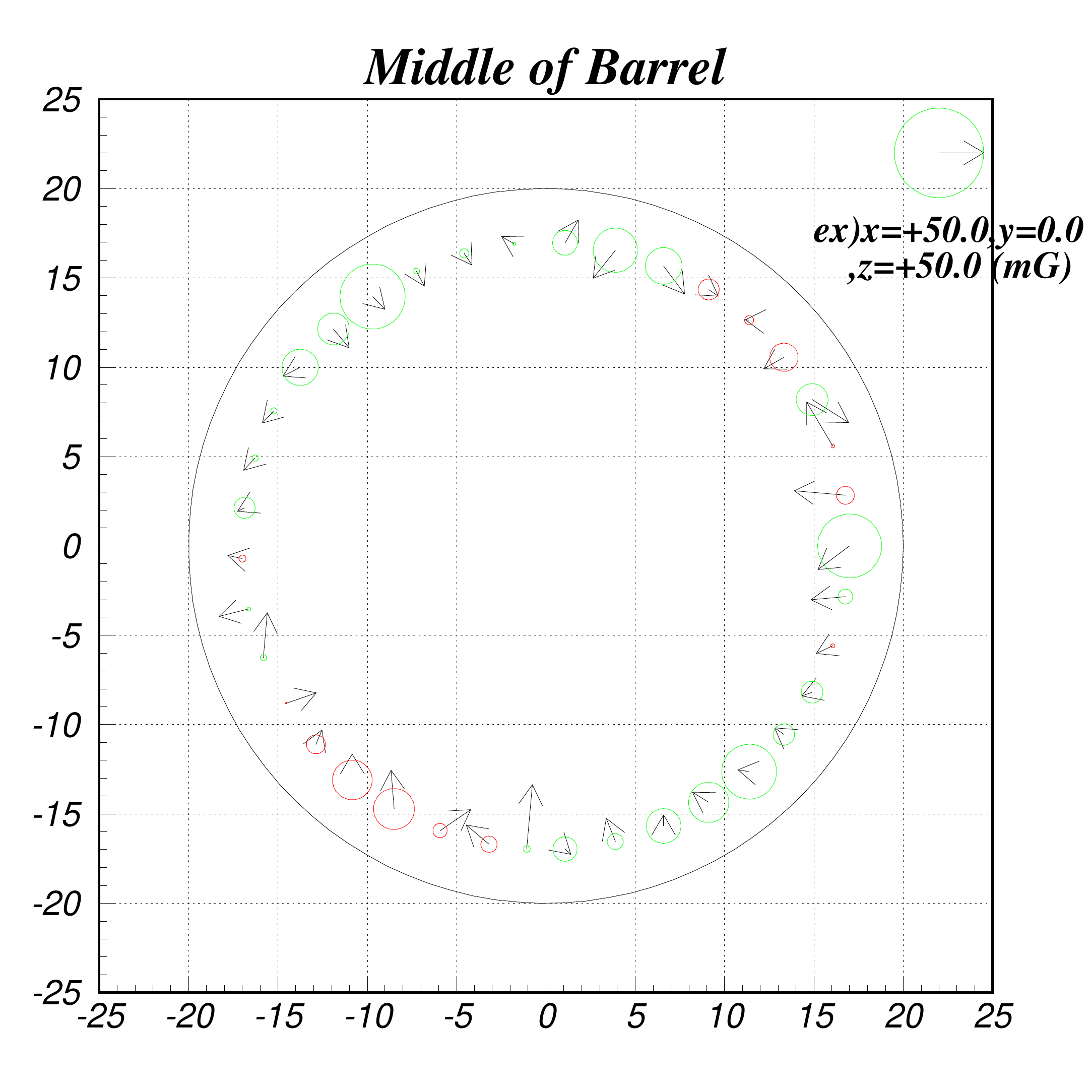}
\includegraphics[scale=0.18]{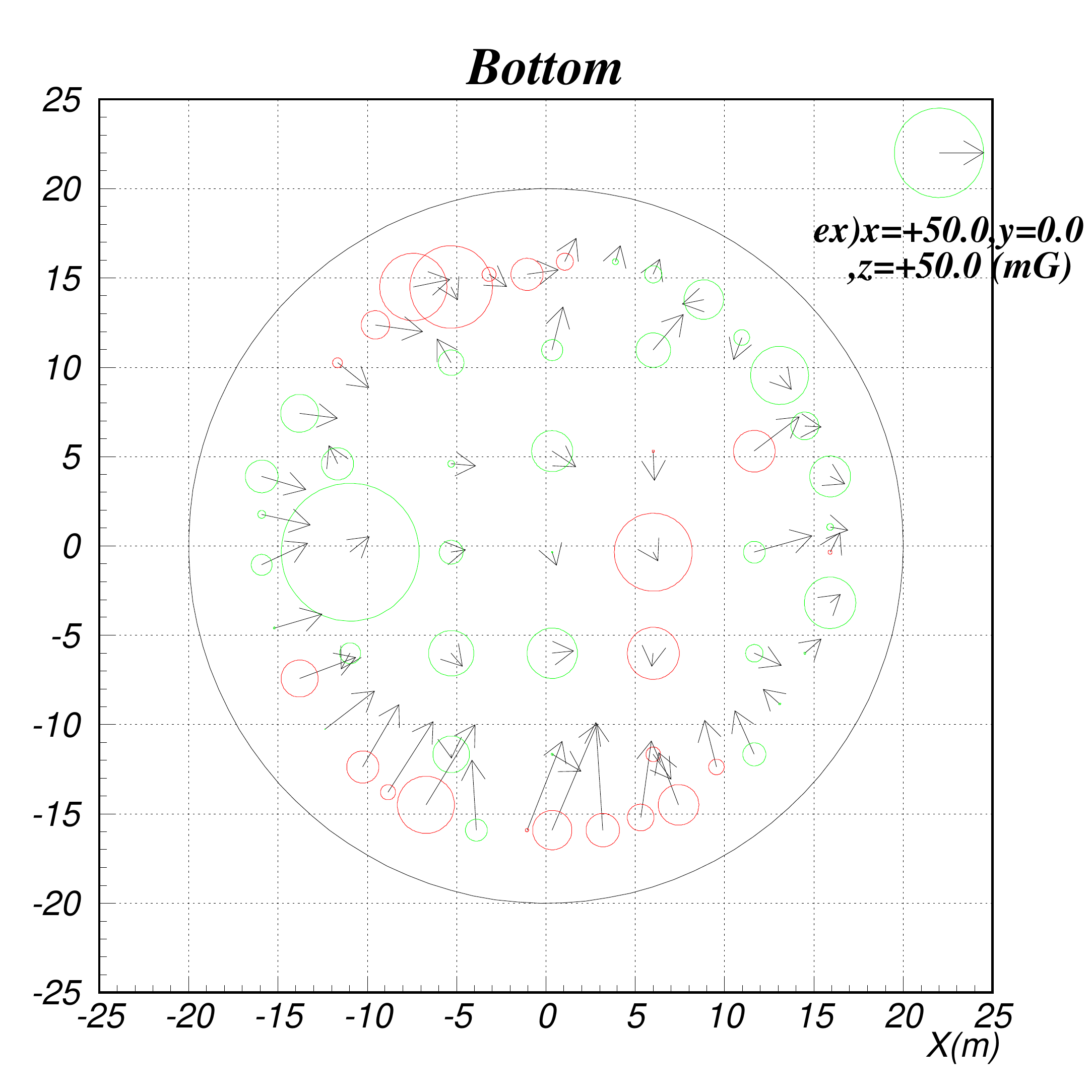}
\caption{Intensities (represented by size of the circle) and directions (represented by an arrow for horizontal; blue solid shows upward and red dashed shows downward) of the remaining magnetic field. The three panels show the results of field measurements at PMT positions at the top, around the center, and at the bottom of the ID. The blue circle and the arrow in the top right show the reference scale for a 50\,mG field in both the x and z direction, respectively.}
\label{fig:mag03}
\end{center}
\end{figure}

 Before and during filling of the tank with water for SK-III,
and with the coils carrying their design currents,
the residual fields at 458 PMT locations around the detector were measured
using a device\footnote{FGM-4DTAM, Walker LDJ Scientific, inc.}
that can simultaneously measure the magnetic field vector along three
orthogonal axes.
The left side of Fig.~\ref{fig:mag01} shows the distribution of measured field
intensities. The average field intensity is 32\,mG,
which corresponds, on average, to 20\,mG along the PMT dynode axis,
and results in 2\% deviation in the collection efficiency of a photoelectron.
The right side of Fig.~\ref{fig:mag01} shows the magnetic field along each 
detector axis. The average vertical (z) component is nearly zero,
but the horizontal components (x and y) show a 10\,mG shift resulting from
overcompensation of the geomagnetic field\footnote{The direction of magnetic north corresponds to ($-0.74, -0.68$).}.
Figure~\ref{fig:mag03} shows field measurements at the top, center and bottom
of the detector. The top and bottom exhibit similar field patterns,
with somewhat higher residual fields than in the barrel.
From these studies we estimate that the effect of those residual fields on the 
collection efficiency of the ID-PMTs is about $1\sim2$\%.

%
\subsection{Water circulation}\label{water-circ}
\begin{figure}[htbp]
\begin{center}
\includegraphics[scale=0.24]{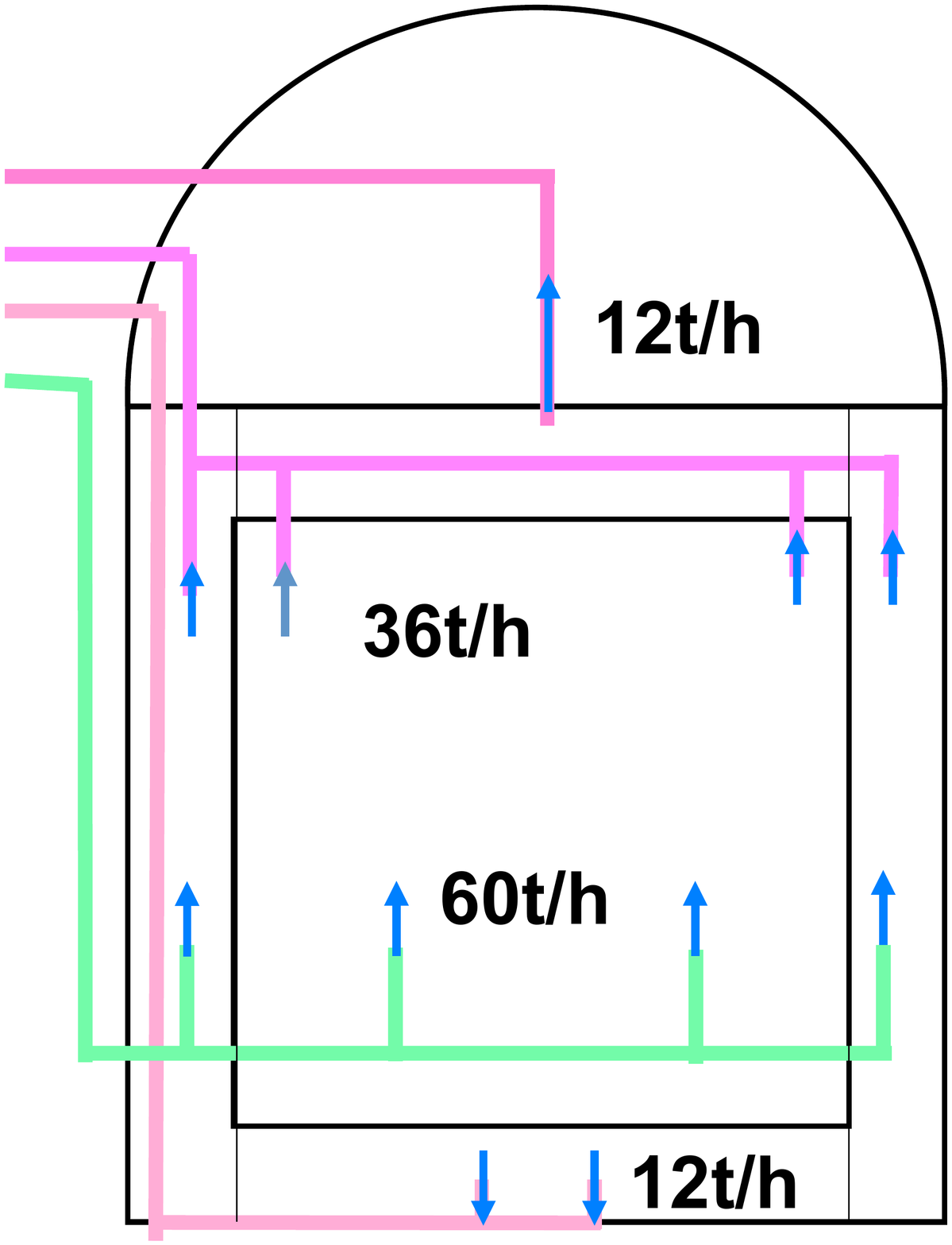}
\caption{Diagram of the water circulation system in SK-IV.
}
\label{fig:watersys}
\end{center}
\end{figure}

 Understanding the condition of the water in the SK detector is
crucial for proper calibration of the detector
because the water condition affects photon propagation in the tank.
The water used in SK is sourced from natural water inside the Kamioka mine.
It is purified by a dedicated system and it is continually recirculated
through parts of that system during normal detector operation.
Figure~\ref{fig:watersys} shows a schematic of the circulation system.
To monitor water temperatures and their time dependence, thermometers with
0.0001$^\circ$ precision\footnote{Denko Co., QT-860S, Yokohama, Japan, Tel:+81-453317089, Fax:+81-453317284.} are placed
at eight positions in both the ID and the OD.

 In the ID, the water circulation pattern can be manipulated by
changing the temperature of the water that is fed into the detector
from the re-circulation system.

\begin{figure}[htbp]
\begin{center}
\includegraphics[width=8cm]{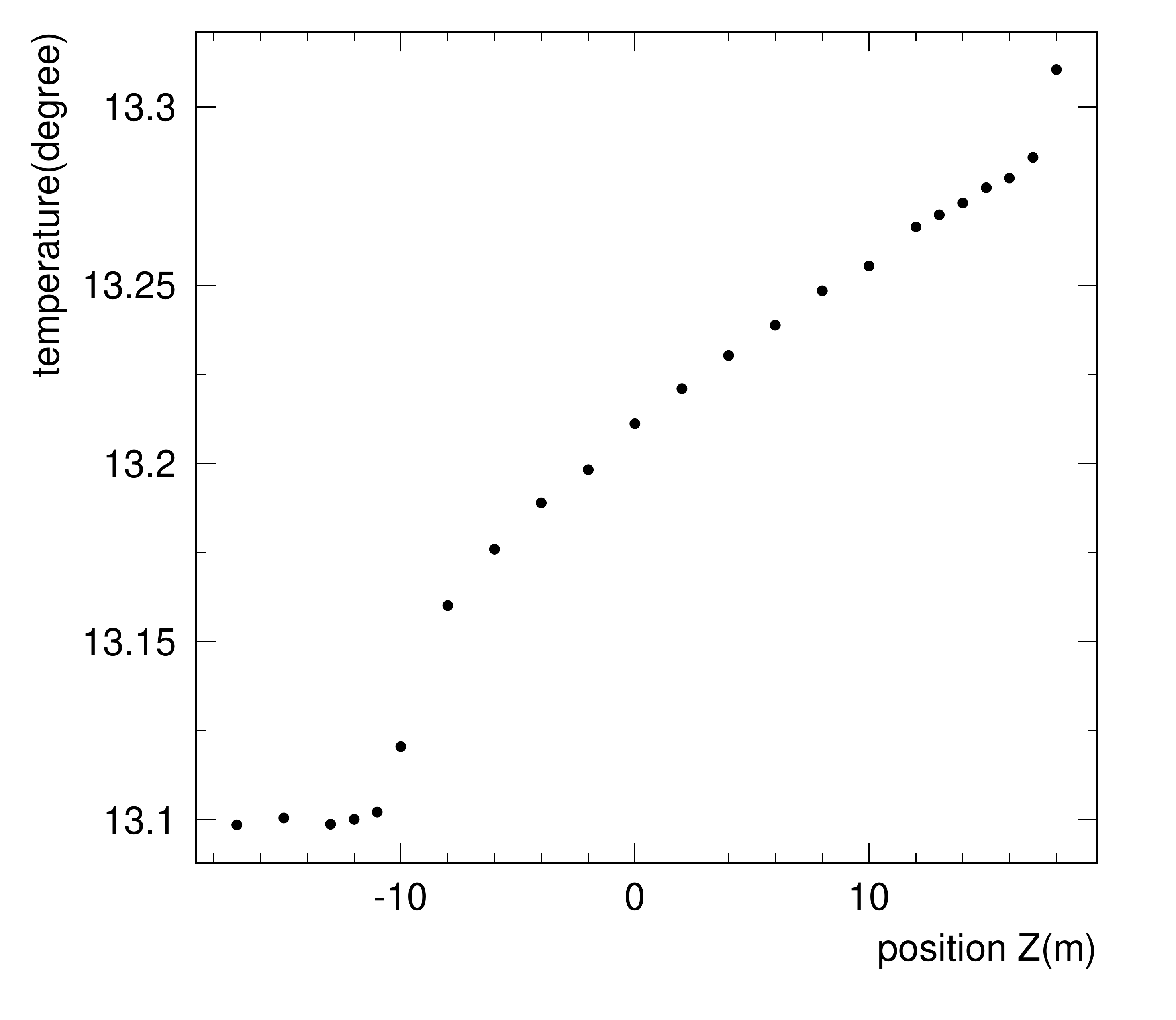}
\caption{
The vertical dependence of the water temperature in the ID.
}
\label{fig:ztemp}
\end{center}
\end{figure}

 In the usual case, we find that water is always convecting
below z=$-11$\,m, which results in uniform temperature throughout that lower
part of the ID. Above that level, water is layered, and its temperature
gradually rises, resulting in a 0.2$^\circ$ difference between the top of the
ID and the convection zone in its lower part (see Fig.~\ref{fig:ztemp}).
The water in this layered region slowly rises through these layers
over the course of one month.
This results in a 5\% difference in water transparency over the detector,
as discussed in Section~\ref{waterparam}.

 While we can expand the convective region that normally resides
at the bottom of the ID to cover the full ID volume,
then the water quality would be uniform.
When convection extends over the whole ID volume,
the water temperature difference becomes less than 0.01$^\circ$.
This small temperature difference clearly indicates full convection
of water in the SK tank and thus indicates
the best uniformity in optical properties of water in the ID.
This water condition period was used for measurement of
relative differences of quantum efficiency in each PMTs
as described in Section~\ref{sec:qe}.

%

\section{Inner detector calibration}\label{sec:inner}
\subsection{PMT and electronics calibrations}\label{sec:pmt}
\subsubsection{Introduction}\label{sec:pmtintro}
 To provide background for this section, a brief description of PMT calibration
is presented here.
The 20-inch diameter PMTs developed by Hamamatsu Photonics K.K.
(R3600-05(A))~\cite{Suzuki:1993zu} are used in the inner detector.
These PMTs have a photo-cathode made of bialkali (Sb-K-Cs),
and has its maximal photon conversion probability in the wavelength 
range of Cherenkov light. The PMT dynodes are of a Venetian-blind type,
and their base circuit is an optimized 11-stage voltage divider.
The high-voltage system for the PMTs was manufactured by CAEN Co.\
and consists of distributors (A933K),
controllers (SY527), and interface modules (V288).

 Since the timing behavior of PMTs depends on the charge of the measured pulse,
we begin discussing ID-PMT calibrations with charge-related issues.
In the definition for the PMT charge calibration, ``gain'' is a conversion
factor from the number of photoelectrons to charge (in units of pC),
and ``QE'' is the product of the quantum efficiency and collection efficiency of
photoelectrons onto the first dynode of the PMT.
Low-energy physics events like solar neutrinos largely consist
of single-photoelectron (single-pe) hits and rely heavily on the QE calibration
for their interpretation, whereas high-energy events like those involving
TeV-scale muons depend more on proper gain calibration. Knowledge of both gain
and QE is important and must be available on a PMT-by-PMT basis.

 Unfortunately, the old ATMs used in SK-I, II, and III did not allow us
to record meaningful single-pe distributions on a PMT-by-PMT basis,
however, a cumulative distribution for all PMTs could be obtained after
the relative gains had been properly calibrated.

 This situation forces us to set up PMT calibration in the following way.
First, we need to determine a suitable high-voltage value to be
applied to each ID-PMT. This determination is described in Section~\ref{sec:hv}.
Next we need to understand the differences in gain between individual ID-PMTs.
Section~\ref{sec:rpc2pe} details this effort and its results.
Once we are able to obtain meaningful cumulative single-pe distribution
for all ID-PMTs, Section~\ref{sec:apc2pe} describes how to use this cumulative
single-pe distribution to calibrate the average gain over all ID-PMTs.
Referencing, in turn, the gain variation for an individual PMT
to the average gain gives the individual gain of each ID-PMT.
In Section~\ref{sec:qe} we use Monte Carlo simulations to extract
a calibration of the QE for each individual PMT.
This new procedure which determines the gain and QE of an ID-PMT's independently
is a major improvement over the procedure used previously.
Section~\ref{sec:qegain} describes the validation of both the gain and QE
calibrations, including verifications of their consistency.
Discussion of charge-related calibration issues is concluded in
Section~\ref{sec:lin}, which describes measurements for assessing the linearity
of charge determinations.
Section~\ref{timing_calib} addresses the ID-PMT timing calibration.

 These calibrations, except for the establishment of 420 reference PMTs,
were performed in the beginning of SK-I, II and III.
In addition, a real-time calibration system monitors crucial parameters
throughout normal operations of the experiment to allow us to consider
variability as well as ensure stability during data-taking.
For this purpose, light sources are permanently deployed near the center of the ID.
During SK data-taking, the lights flash in turn at approximately 1-s intervals.
As detailed in subsections ~\ref{sec:hv}, ~\ref{timing_calib}, and ~\ref{waterparam},
they monitor ID-PMT gains and timing as well as optical parameters of ID water.

\subsubsection{Determination of the high-voltage setting for each PMT}\label{sec:hv}
\begin{figure}[htbp]
\begin{center}
\includegraphics[width=10cm]{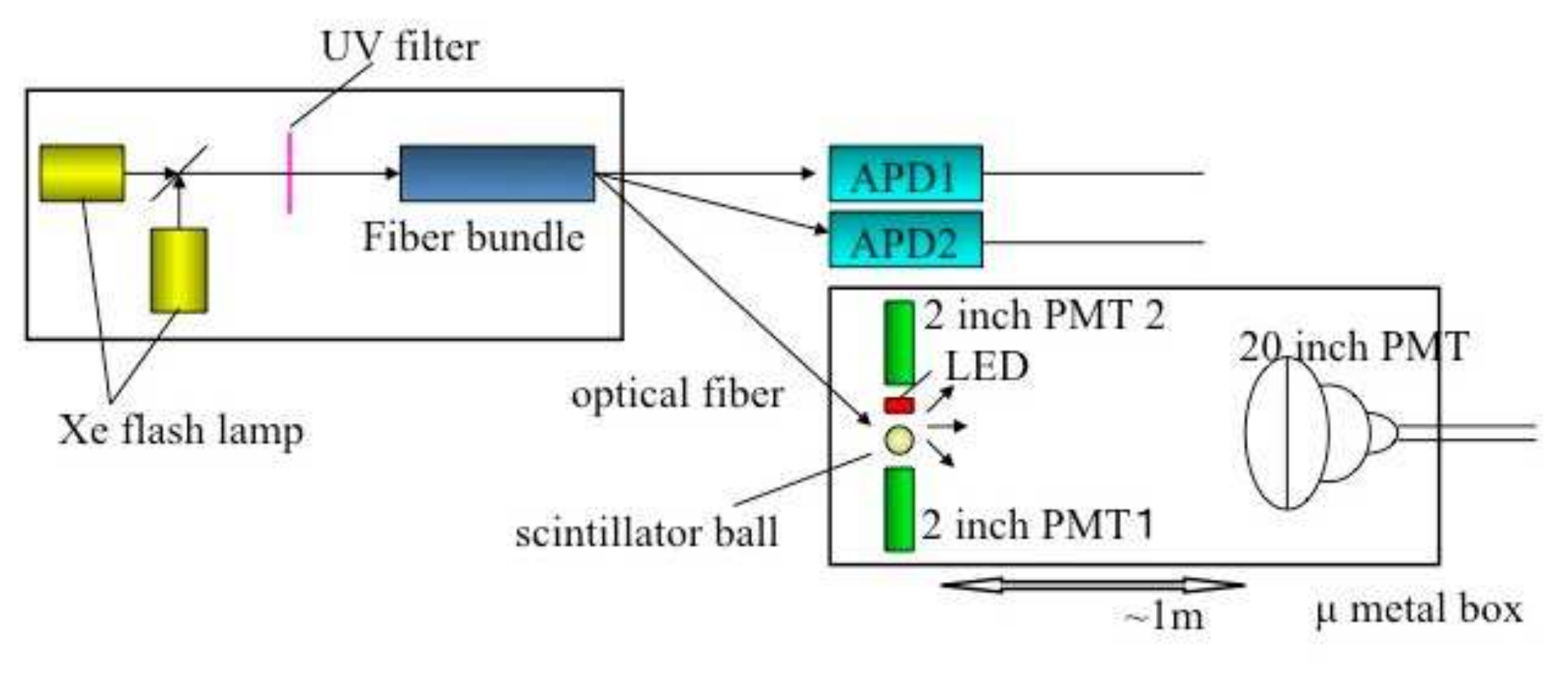}
\caption{Schematic view of the setup for the pre-calibration. A Xe flash lamp, placed inside a box, emitted light that was guided by optical fibers through a fiber bundle to two avalanche photodiodes and a scintillator ball located in another $\mu$-metal shielded
dark box, where a 20 inch PMT was exposed to the light from the scintillator ball.
Two 2-inch PMTs monitor the light output of the scintillator ball.}
\label{fig:pre00}
\end{center}
\end{figure}

\begin{figure}[htbp]
\begin{center}
\includegraphics[scale=0.33]{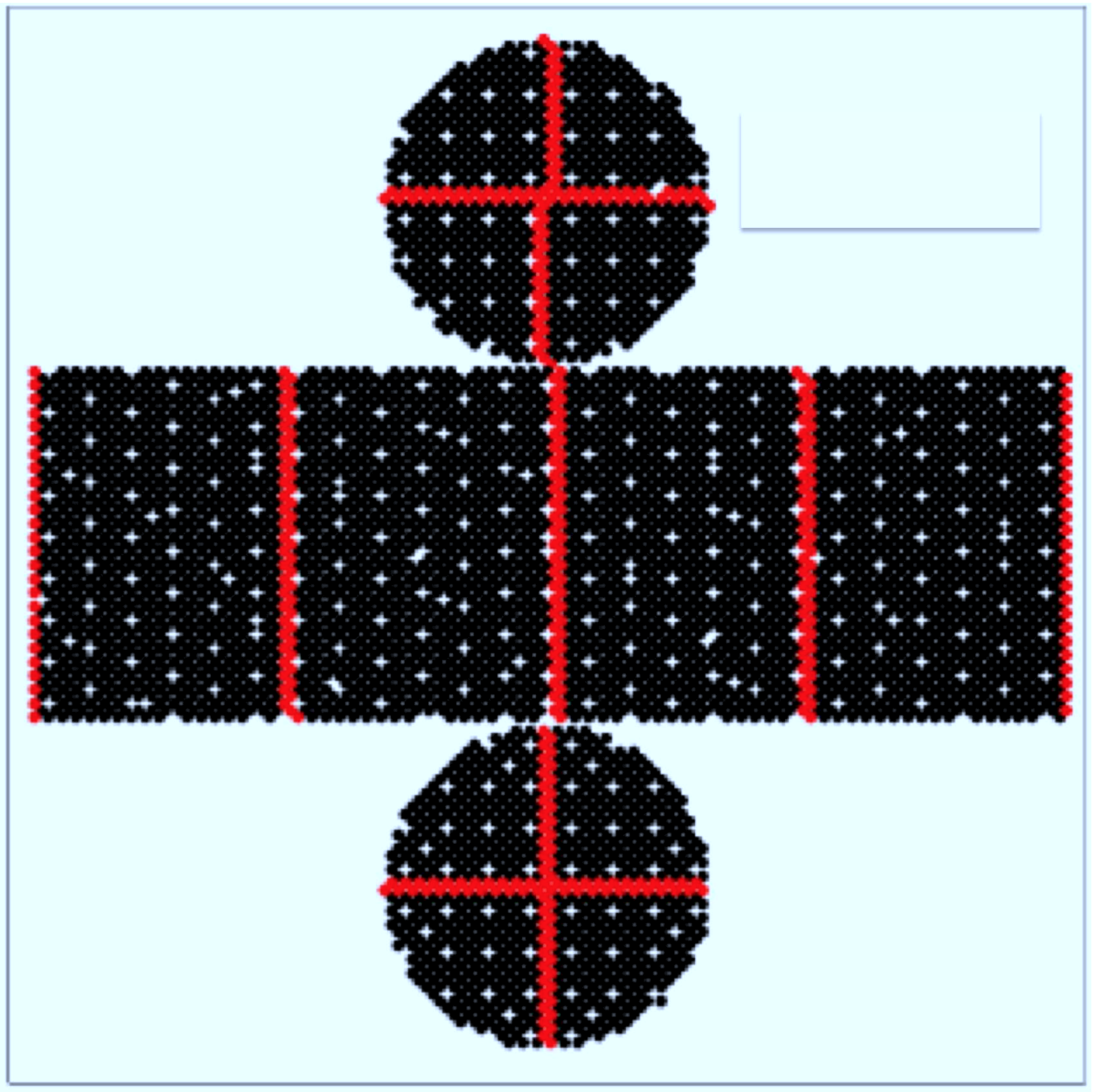}
\includegraphics[scale=0.22]{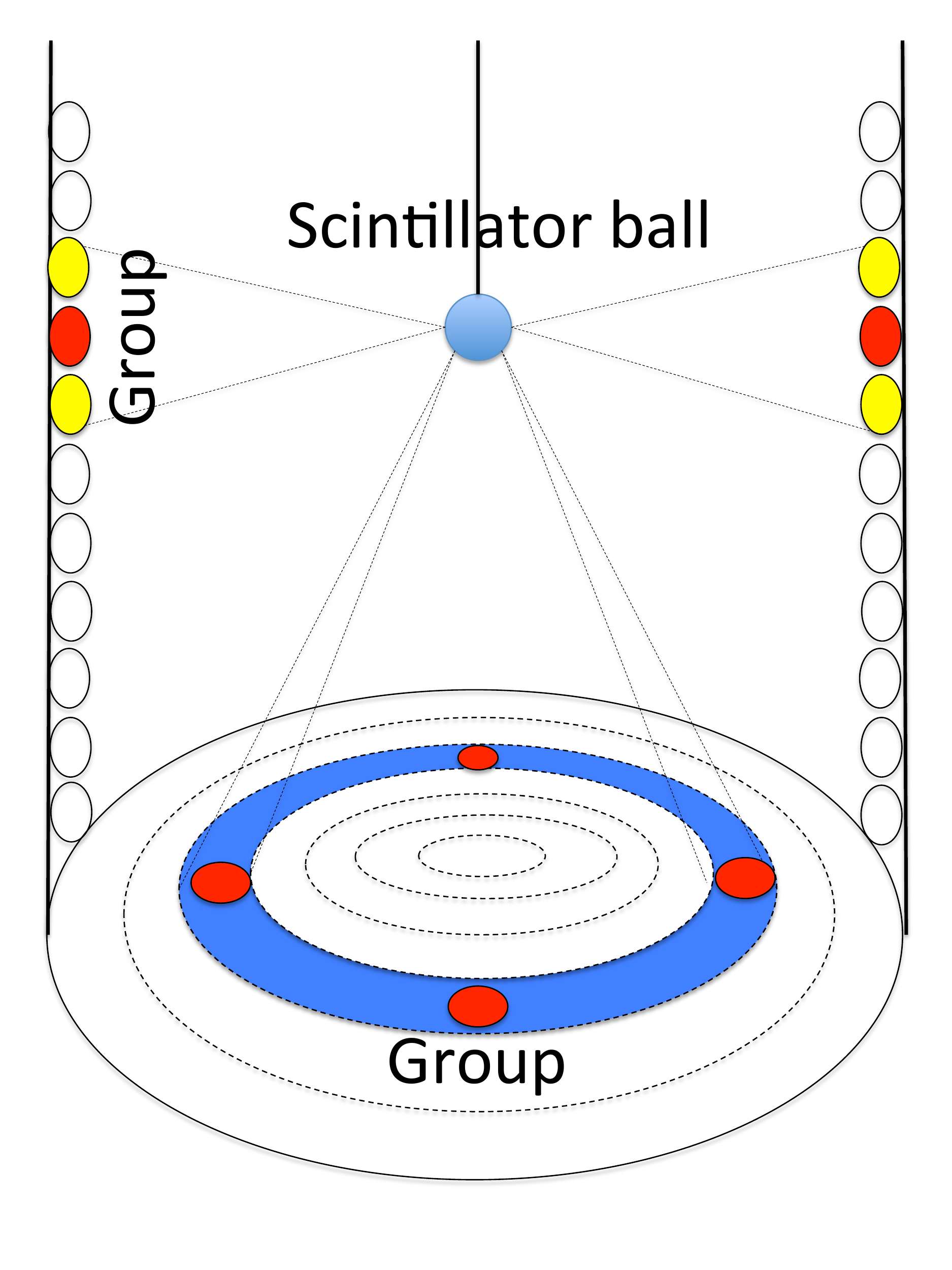}
\caption{The location of ``standard PMTs'' inside the SK inner detector (left). The red points indicate the locations of the standard PMTs. These PMTs served as references for other PMTs belonging to the same group with similar geometrical relationship to the light source (right). 
}
\label{fig:std}
\end{center}
\end{figure}

 To establish the hith-voltage (HV) setting for each PMT, we require that all
PMTs give the same output charge for the same incident light intensity.
For this purpose,
an isotropic light source is placed at the center of the SK tank.
Since the SK tank is a large cylinder about 40\,m in both diameter and height,
we expect the amount of light reaching each PMT from that source
to be about a factor of two different between the closest and farthest PMT.
Correcting for only this geometrical difference is insufficient, because it
ignores important corrections, such as photon propagation in the water,
which depends on water quality and reflections from ID surfaces.
To avoid such problems, 
we prepared 420 ``standard PMTs'' for which HV values were determined
individually before installation during SK-II and SK-III, these determinations
used the pre-calibration system shown in Fig.~\ref{fig:pre00}.
The standard PMTs were later mounted in the ID 
and served as references for other PMTs that were grouped based on similar
geometrical relationships to the light source at the detector center.
(See also Fig.~\ref{fig:std}.)

 In the pre-calibration system, the light from a Xe lamp was passed through a UV filter and
then injected simultaneously into three optical fibers.
One fiber went to a ``scintillator ball'' placed in a dark light-tight box.
The other two fibers went to avalanche photodiode (APD) modules,
which monitored the light intensity of the Xe lamp.
The scintillator ball\footnote{Manufactured by CI Kogyo, Saitama, Japan, Tel:+81-429481811, Fax:+81-429491395.} is
a 5 cm diameter acrylic ball containing 15\,ppm of POPOP as a wavelength
shifter and 2000\,ppm of MgO as a diffuser to make the light emission
from the ball as uniform as possible. Its uniformity in phi was measured to be
better than 1\%.
The dark box was made of $\mu$-metal to shield against the geomagnetic field.
The residual field inside the dark box was measured to be less than 20\,mG.
Two 2-inch PMTs were also mounted inside that box to monitor the light intensity
from the scintillator ball.
The HV values of the 420 PMTs were adjusted so that the observed ADC counts 
returned the target ADC value for the reference about 30 photoelectrons
for the light intensity from the scintillator ball.

\begin{figure}[htbp]
\begin{center}
\includegraphics[width=8cm]{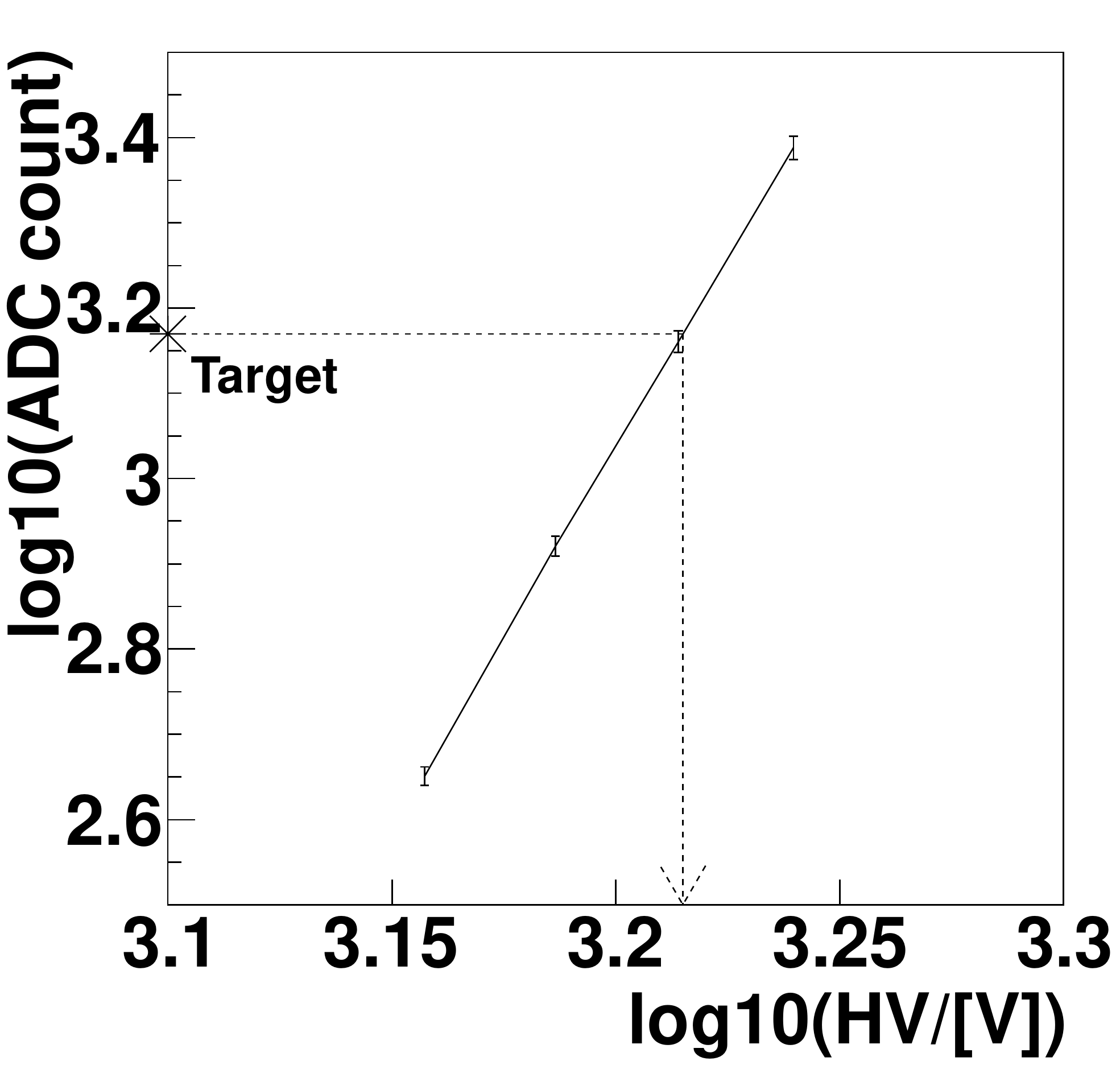}
\caption{Typical gain curve. The horizontal axis is HV value and the vertical axis is ADC counts. Both axes are in log scale. After measurements with four different HV values, the HV for the target ADC count was calculated by fitting the four points by an analytical formula $y=\alpha x^{\beta}$, where $\alpha$ and $\beta$ are constants for each PMT. An additional measurement with that HV value was performed for confirmation.}
\label{fig:pre01}
\end{center}
\end{figure}

\begin{figure}[htbp]
\begin{center}
\includegraphics[width=8cm]{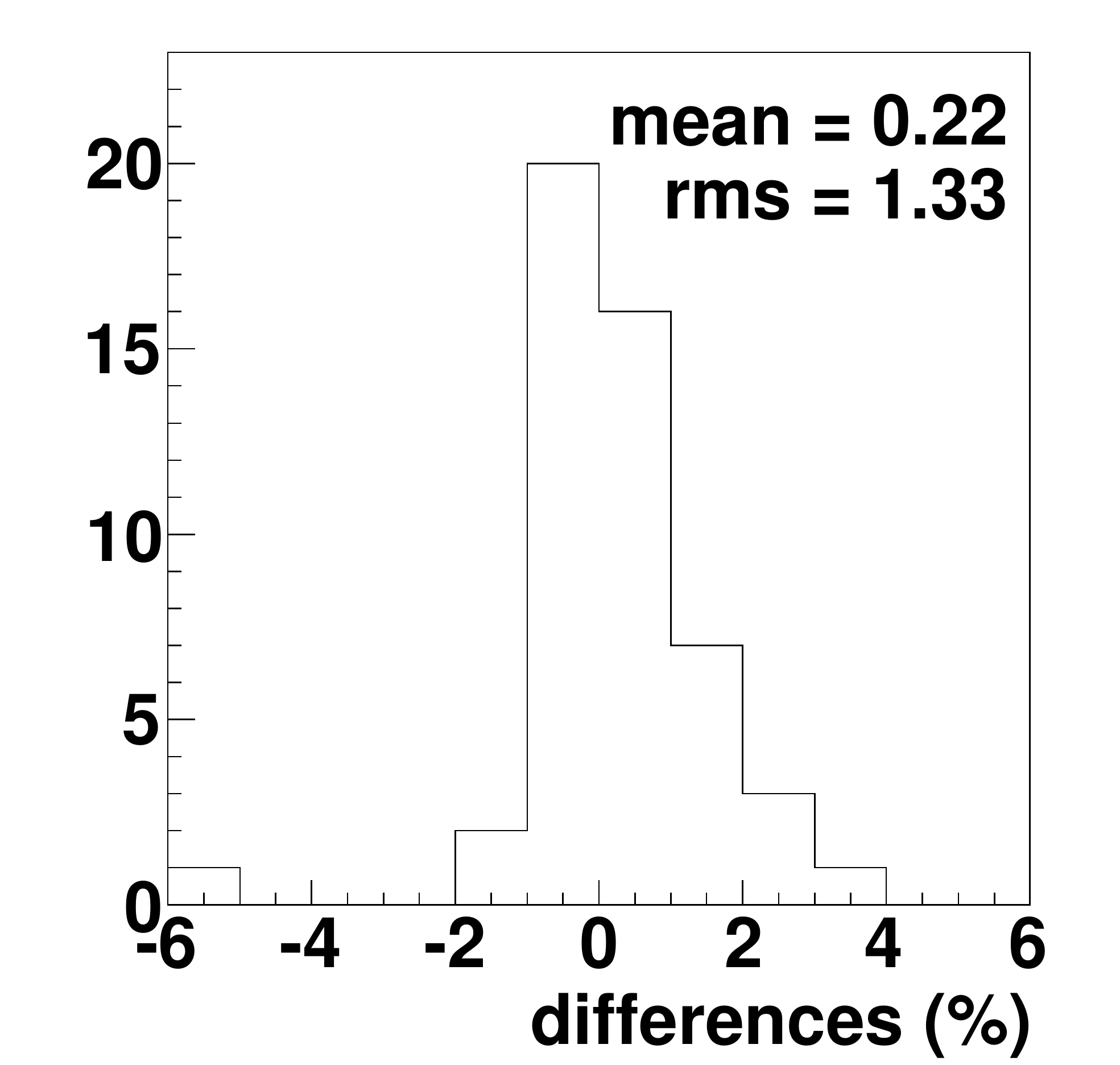}
\caption{Distribution of the observed charge differences between the first and second measurements in pre-calibration for checking reproducibility. 50 out of 420 standard PMTs had been
selected at random for this check.
}
\label{fig:pre02}
\end{center}
\end{figure}

 These HV adjustments were completed over a two-week period inside the Kamioka mine.
To monitor the stability of the adjustment procedure,
the same PMT was tested every morning and every evening,
and variations were found to be smaller than 1\%.
The temperature inside the dark box was also monitored, and its variation
was less than 0.1\,degree\footnote{This could be achieved since all work
was done inside the Kamioka mine where ambient temperature is quite stable.}.
The measurements were done by first applying four different HV values.
Figure~\ref{fig:pre01} shows a typical gain curve.
Applying a linear fit to this plot, the HV value that gave the target ADC
counts was determined.
Finally, the obtained HV was applied, and it was confirmed that the observed
ADC counts were within 1\% of the target value.
Reproducibility was also checked. At the end of the measurement,
we selected 50 PMTs randomly from the 420 PMTs and repeated the above procedure
on them. Figure~\ref{fig:pre02} shows fractional differences between
the results of the first and second HV adjustments.
The reproducibility was within 1.3\% RMS.

\begin{figure}[htbp]
\begin{center}
\includegraphics[width=8cm]{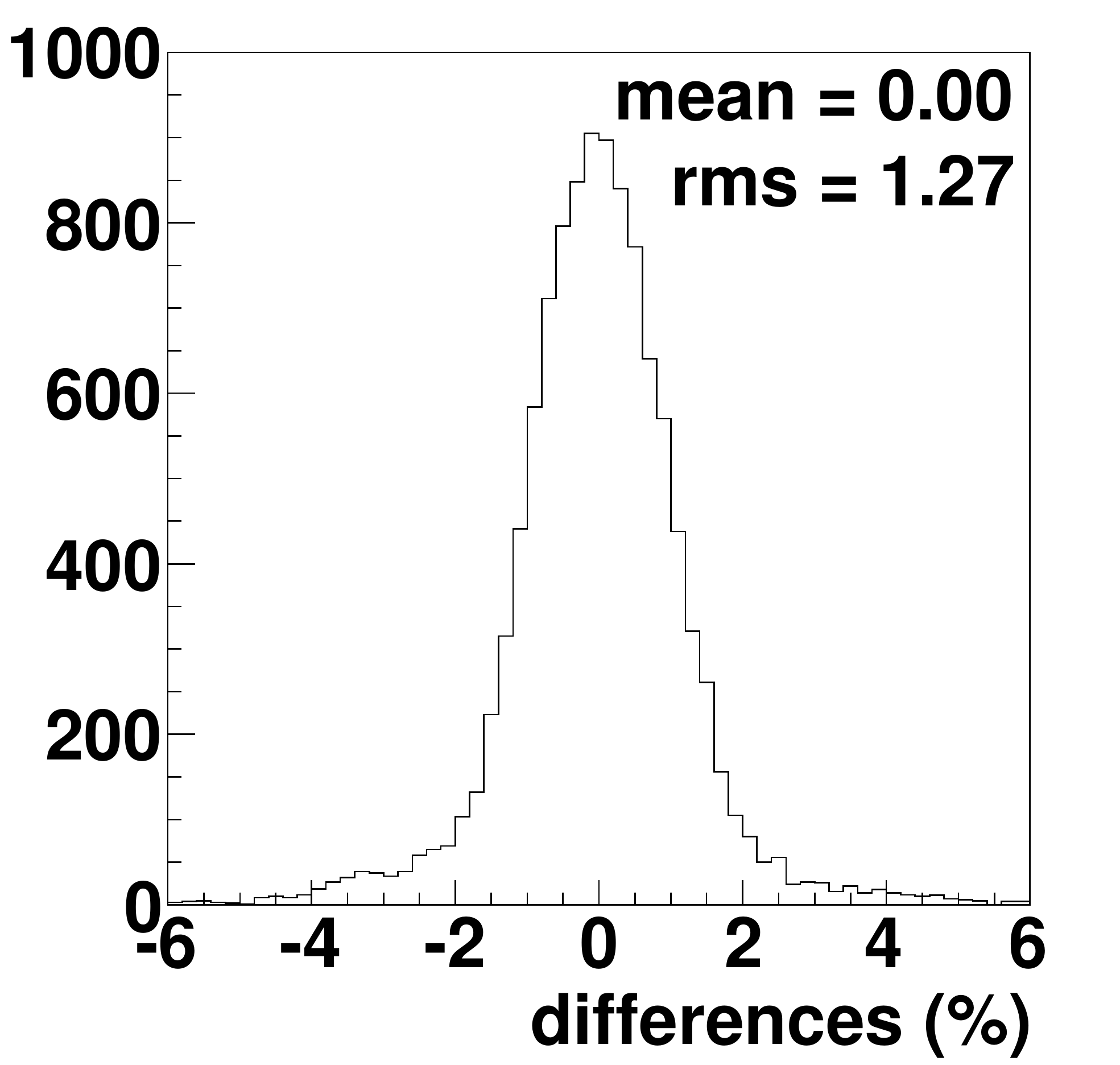}
\caption{The observed percent charge differences for all ID-PMTs
from their respective reference value.}
\label{fig:hv01}
\end{center}
\end{figure}

 The 420 standard PMTs were placed as shown by the red points in Fig.~\ref{fig:std}
to serve as references for other PMTs that would have similar geometrical
relationships to the light source at the center of the ID.
Examples of such groupings are shown on the right in Fig.~\ref{fig:std}.
The HV settings of those other PMTs were adjusted so that the observed
charges from the light flashes of the light source\footnote{The accuracy of the position of the light source was within $\pm1$~cm at the detector center.}
were matched to the average charges for those standard PMTs having
a similar geometrical relationship to the light source, as shown in Fig.~\ref{fig:std}.
This HV adjustment was done at the beginning of SK-III,
just after installing all PMTs into the detector.
The same scintillator ball/Xe lamp combination was used in this step as had
been used in the pre-calibration.
After determining proper HV settings for all PMTs, we checked reproducibility.
Figure~\ref{fig:hv01} shows the percent differences in observed charge with
respect to the reference value. Good reproducibility, within 1.3\% RMS, was found,
this was consistent with the result from the pre-calibration.
The scintillator ball/Xe lamp light source remains permanently at the center of
the ID for real-time as well as long-term monitoring of ID-PMT gains.


\clearpage 
\subsubsection{Relative differences in gain}\label{sec:rpc2pe}
 To interpret the output charge from the PMTs in number of
photoelectrons, we need to determine the gain for each PMT
and is done in two steps.
The first step is to derive the relative difference among PMTs (described in
this section). Based on this first step we can then determine the average
gain over the whole detector (described in the next section). Once we know
the average and the deviation from that average for each individual ID-PMT,
we can extract the gain of each individual ID-PMT.

 To measure of the relative gain difference,
a stable light source emitting constant-intensity flashes
is deployed at a certain position in the tank.
There are two measurements: the first uses high-intensity
flashes from which every PMT gets a suitable number of photons,
creating an average charge $Q_{obs}(i)$ for each ID-PMT $i$.
The second measurement uses low-intensity flashes
in which only a few PMTs are hit in each event, therefore, we can be reasonably
sure that each of these is a single-pe hit.
We count the number of times $N_{obs}(i)$ that PMT $i$ records
a charge that is greater than the threshold value.
Since the location of the light source is not changed between
the two measurements, the complicating factors in estimating
those two intensities $Q_{obs}(i)$ and $N_{obs}(i)$ are almost identical:

\begin{eqnarray}
  Q_{obs}(i) & \propto & I_s \times a(i) \times \varepsilon_{qe}(i) \times G(i) \label{eq:Qobs},\\
  N_{obs}(i) & \propto & I_w \times a(i) \times \varepsilon_{qe}(i), \label{eq:nhit}
\end{eqnarray}
where $I_s$ and $I_w$ are the average intensities of high and low intensity
flashes, respectively, $a(i)$ is the acceptance of ID-PMT $i$,
$\varepsilon_{qe}$ denotes its QE, and $G(i)$ its gain.
The threshold is sufficiently low that the relative changes in gain,
which we want to track, have little effect on $N_{obs}(i)$,
for example, 10\% gain change makes the $N_{obs}(i)$ just 1.5\% change.
The low threshold enables us to ignore, in the above calculations,
differences in probability for having a charge below the discriminator
threshold among PMTs.
The gain of each PMT can then be derived by taking the ratio of
Eq.~(\ref{eq:Qobs}) and (\ref{eq:nhit}), except for a factor common to all PMTs:
\begin{equation}
G(i) \propto \frac{Q_{obs}(i)}{N_{obs}(i)}.
\label{eq:rel}
\end{equation}
Then the relative gain of each ID-PMT can be obtained by normalization
with the average gain over all PMTs.\footnote{The common factor $I_s/I_w$ is also eliminated by this normalization. In the actual measurement, $N_{obs}$ was corrected by occupancy.}

 To perform this calibration we need a means to change the
intensity of the flashes of the light source. The light source is 
nitrogen-laser-driven dye laser (Section~\ref{timing_calib}).
To manipulate the overall intensity of the light delivered into the ID,
we used a filter wheel with neutral density filters between the dye laser,
and the optical fiber that feeds light into the diffuser ball.

\begin{figure}[htbp]
\begin{center}
\includegraphics[width=8cm]{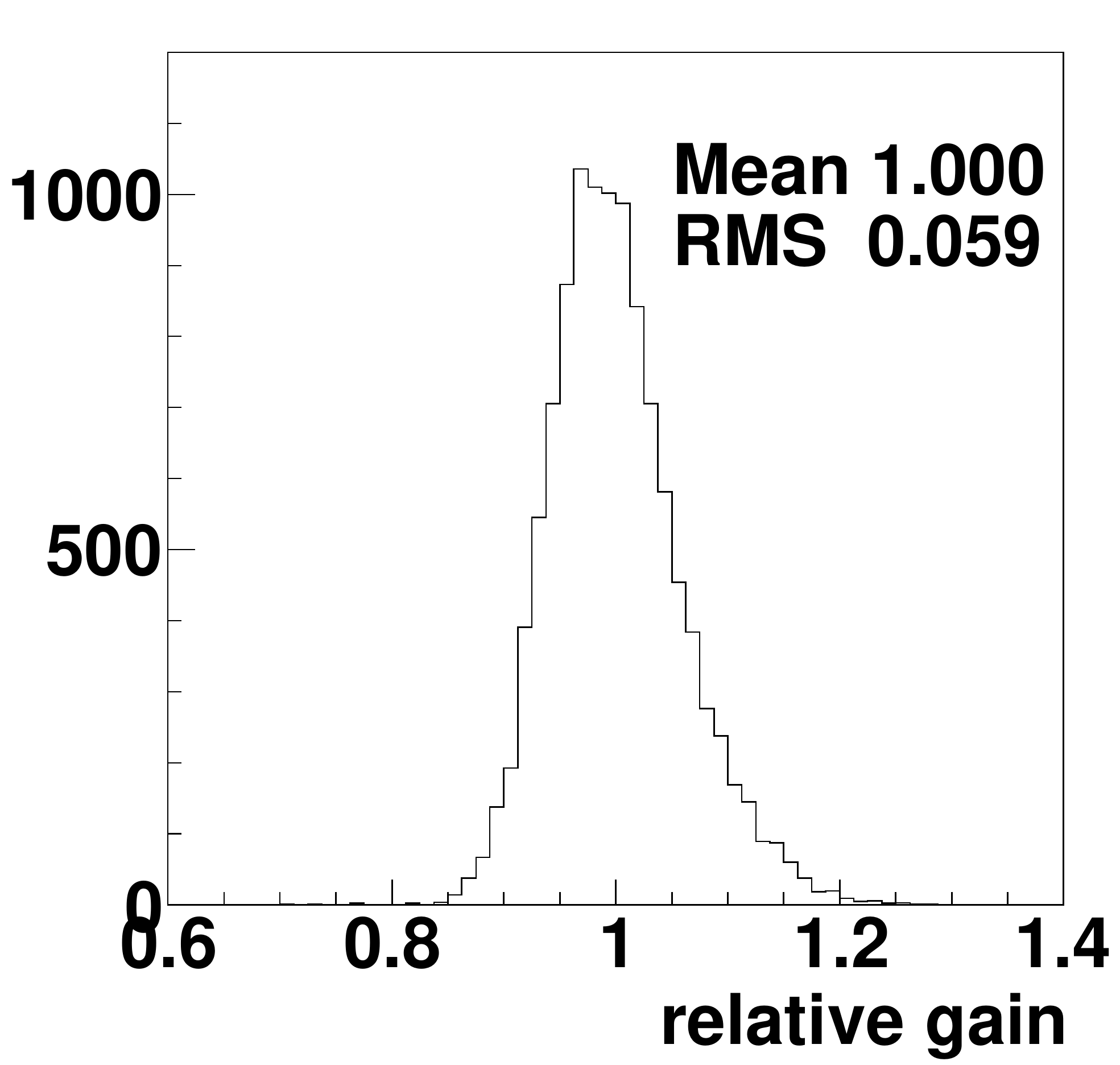}
\caption{Distribution of relative gain of PMTs.}
\label{fig:rel01}
\end{center}
\end{figure}

 Figure~\ref{fig:rel01} shows the ratio (\ref{eq:rel}) for each PMT,
the RMS of the distribution was found to be 5.9\%.
Since the HV value for each PMT was determined to make $Q_{obs}$ be the same,
we infer that this deviation is due to differences in QE among PMTs.
The observed ratio in Eq.~(\ref{eq:rel}) for each PMT, normalized by the average
over all PMTs, contributed to a table of relative gain differences among PMTs.
These factors for relative gain differences of each PMT are used as fine
corrections in conversion from output charge to number of photoelectrons
observed.

\subsubsection{Absolute gain conversion factor}\label{sec:apc2pe}
As pointed out previously, the relative gain for a PMT is usually obtained
from the average of its single-pe distribution. Problems with pedestal
subtraction in the ATMs before SK-IV prevented us from doing this on a
PMT-by-PMT basis.
Given the continuous distribution of relative gain corrections
obtained in the previous section, we can build the
cumulative single-pe distribution for all ID-PMTs.
Applying the relative gain correction aligns the single-pe spectra
of all ID-PMTs so that it makes sense to add them,
it also effectively smoothes the sampling of this distribution
enough to overcome problems encountered at the single-PMT level.
While some additional smearing is introduced by
the intrinsic resolution of the relative gain calibration,
the resulting cumulative single-pe distribution largely represents
the average single-pe response of the detector.
In particular, we can extract from it the absolute gain
of all ID-PMTs, which had been normalized out when we calibrated
the relative gains. In this section, we describe the data we used and present the results
of absolute gain calibration.

\begin{figure}[htbp]
\begin{center}
\includegraphics[scale=0.28]{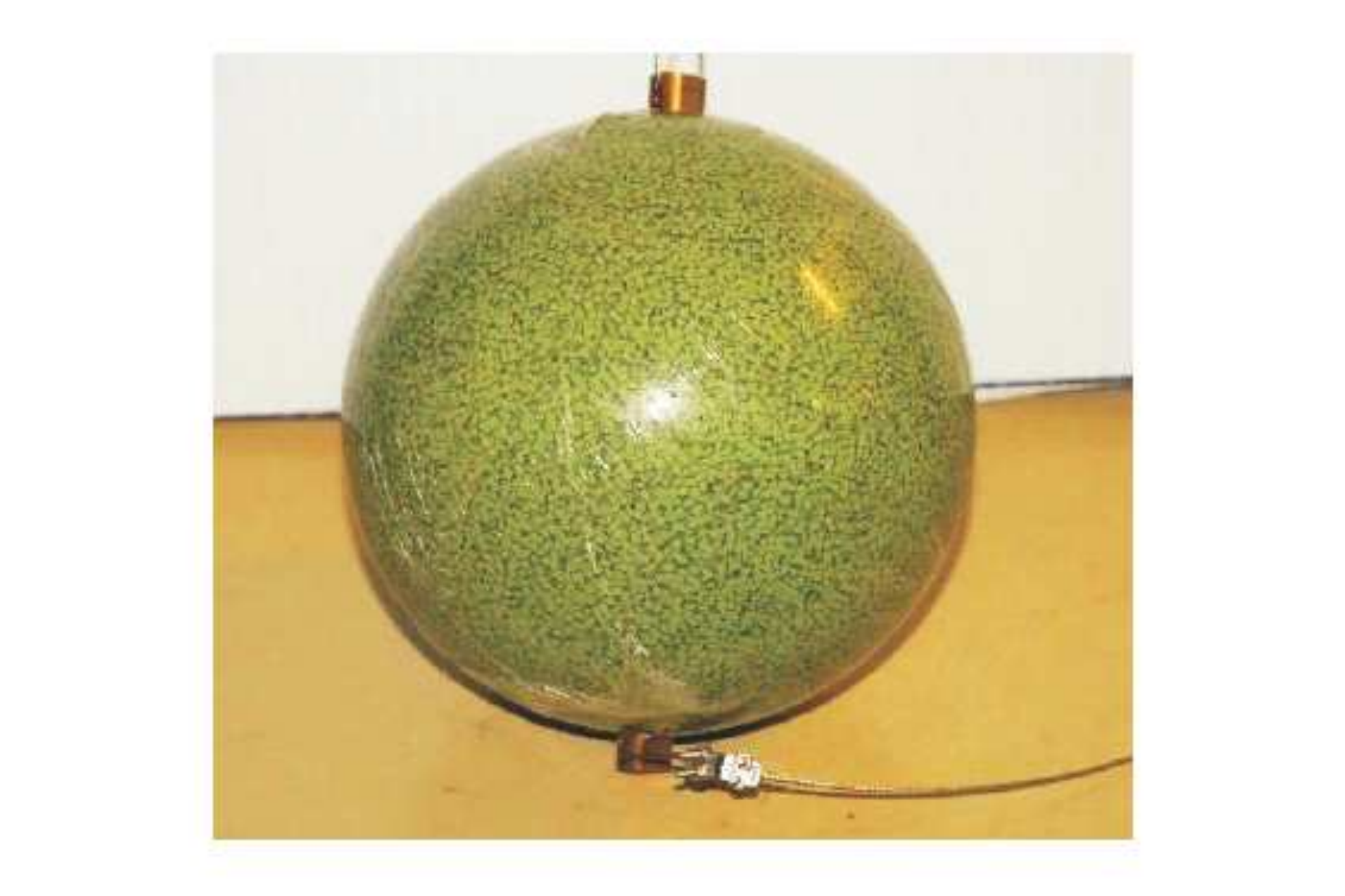}
\caption{Picture of
the nickel source which was manufactured by CI Kogyo.
The ball was made of 6.5~kg of NiO and 3.5~kg of polyethylene.
The Cf source was inserted into the center of the ball and held there by a brass rod.}
\label{fig:nic}
\end{center}
\end{figure}

For this measurement, a uniform and stable source of single-photoelectron
level light is required. We used a ``nickel source'' that isotropically emits
gamma rays. The gamma rays have about 9~MeV from thermal neutron capture on
nickel from the reaction $^{58}$Ni($n,\gamma$)$^{59}$Ni. A $^{252}$Cf source
provides neutrons.
More details can be found in Section~8.7 of~\cite{Fukuda:2002uc}.
The cylindrical geometry and inhomogeneous nickel distribution of the old
nickel source led us to build a new one with significantly improved symmetry.
(Figure~\ref{fig:nic})
Deployed at the center of the ID, this source delivers 0.004\,pe/PMT/event
on average, a level at which more than 99\% of observed signals are due to single-pe.

\begin{figure}[t!]
\begin{center}
\includegraphics[scale=0.46]{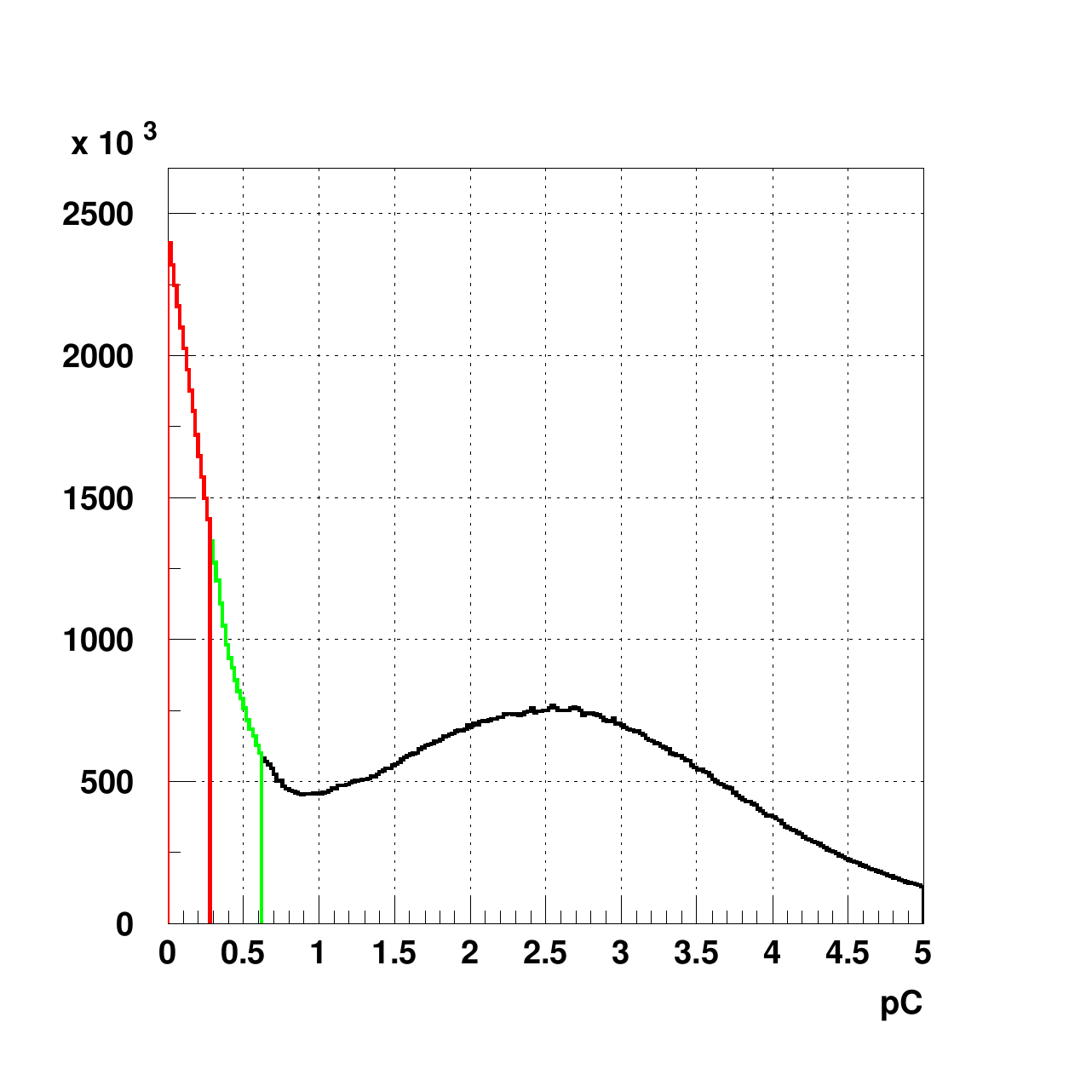}
\includegraphics[scale=0.33]{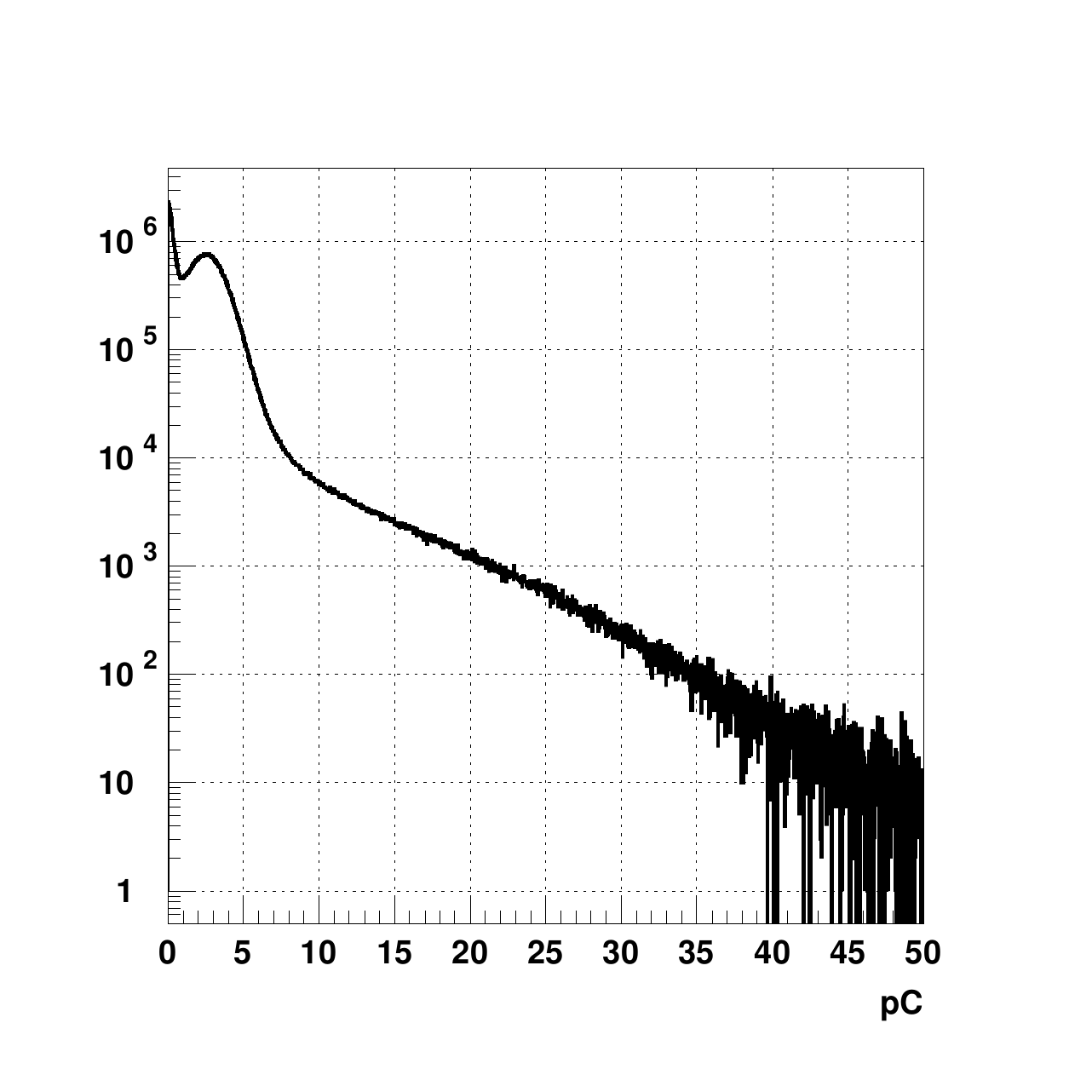}
\caption{The single-pe distributions in pC unit for nickel source data in SK-III. The right plot shows the same histogram in logarithmic scale. The solid line in the left figure shows the data with normal PMT gain, the dashed line shows the data with double gain and half threshold, and the dotted line is linear extrapolation.}
\label{fig:pc2pe01}
\end{center}
\end{figure}

The nickel source measurement was done at the beginning of SK-III.
The resulting charge distribution is shown in the histogram in Fig.~\ref{fig:pc2pe01}
that was obtained after correcting for relative gain differences, as described in Section~\ref{sec:rpc2pe},
and accumulating data from all ID-PMTs. To minimize the effect
of dark hits, a similar distribution was made using off-time (the timing window
in which we do not expect a signal) data and subtracting it
from on-time (timing window in which we do expect a signal) data.
To evaluate the distribution below the usual threshold of 0.25\,photoelectron,
data with higher PMT gain and lower discrimination threshold were obtained.
The dashed histogram in Fig.~\ref{fig:pc2pe01} shows the data with double the
usual PMT gain and half the usual discrimination threshold.
Since it was not possible to obtain data in the region less than 0.3\,pC,
we used a straight-line extrapolation into this low-charge region.
The systematic uncertainty introduced by this assumption
below 0.3\,pC becomes negligible after considering the true discrimination
threshold and the small amount of charge.
The value averaged over the whole pC region was defined as the conversion factor
from pC to single-pe; the value of this conversion factor was 2.243\,pC to single-pe.
At the beginning of SK-IV, we repeated this measurement and found the new
conversion factor to be 2.658\,pC per photoelectron.
This difference comes from a long-term increase in the PMT gain.
No clear reason has been identified for this increase, but it is accounted
for in physics analyses.

\begin{figure}[t!]
\begin{center}
\includegraphics[scale=0.6]{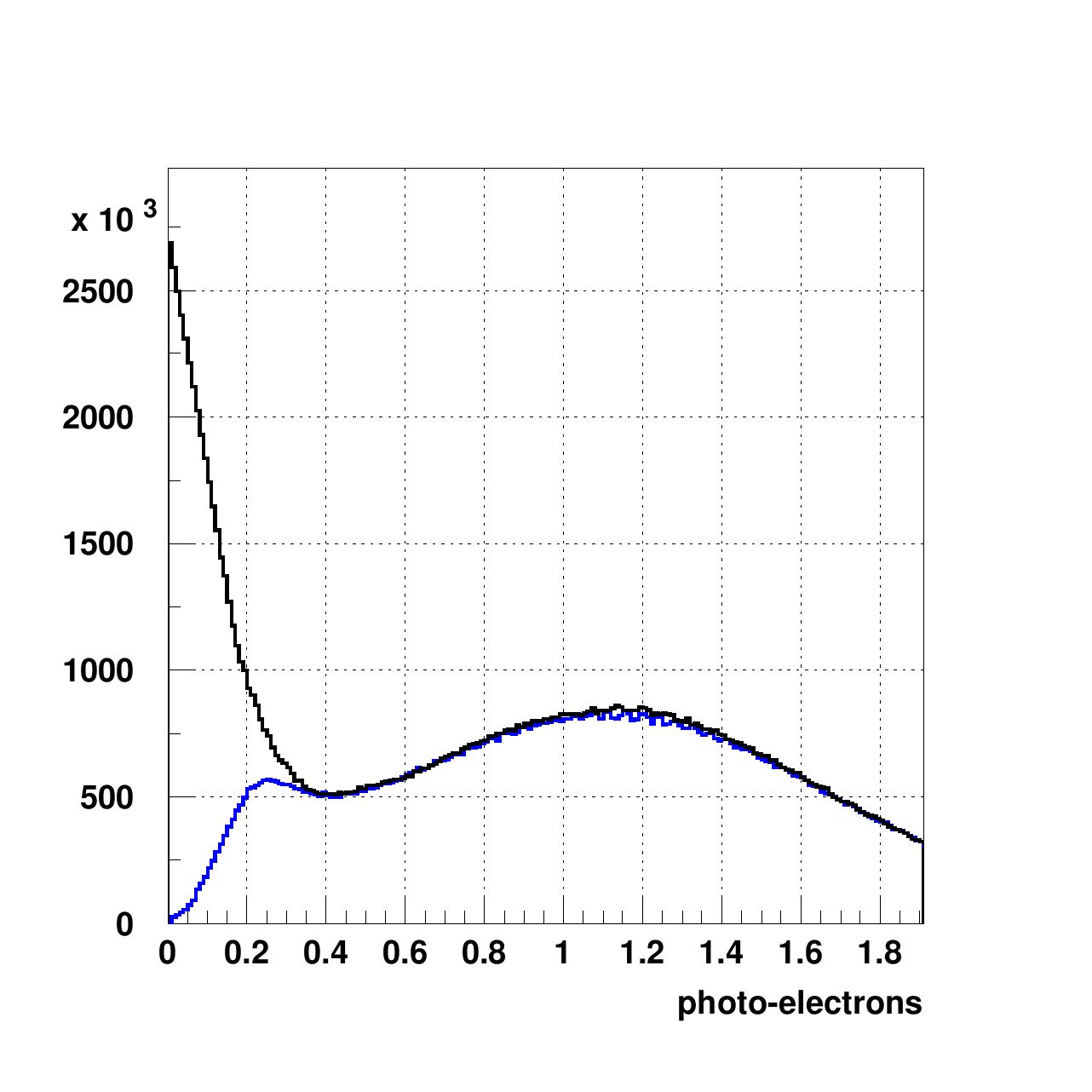}
\caption{The single-pe distributions for our MC simulation (solid line).
This distribution was implemented while we were still using the ATM-based
electronics. 
The dashed line shows the distribution of number of photoelectrons 
from the nickel data in SK-IV. The difference between them is due to the threshold
function of the QBEE, and the ratio is also put into our MC simulation.}
\label{fig:pc2pe02}
\end{center}
\end{figure}

The single-pe distribution, as constructed above, is also implemented in MC
simulations. The solid line in Fig.~\ref{fig:pc2pe02} is
the same as the one we pieced together in Fig.~\ref{fig:pc2pe01},
with the axis converted from pC to photoelectron.
For simulations of multiple photons  in ID-PMTs, we sum values
drawn from this distribution. The nickel-source data are also used
to extract the threshold behavior for MC simulations.
The dashed histogram in Fig.~\ref{fig:pc2pe02} is the experimentally-observed distribution and has the threshold folded into it.
In MC simulations, we use the ratio of the observed (dashed)
and partly observed, partly extrapolated (solid) histograms in Fig.~\ref{fig:pc2pe02}
to implement single-hit threshold behavior.

\subsubsection{Relative differences in QEs}\label{sec:qe}
Values for QE differ from PMT to PMT.
Here we describe how we determine the relative
QE for each PMT.
If the intensity of a light source is low enough,
the observed hit probability should be proportional to the value of QE for the
PMT, as can be seen from Eq.~(\ref{eq:nhit}).
While we can count the number of hits measured by each PMT,
we cannot easily determine how many photons reached it.
Therefore, we used MC simulation to predict the number of photons
arriving at each PMT, and took the ratio of the observed number of hits to
predicted number of hits.

For this measurement, we use the nickel source used in absolute gain
measurements (Section~\ref{sec:apc2pe}).
In addition, the uniformity of water quality throughout the tank is quite
important, since any non uniformity in water properties causes
the hit probability to depend on the PMT position not just because of relative
geometry, but also because of the exact conditions along the photon path.
As discussed in Section~\ref{water-circ}, this condition can be identified by
measuring the temperature profile throughout the ID.
We conducted this calibration when the water convected over the whole ID
volume.\footnote{the data were taken on October 12, 2006 when was just after SK-III started.}
It was also confirmed that no significant top-bottom asymmetry
of water quality existed because no differences in the standard PMTs
appeared between top and bottom.
We used nickel source data from that day for this measurement.

\begin{figure}[htbp]
\begin{center}
\includegraphics[width=12.0cm]{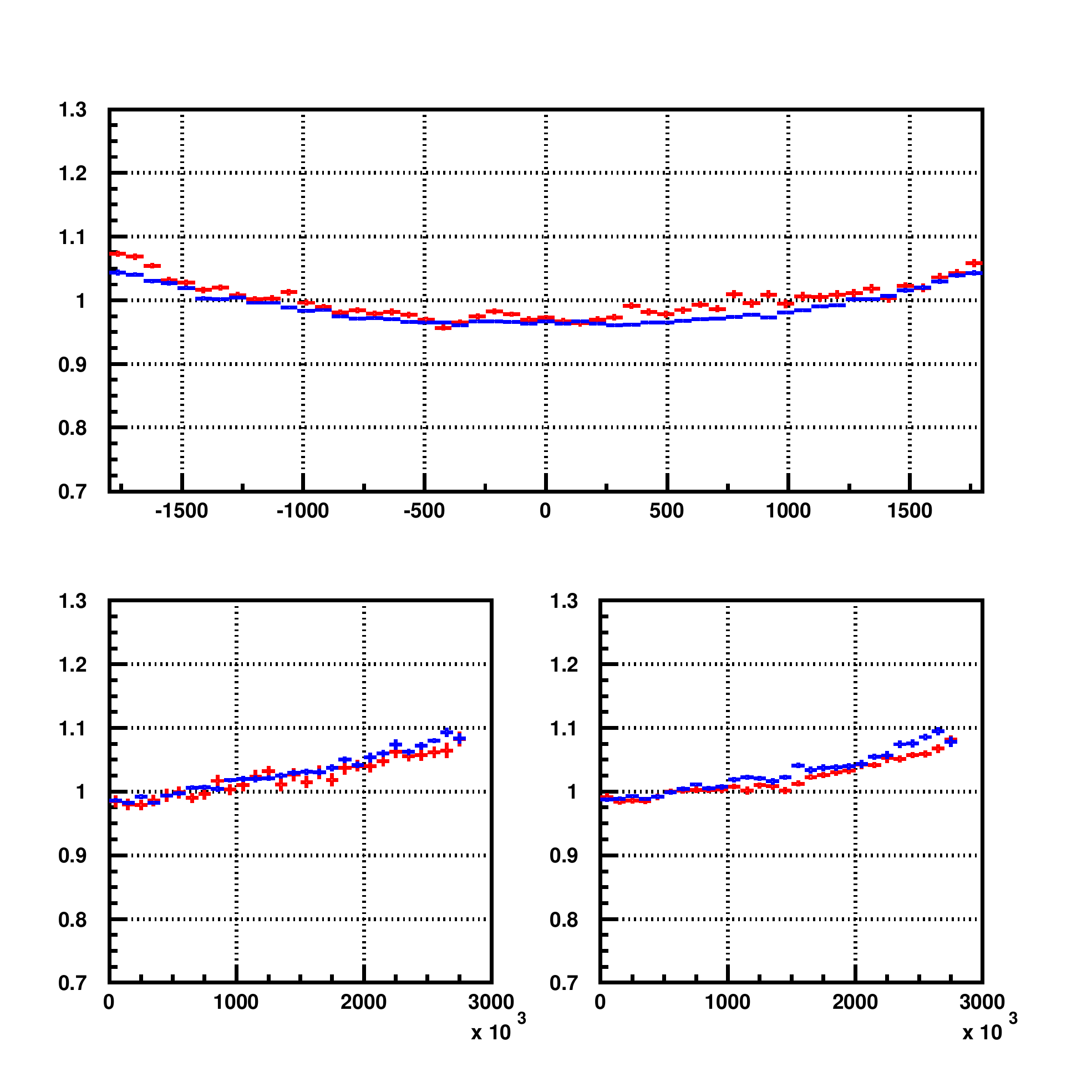}
\caption{Hit probability as a function of PMT position. The vertical axis shows the number of hits normalized by average value of all the PMTs. The upper figure shows the barrel PMTs where the horizontal axis denotes the z (cm) position of PMTs. The lower figures show  top (left) and bottom (right) PMTs, where the horizontal axis shows the square of the distance from the center (cm$^2$). The red thick line shows the data, and the blue thin line shows MC. The MC was not corrected by QE differences in each PMTs.}
\label{fig:qe01}
\end{center}
\end{figure}

Figure~\ref{fig:qe01} shows the position dependence of the hit probability
with the following corrections:
\begin{equation}
N_{obs}(i) \times R(i)^2 / a(\theta(i)),
\end{equation}
where $i$ again indexes the ID-PMTs, $R(i)$ is the distance from the source 
position to the PMT position, and $a(\theta)$ is the acceptance as a function
of incident angle $\theta$ \cite{Hosaka:2006}.
Even after this correction, some position dependence still remains
because of reflection from neighboring surfaces,
plus scattering and absorption by the water.
These further corrections were estimated through Monte Carlo simulations,
which considered absorption and scattering in water
and reflection from the surfaces of the PMTs and the black sheet.
Once these light propagation effects, as estimated by the MC simulations,
are considered, the remaining difference is attributed
to the QE of individual PMTs. To remove the dependence of the absolute light
intensity from the calibration, we normalize taking the average
of these residuals to obtain a measure for the relative QE of the ID-PMTs.
This quantity is tabulated for use in the MC simulations.

\begin{figure}[htbp]
\begin{center}
\includegraphics[width=12.0cm]{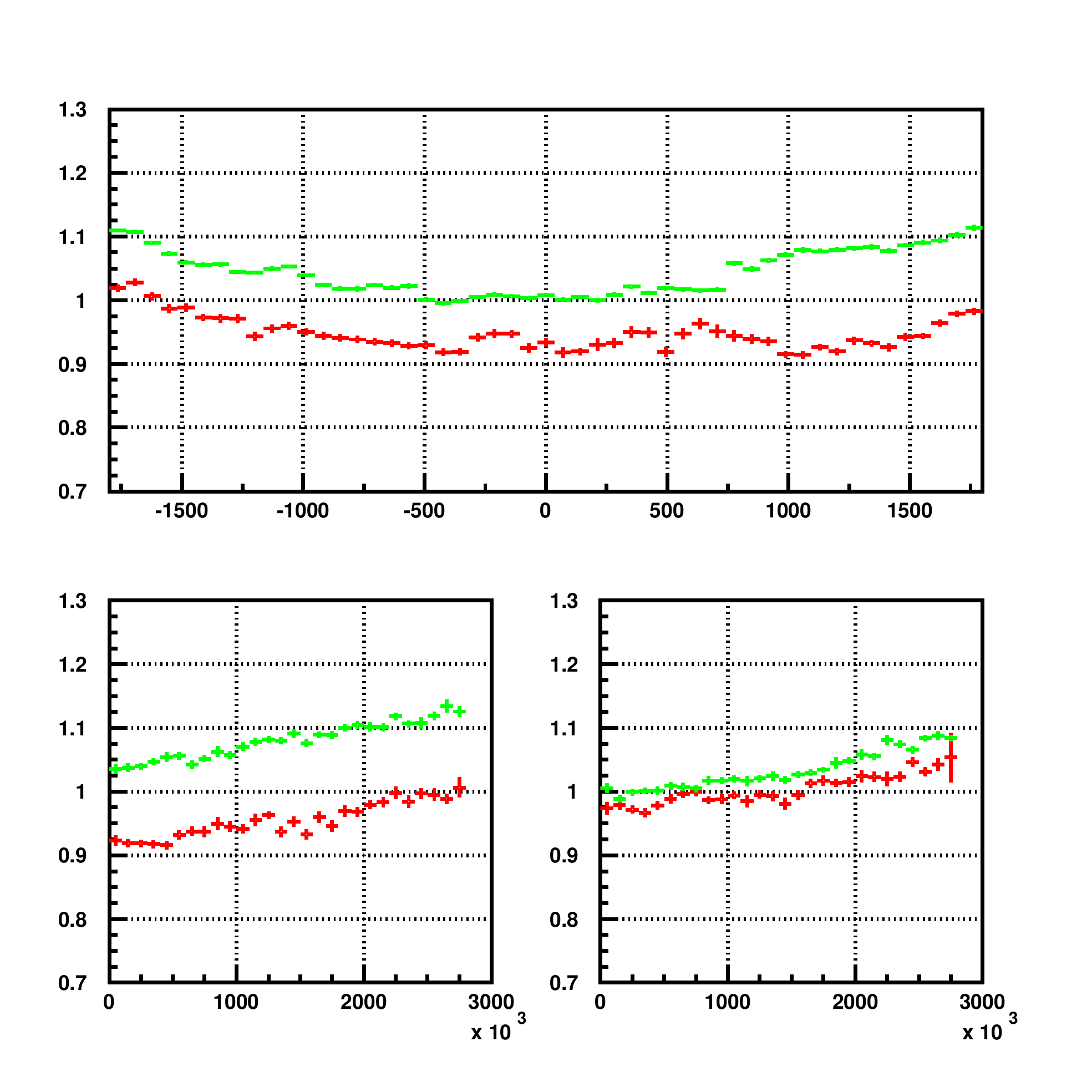}
\caption{Normalized hit probability as a function of PMT position, similar to Fig.~\ref{fig:qe01}. The red thick line shows the PMTs used in SK-II, and the green thin line shows the PMTs newly installed from SK-III.}
\label{fig:qe02}
\end{center}
\end{figure}

The cause of these relative differences among PMTs is interesting.
One cause is, of course, an irreducible variability in the manufacturing
process for the PMTs. We also found some systematic changes related
to the time of manufacturing of the PMTs.
To see this, the PMTs were categorized into two groups: those used in SK-II,
and the newly produced PMTs for the reconstruction of SK after the accident.
The latter were first installed in SK-III. Figure~\ref{fig:qe02}
shows a plot similar to Fig.~\ref{fig:qe01}, but divided into the two
categories; a clear difference can be seen. First, among the PMTs used in SK-II,
the bottom PMTs have $4.5\%$ higher QEs than the top ones.
These PMTs came from different production runs:
the PMTs installed in the bottom region were produced 
in 1996 and 1997, while those in the top and barrel were
produced between 1992 and 1995. The newer PMTs have higher QEs.
An even clearer difference can be seen if we compare PMTs for
SK-II and SK-III. The origin of this systematic difference was
traced to the fact that the manufacturer, Hamamatsu Photonics K.K., 
continued to improve the transparency of the glass envelope.
All the tendencies we observed in 
the QE measurements are supported by
independent measurements of glass properties at the manufacturer.

These findings also have implications for the interpretation of the width
of the relative gain distribution in Fig.~\ref{fig:rel01}. As described
in Section~\ref{sec:hv}, we balanced the response of the ID-PMTs by adjusting
their HV so that, at some light level, each PMT $i$ responded with the
same charge $Q_{obs}(i)$, which can be calculated from its gain
and QE in Eq.~(\ref{eq:Qobs}). According to that equation,
the charge produced by the PMT is proportional to the product of QE and gain.
Because we know that there are systematic differences in QEs between the
PMTs in the detector, we also know that the way we adjusted the HV for an
individual tube compensated for a lower (higher) QE by requiring a higher
(lower) gain (higher (lower) HV) to deliver the same charge at the same light
level 
Later, as described in Section~\ref{sec:rpc2pe}, we took the ratio
of a similar set of high-light-level data to low-light-level data for each
tube to eliminate acceptance and QE effects, thereby obtaining the PMT
relative gain, the distribution of these relative gains is shown in Fig.~\ref{fig:rel01}.
In the next section we show that the procedures discussed above
produce a well-balanced detector response.  

\subsubsection{Overall symmetry of detector response}\label{sec:qegain}
\begin{figure}[htbp]
\begin{center}
\includegraphics[scale=0.46]{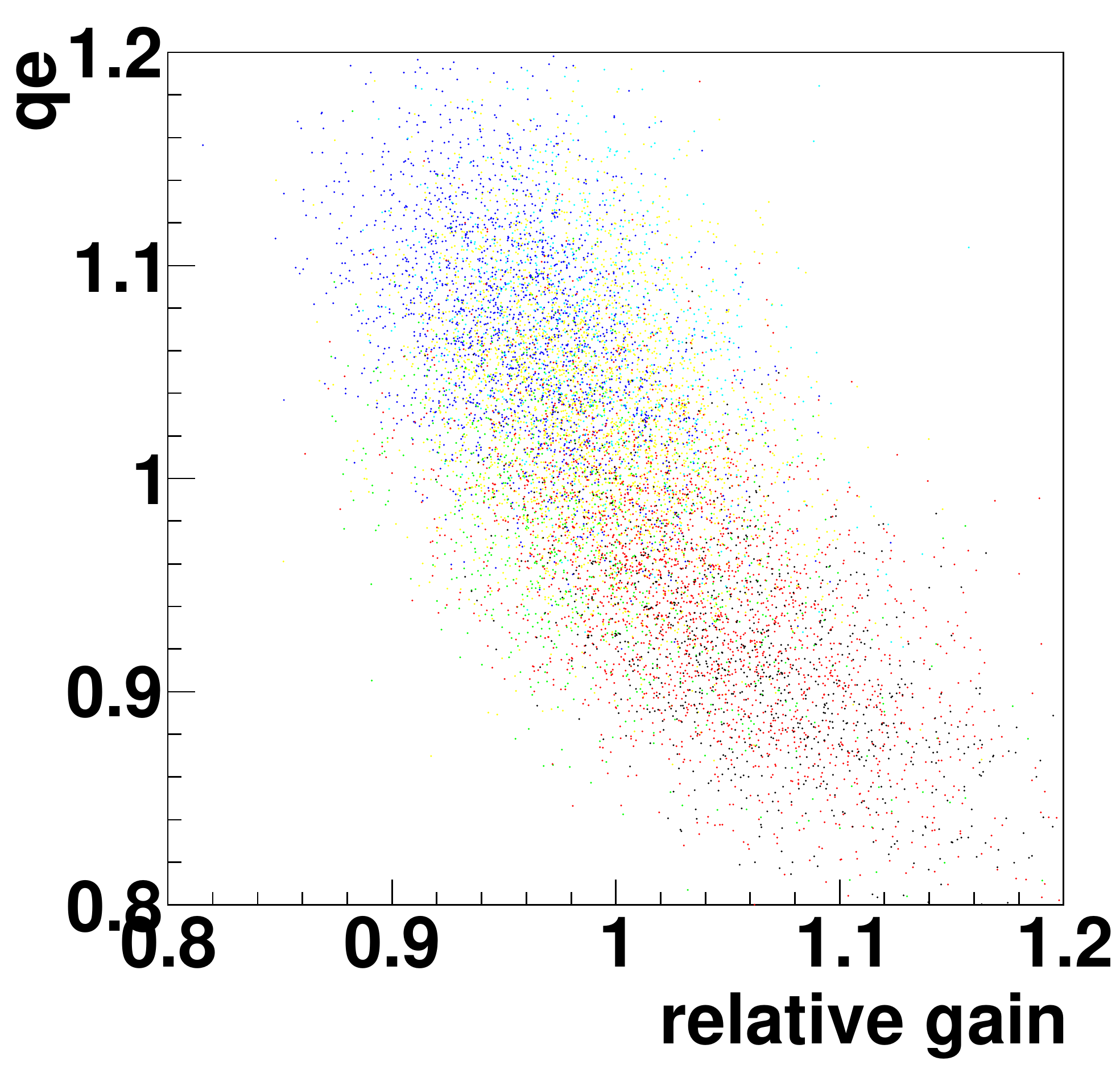}
\caption{The relationship between the relative gains and the QEs. The color shows the PMT production period: black 1992-1993, red 1994-1995, green 1996-1997, sky-blue 2003, blue 2005, and yellow 2006.}
\label{fig:relqe}
\end{center}
\end{figure}

 From the discussion in the previous section, we expect an anti-correlation
between the relative gains and the QEs that our procedures established for the
ID-PMTs and this anti-correlation Fig.~\ref{fig:relqe}.
From the previous discussion, we also know that the distribution of PMTs
with intrinsically higher QE is not uniform over the detector,
since the water pressure at the bottom of the tank is significantly higher than
at the top, we systematically installed the new and presumably less-stressed PMTs
at the bottom and concentrated the older PMTs
at the top of the detector. In this section we describe a dedicated experiment
that was designed to assess whether independent calibrations of the two anti-correlated
quantities combined with this non uniform distribution inside the detector
result in any residual non uniformity.

\begin{figure}[htbp]
\begin{center}
\includegraphics[scale=0.13]{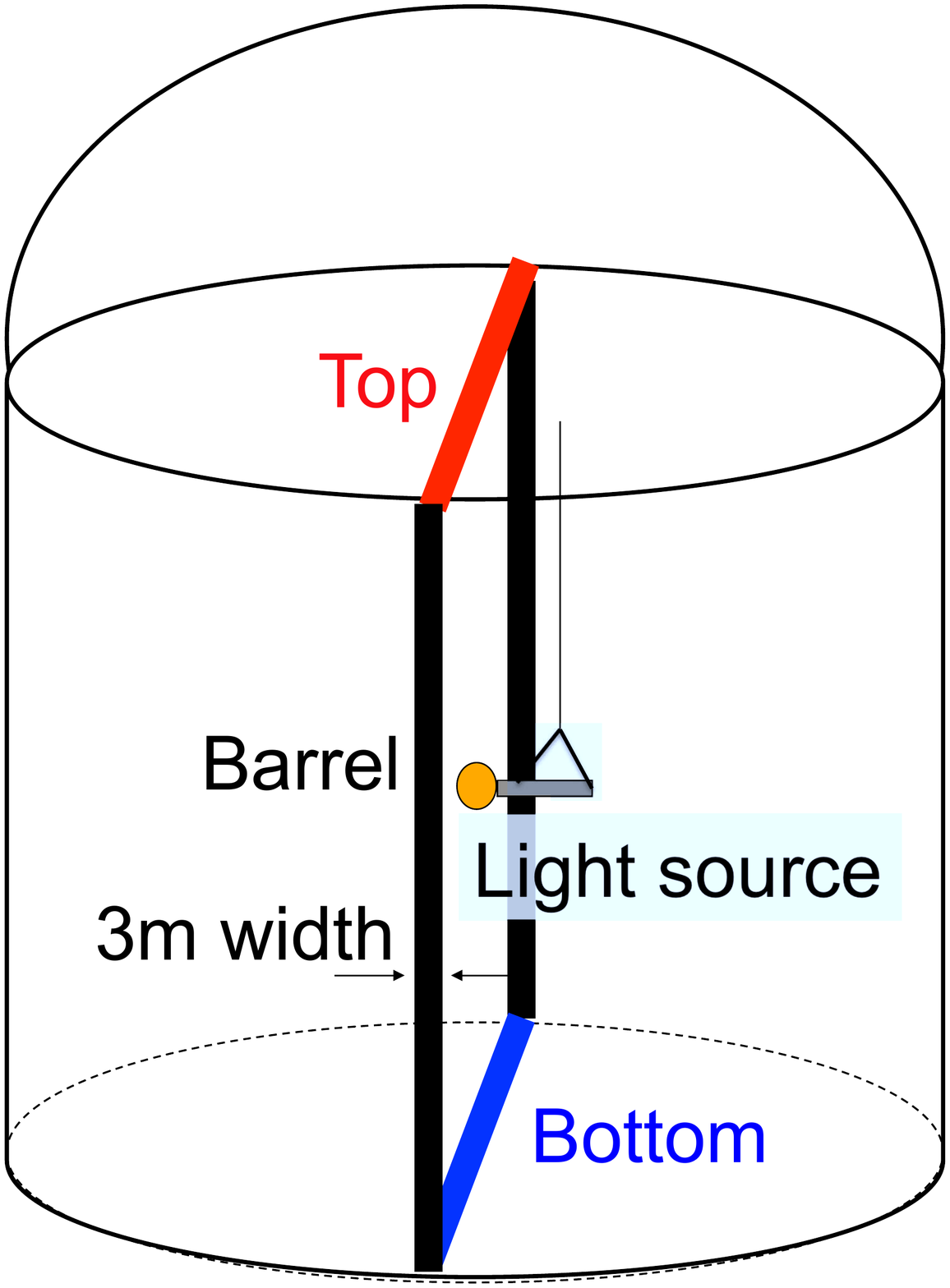}
\includegraphics[scale=0.46]{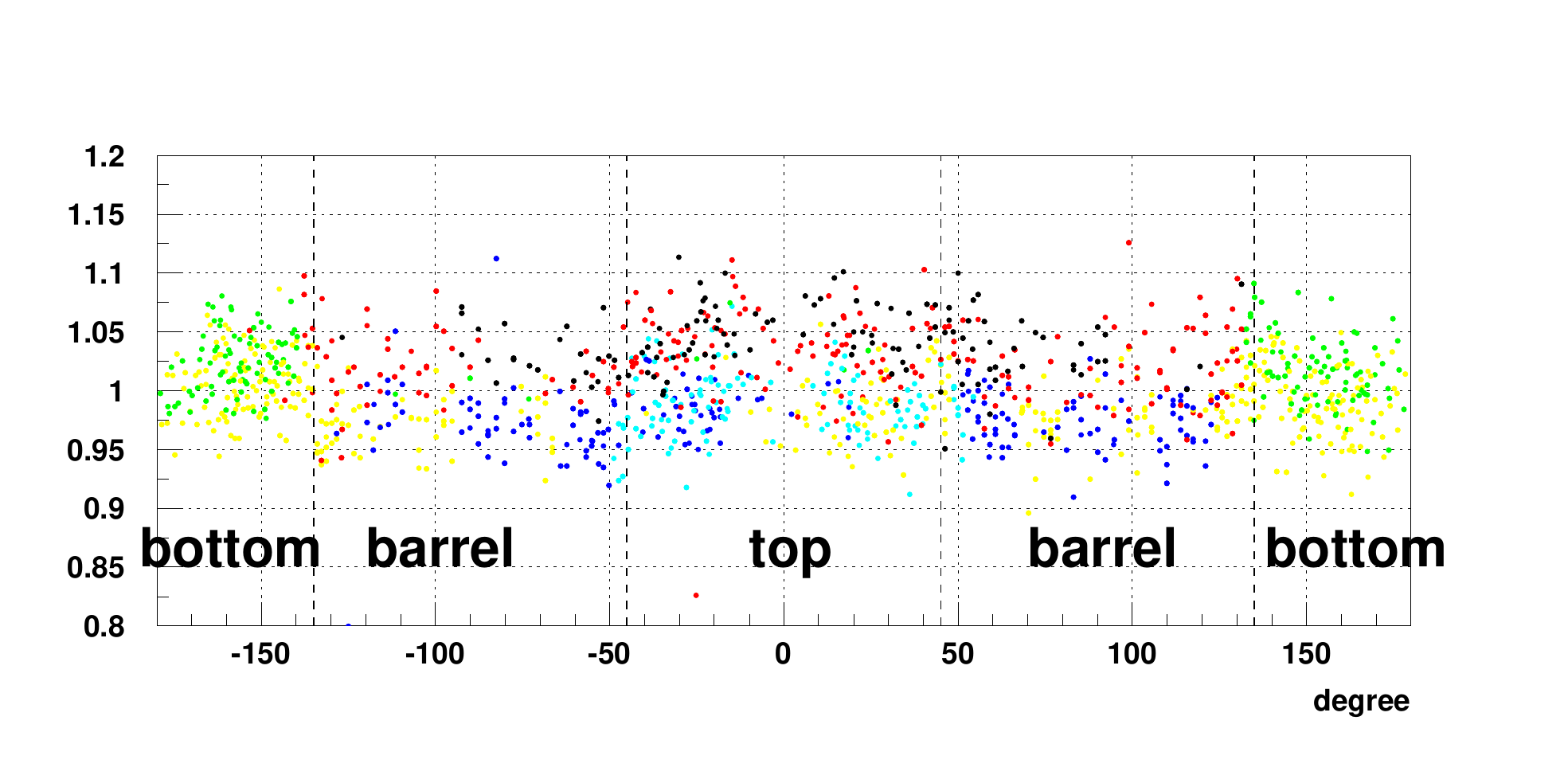}
\caption{The observed corrected charge divided by the gain and QE factor
and normalized by its average over all PMTs.
The left side of the figure shows the setup.
The horizontal axis in the right figure shows the zenith
angle from the top-center point.
The color of the points in the right scatter plot corresponds to the 
PMT production period described in the caption of the Fig.~\ref{fig:relqe}.
}
\label{fig:qe03}
\end{center}
\end{figure}

 The left side of Fig.~\ref{fig:qe03} shows the setup for this experiment.
The light source was the same one used to determine the HV settings:
the scintillator ball driven by Xe flash lamp. As pointed out in
Section~\ref{sec:hv}, the light from the Xe flash lamp is injected
into the scintillator via an optical fiber.
In this measurement, we are mostly concerned with a possible top-bottom
asymmetry in the detector response, therefore, we mounted the scintillator ball
so that the fiber entered the ball in the horizontal rather than
the usual vertical direction (the fiber normally enters straight from above).
In this way, any residual forward/backward asymmetry in the light output
from the scintillator ball would affect the wall region of the ID rather than
the top and bottom. In contrast, during the original HV-setting procedure such
an asymmetry would affect the top and bottom regions.

 The right side of Fig.~\ref{fig:qe03} shows the results from this measurement.
This data was obtained the day before the nickel-source measurement.\footnote{October 11, 2006}
The data confirmed that convection occurred and no significant top-bottom
asymmetry appeared, as described in Section~\ref{sec:qe}.
For the plot in Fig.~\ref{fig:qe03}, we used PMTs that viewed the source
from positions close to perpendicular to the fiber-injection axis of the
scintillator ball. As before, the colors reflect the PMT production period:
the abscissa contains the angle around this injection axis.
For the selected PMTs, the figure shows the following quantity,
normalized to the overall average charge:
\begin{equation}
Q_{obs}(i) \times R(i)^2 / a(\theta(i)) / (\varepsilon_{qe}(i) \times G(i)).
\end{equation}
Here, 
$\varepsilon_{qe}(i)$ and $G(i)$ are as in Eq.~(\ref{eq:Qobs});
the determinations for each ID-PMT are described in Sections~\ref{sec:rpc2pe}
and \ref{sec:qe}, respectively.
The light level used here was equivalent to 30 to 70 photoelectrons\ at the ID-PMTs.
The residual top-bottom asymmetry was 0.8\%. 

\subsubsection{Linearity of charge measurements}\label{sec:lin}
\begin{figure}[htbp]
\begin{center}
\includegraphics[scale=0.5]{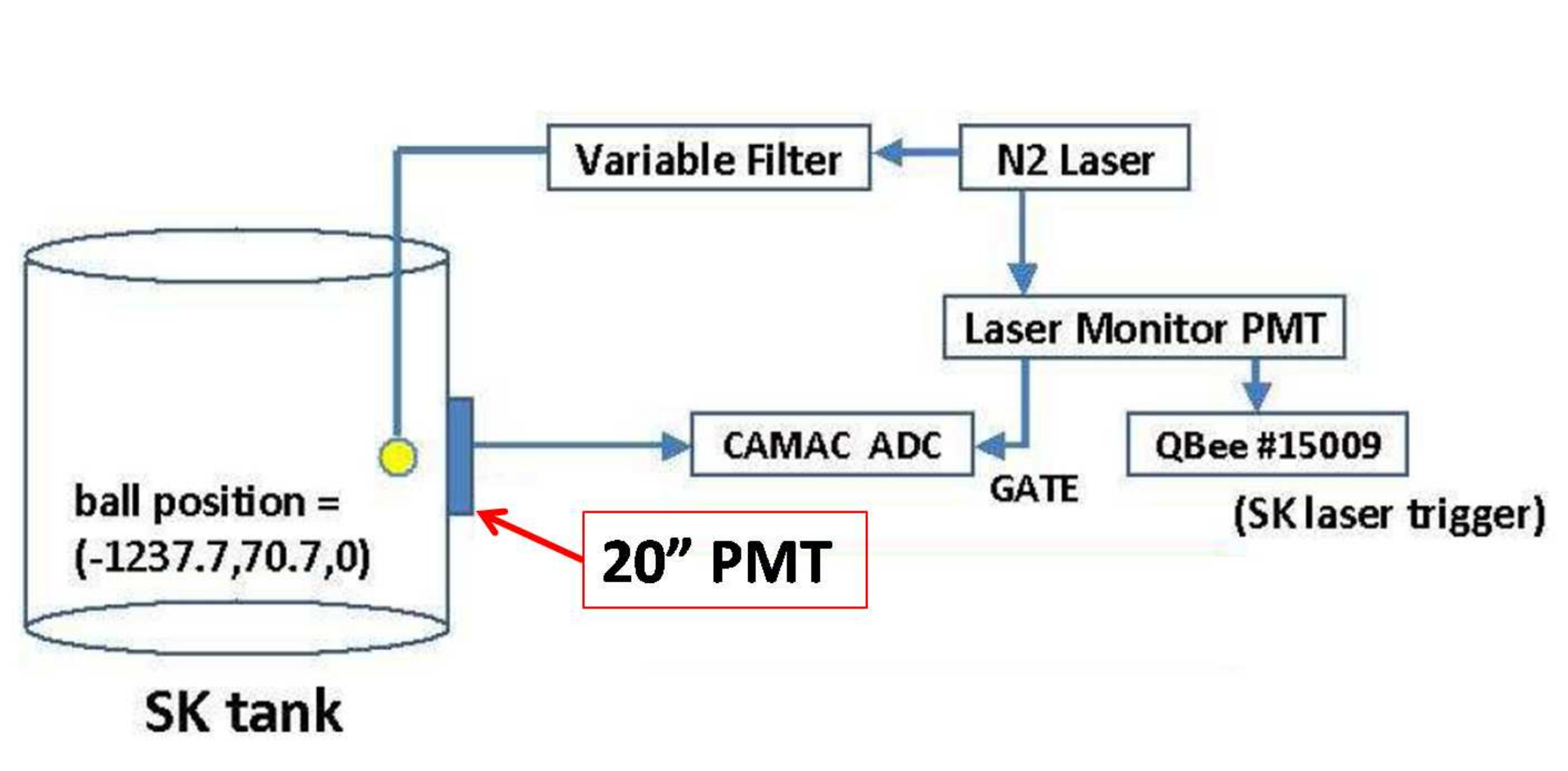}
\caption{Schematic view of the nonlinearity calibration}
\label{fig:nonlin01}
\end{center}
\end{figure}

As discussed in Section~\ref{sec:elec} and Table~\ref{tab:range},
the new QBEEs have 
three different ranges for charge digitization. Testing of the on-board QTC 
chip verified that, within each of these ranges, the linearity of the 
electronics is better than 1\%. However, the overall charge response of the
readout system must also include the PMTs. To test the linearity of the whole
system, we set up a dedicated calibration run, as illustrated in
Fig.~\ref{fig:nonlin01}. The laser system, diffuser ball, and 
laser monitor PMT are the same as used in the timing calibration
(see Section~\ref{timing_calib} for more details). 

The goal of this measurement was to cover the widest possible range of light
intensities, all the way into saturation. Positioning the diffuser ball in 
a highly off-center position towards the barrel of the ID resulted in a highly 
non-uniform light distribution in the detector, and extended the range of 
intensities seen by the ID-PMTs. To reliably monitor such a wide range of
intensities, we deliberately reduced the gains of some ID-PMTs
to establish an intensity ladder as the intensity moved through
overlapping ranges of good linearity for these PMTs, and their electronics.
Ten PMTs in the barrel part of the ID and close to the diffuser ball were selected 
to monitor the light injected into the tank. These ten PMTs were read out 
not with the usual QBEE electronics but with a well-understood, high-resolution
CAMAC ADC. The ten tubes were set to five different gains, resulting
in five monitor ranges with suitable overlap and redundancy.
The overall linearity of this intensity monitor ladder and its CAMAC readout 
electronics was found to be better than 1\%. 

Thirty different laser intensities were
injected into the diffuser ball with the 
help of a neutral-density filter wheel. In this way, the electronics recorded
the charge output for 30 different, but a priori unknown, light intensities 
impinging on the ID-PMTs. Generally, those closer to the off-center diffuser 
ball would be exposed to higher intensities, and those further away would be
exposed to lower 
intensities. The relative intensities for those 30 sets of different-intensity 
laser shots were established with the help of the ten monitor PMTs. Since we 
did not want to involve MC estimates for the contributions from scattered and 
reflected light, we fixed the intensity scale for each individual ID-PMT by 
choosing the intensity for which its recorded charge was closest to 30\,photoelectrons
The justification for this value is that the HV settings for each tube were 
derived from the reference tubes, the HV value of which, in turn, had been 
established at light level resulting in 30\,photoelectrons equivalent charge output. 
Therefore, the linearity check was anchored at roughly the same intensity
as that at which HV settings were established. Using the relative intensities 
derived from the ten monitor PMTs, we can calculate an expected charge 
for each of the other 29 intensities that the laser system delivered. We can
do that for each ID-PMT, starting from its anchor, i.e. from the intensity
at which the charge response of a particular PMT was closest to 30\,photoelectrons

\begin{figure}[htbp]
\begin{center}
\includegraphics[scale=0.6]{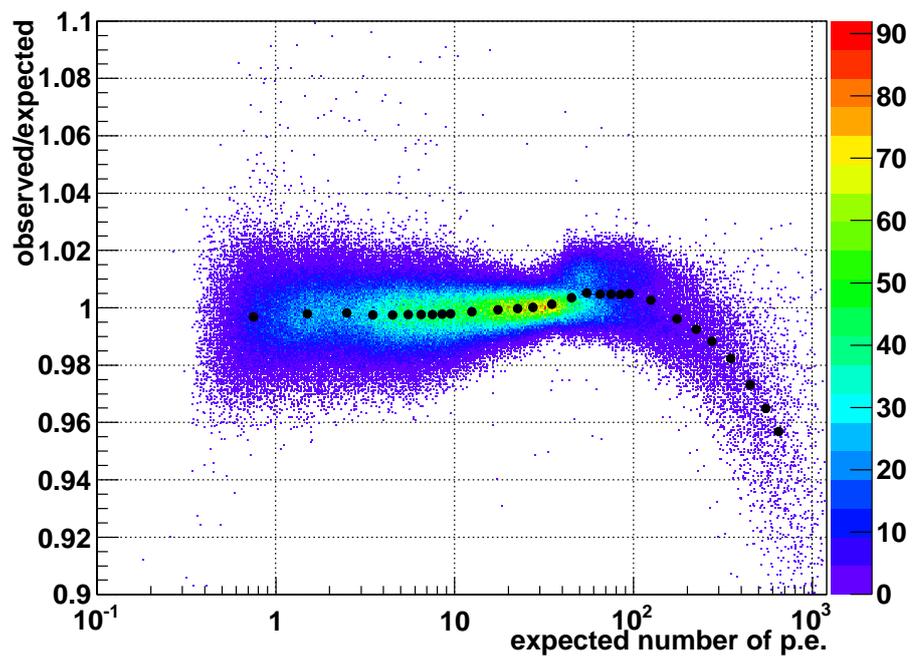}
\caption{
Linearity curve of all the PMTs. The points show the average for each expected number of photoelectrons region. 
}
\label{fig:nonlin02}
\end{center}
\end{figure}

Figure~\ref{fig:nonlin02} shows the result of the linearity measurements.
When less or near 200\,photoelectrons the response of the ID-PMTs is linear within 1\%,
and above 1000\,photoelectrons the deviation from linearity increases to more than 10\%.
As indicated earlier, the
linearity of the new electronics (QBEEs) is better than 1\% over the whole 
range. According to the specifications for the 20-inch PMTs, a 5\% anode 
non-linearity is expected about 250\,photoelectrons, and roughly agrees
with the measurements. Taking the average over the PMTs contributing to a 
certain charge range (indicated by red dots in Fig.~\ref{fig:nonlin02}), this 
non-linearity is considered in MC simulations.

%

\subsubsection{Timing calibration} \label{timing_calib}

The time response of each readout channel, including PMTs and
readout electronics, has to be calibrated for precise
reconstruction of the event vertices and track directions.
The response time of readout channels can vary due to
differences of transit time of PMTs,
lengths of PMT signal cables, and processing time of
readout electronics.
In addition, 
the response time of readout channels (the timing of discriminator output)
depends on the pulse heights of PMTs, since the rise time of a large 
pulse is faster than
that of a smaller one. 
This is known as the ``time-walk'' effect.
The overall process time and the time-walk effect can be calibrated by
injecting fast light pulses into PMTs and by varying the intensity
of light.

\begin{figure}[htbp]
  \begin{center}
    \includegraphics[width=0.4\hsize]{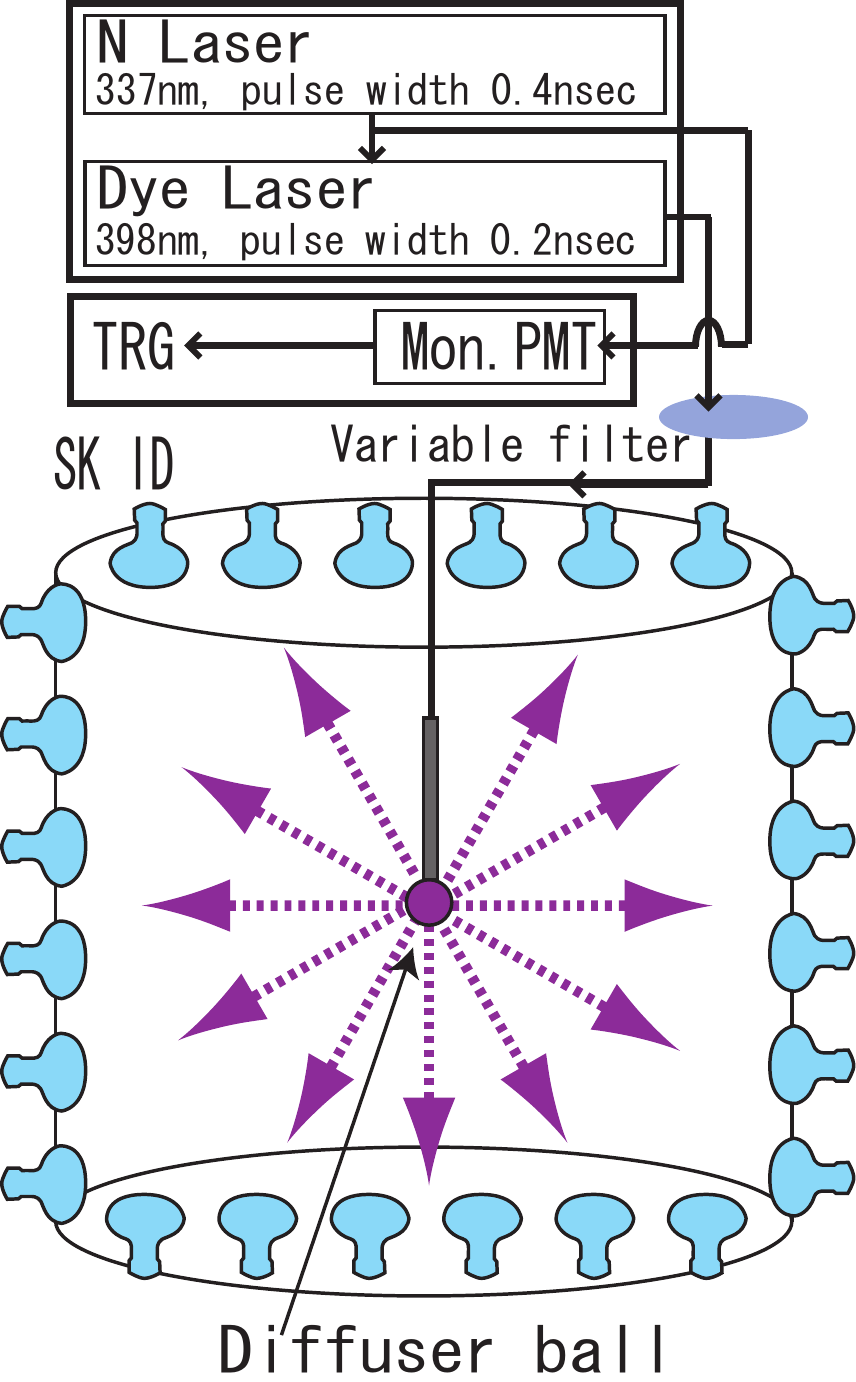}
    \caption{Schematic view of the timing calibration system.
      \label{fig:tqcalib_schem}}
  \end{center}
\end{figure}

\begin{figure}[htbp]
  \begin{center}
    \includegraphics[width=0.4\hsize]{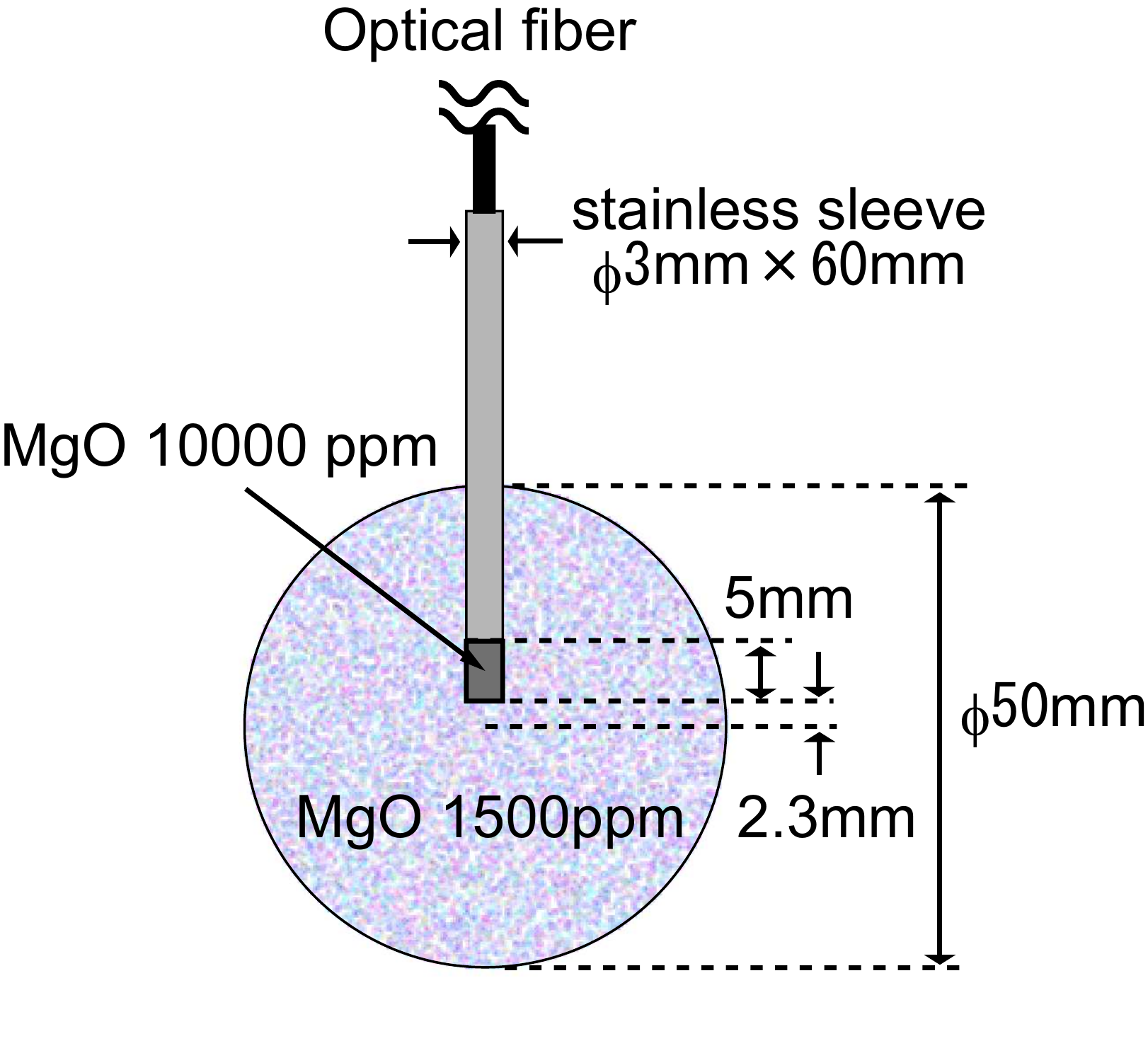}
    \caption{Cross section of the diffuser ball.\label{fig:tq_diffuser}}
  \end{center}
\end{figure}

Figure~\ref{fig:tqcalib_schem} shows a schematic diagram of the SK timing
calibration system.
The SK uses a nitrogen laser as a light source for timing calibrations.
The nitrogen laser (USHO KEC-100) is a gas flow laser that emits fast pulsing
light of 0.4\,ns FWHM at a wavelength of 337\,nm.
The laser output is monitored by a fast response 2-inch PMT (Hamamatsu H2431-50,
rise time: 0.7\,ns, T.T.S: 0.37\,ns). 
This monitor PMT is used to define the time of laser light injection.
The wavelength of the laser light is shifted to 398\,nm by a dye,
where the convoluted response with
Cherenkov spectrum, light absorption spectrum and quantum
efficiency of the PMTs is almost maximum.
The pulse width of the dye is 0.2\,ns FWHM.
The light intensity is varied by a variable optical filter.
We use the optical filter to measure the time responses of readout
channels at various pulse height.
The filtered light is guided into the detector by an optical fiber (400\,$\mu$m core)
and injected into a diffuser ball located near the center of the tank 
to produce an isotropic light.
Figure~\ref{fig:tq_diffuser} shows a cross section of the diffuser ball.
Directional variations in the photon emission time of the diffuser ball were
measured to be less than 0.2\,ns.

\begin{figure}[htbp]
  \begin{center}
    \includegraphics[width=0.75\hsize]{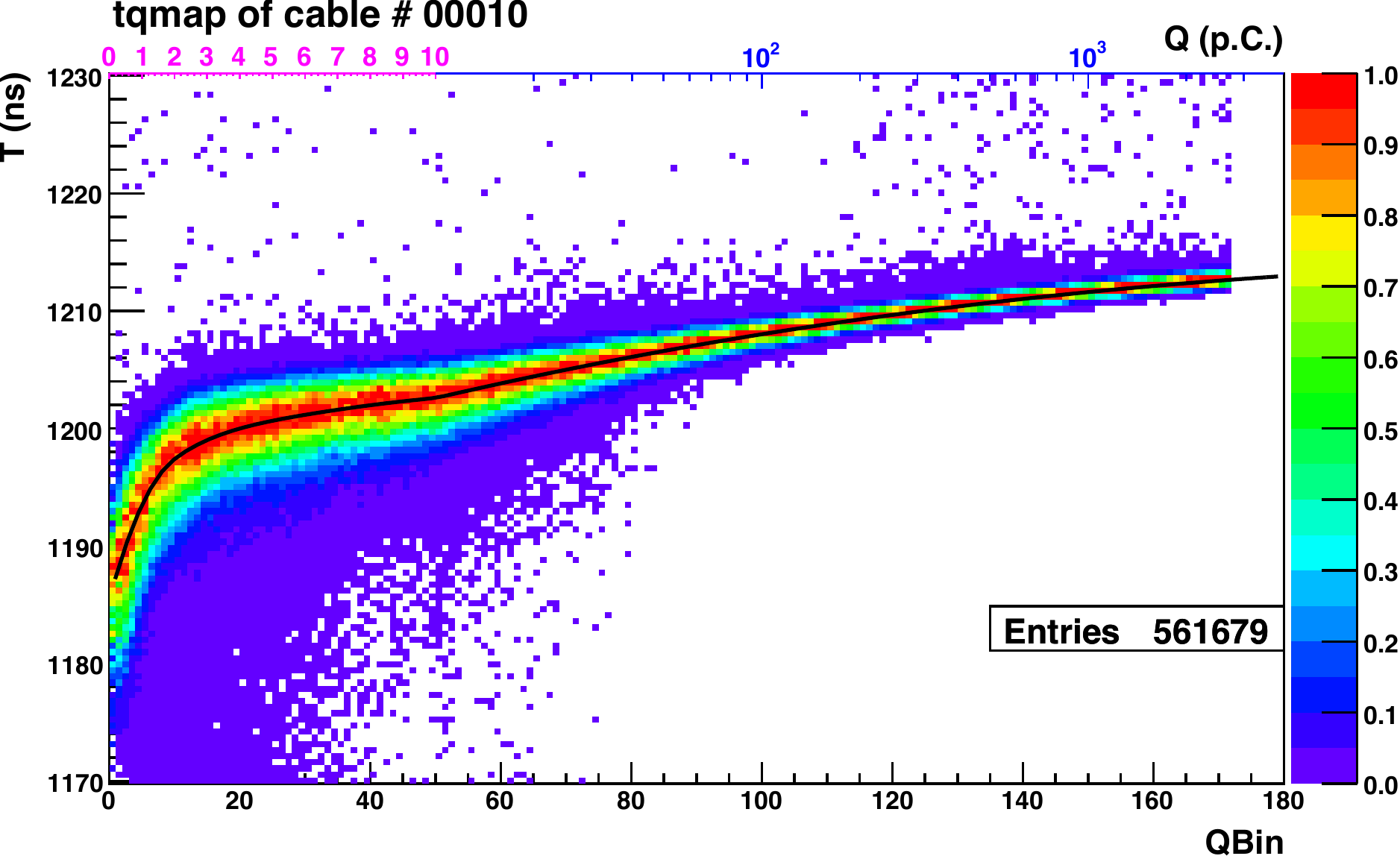}
    \caption{Typical TQ distribution for a readout channel.
      The horizontal axis is charge (Qbin) of each hit,
      and the vertical axis is time-of-flight-corrected timing (T) of the hits.
      Larger (smaller) T corresponds to earlier (later) hits in this figure.
      \label{fig:tqmap}}
  \end{center}
\end{figure}

Timing calibrations for SK ID readout channels were conducted based on two-dimensional,
timing versus pulse height (charge), correlation tables that are called ``TQ distributions''.
Figure~\ref{fig:tqmap} shows a typical scatter plot of the TQ distribution for one readout channel.
The calibration constants, called the ``TQ-map'', are derived by fitting 
the TQ distribution to polynomial functions.
A TQ-map includes overall process time and the time-walk effect;
each readout channel has its own TQ-map.

In the laser event selection, we require that the monitor PMT is fired.
The fired timing defines the reference timing of the laser flashed.
For laser events, we apply a timing correction, called
time-of-flight (TOF) correction, that subtracts
time of flight from the diffuser ball to the respective
PMT position using group velocity of light
with the measured wavelength, $\sim$398~nm.
Using the TOF-corrected hit timing, ``laser hits'' are defined as hits
in a time window $\pm50$~ns around the hit timing of the monitor PMT.

The selected laser hits of each readout channel are divided into
180~bins of charge, called ``Qbin''s.
Each Qbin is defined as the amount of charge from the PMT in pC;
they are defined on a linear scale from 0 to 10~pC (0.2~pC/Qbin) and on
a logarithmic scale from 10 to 3981~pC ($50 \log({\rm pC})$/Qbin).
After the TQ distributions are divided into 180 Qbins,
the timing distribution is smoothed by a Gaussian to minimize
statistical fluctuations.
Although the timing distribution in each Qbin is almost Gaussian, the timing
distributions have a small asymmetric feature because of
an asymmetric time response of PMT and
contributions from direct and indirect light;
direct light causes early hits, while indirect light causes late hits
due to reflection and scattering.
In order to take the asymmetric feature into account,
the timing distribution in each Qbin is fitted to an asymmetric Gaussian, which
provides the peak timing and standard deviation.
The peak timing and standard deviations for respective charges are fitted by
polynomial functions depending on Qbin:
{\small
\begin{eqnarray}
  polN(x) &\equiv& p_{0} + p_{1}x + p_{2}x^2 + ... + p_{N}x^N \label{eq:tqcurv_0} \\
  {\rm Qbin} \leq 10: F_{1}(x) &\equiv& pol3(x), \label{eq:tqcurv_1}\\
  {\rm Qbin} \leq 50: F_{2}(x) &\equiv& F_{1}(10) + (x-10) \cdot
  \left[ F'_{1}(10) + (x-10) \cdot pol3(x-10) \right], \label{eq:tqcurv_2}\\
  {\rm Qbin} >  50: F_{3}(x) &\equiv& F_{2}(50) + (x-50) \cdot pol6(x-50).
  \label{eq:tqcurv_3} 
\end{eqnarray}
}
where $F'_{1}$ in Eq.~(\ref{eq:tqcurv_2}) is a derivation of $F_{1}$, that is introduced
for continuity between $F_{1}(x)$ and $F_{2}(x)$ at ${\rm Qbin}=10$.
In Eq.~(\ref{eq:tqcurv_2}) and Eq.~(\ref{eq:tqcurv_3}), $F_{1}(10)$ and $F_{2}(50)$ are
introduced to satisfy the boundary conditions to connect $F_{i}(x)$ ($i=1, 2, 3$) at
${\rm Qbin}=10$ and ${\rm Qbin}=50$.
$F_{1}(x)$ and $F_{2}(x)$ have 4 fit parameters each, and 7 fit parameters in $F_{3}(x)$.
Thus, the number of fit parameter is 15 in total.
The parameters resulting from the fit are saved in a database as the TQ-map
and are used to correct the time response of each readout channel.

\begin{figure}[htb]
  \begin{center}
    \includegraphics[width=0.7\hsize]{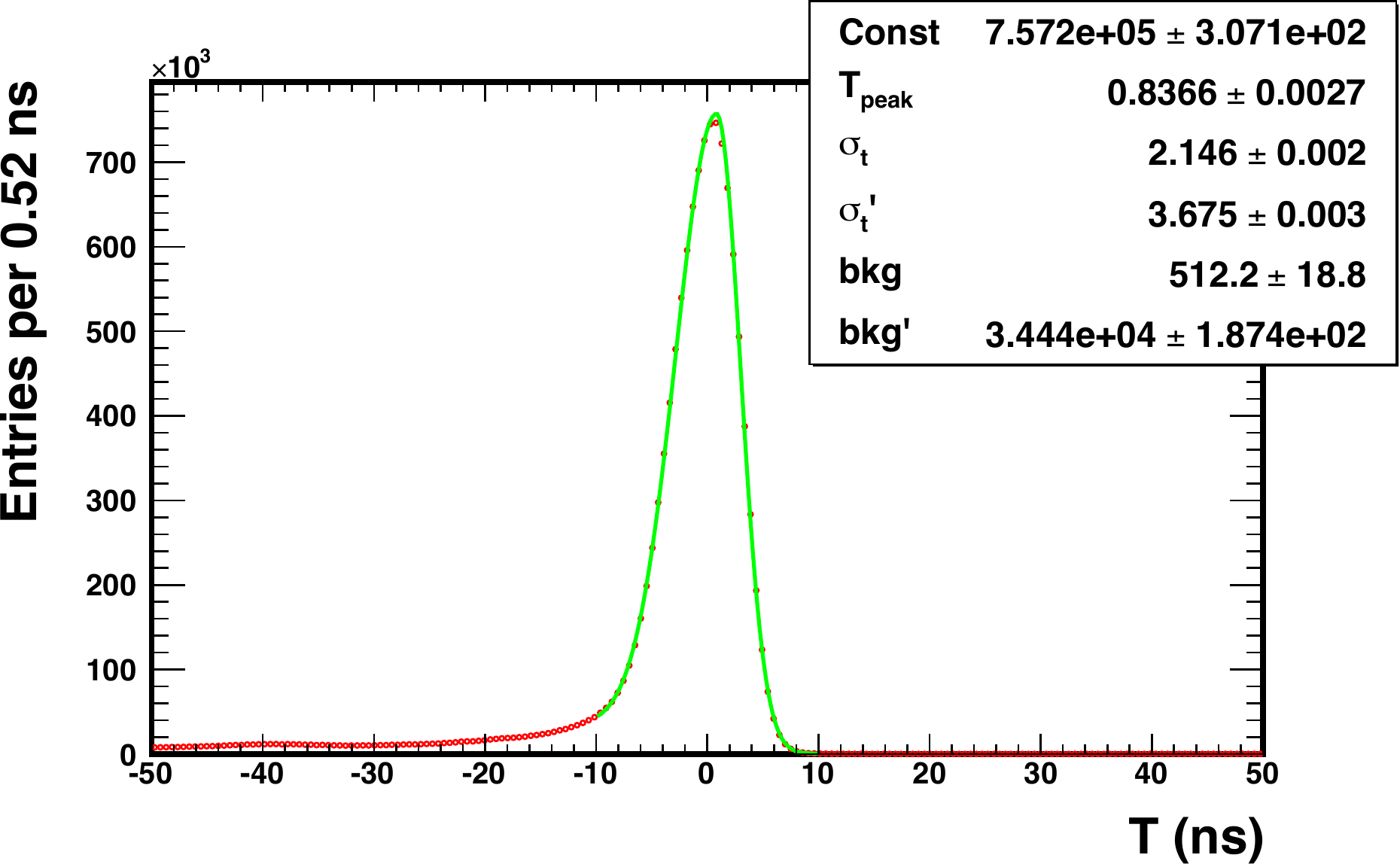}
    \caption{
      Timing distribution added over all the readout channels in ${\rm Qbin}=14$
      ($\sim1$~photoelectron).
      The result of the fit to an asymmetric Gaussian is shown by the solid curve.
      \label{fig:tq_sigfit}
    }
  \end{center}
\end{figure}

The timing resolution of the SK detector is evaluated using the same data set as
used for the TQ-map evaluation.
To evaluate the timing resolution, 
all PMT timing distributions, corrected by their TQ-maps, are accumulated in each Qbin
and the timing distributions in Qbins are fitted by an asymmetric Gaussian
that is defined,
\begin{eqnarray}
  f(t; t > T_{peak}) \equiv
  A_{1} \cdot \exp(-(t - T_{peak})^2 / \sigma_{t}^2 ) + B_{1},\\
  f(t; t \leq T_{peak}) \equiv
  A_{2} \cdot \exp(-(t - T_{peak})^2 / \sigma_{t}'^2 ) + B_{2},
\end{eqnarray}
where $A_{i}$, $B_{i}$ ($i=1,2$), $\sigma_{t}$ and $\sigma_{t}'$ are fit parameters
(note that, in these equations, a larger $t$ corresponds to earlier hits).
The fit parameters need to satisfy a boundary condition,
$A_{1}+B_{1}=A_{2}+B_{2}$, to connect two Gaussian functions at $t=T_{peak}$.
As an example, Fig.~\ref{fig:tq_sigfit} shows the timing distribution and the function resulting
from the fit for ${\rm Qbin}=14$.

\begin{figure}[htbp]
  \begin{center}
    \includegraphics[width=0.7\hsize]{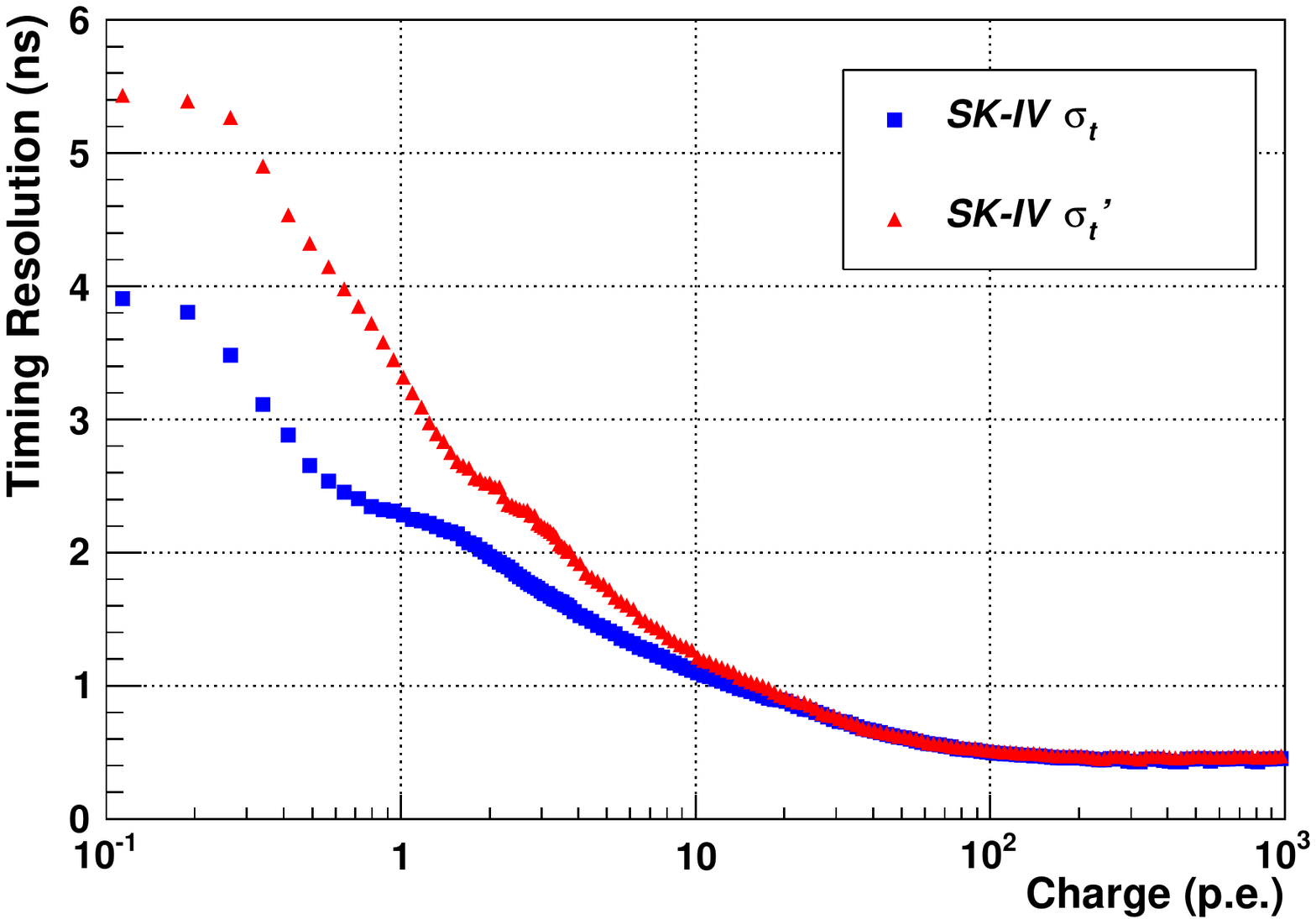}
    \caption{
        Timing resolution $\sigma_{t}$ (square) and $\sigma_{t}'$ (triangle) as a function
        of charge (photoelectron) for SK-IV.
      \label{fig:tq_tresolution}
    }
  \end{center}
\end{figure}

Figure~\ref{fig:tq_tresolution} shows the timing resolution, i.e. the standard deviation
$\sigma_t$ and $\sigma_t'$, as a function of charge.
They are implemented in the SK detector simulation.
By implementing both $\sigma_{t}$ and $\sigma_{t}'$, SK detector simulation reproduces
timing distributions of data, e.g. LINAC calibration\footnote{The details of SK LINAC calibration can be found in reference~\cite{Fukuda:2002uc}.} data,
better than by implementing only $\sigma_{t}$.

The task of real-time monitoring is to monitor the deviation of the time-response of
readout channels during  SK detector operation, and to apply a correction to the
TQ-map as necessary.

For the real-time monitoring of the time response of readout channels, SK employs a
nitrogen laser (Laser Science Instruments, VSL-337 and DLM-120 dye laser system).
This laser system uses  sealed nitrogen as a gain medium and is better suited for continuous operation than the gas-flow laser that used for the TQ-map evaluation.
This laser also emits fast pulsing light of less than 4\,ns FWHM, and the wavelength of
the light is 337.1\,nm.
The wavelength of the laser light is shifted by a dye system to 398\,nm.
The light output of the dye system is injected into the same diffuser ball as used for TQ-map evaluation.

For real-time monitoring, SK uses a fixed light intensity that is adjusted
to obtain an averaged SK ID occupancy of $\sim99$\% and an averaged charge
of $\sim20$~photoelectrons per readout channel at maximum.
The frequency of the flashing laser light is $\sim0.03$~Hz, and a typical 24-hour
operation provides a  statistical accuracy of $\sim0.05$\,ns for monitoring
the time response of the readout channels.

\begin{figure}[htbp]
  \begin{center}
    \includegraphics[width=0.65\hsize]{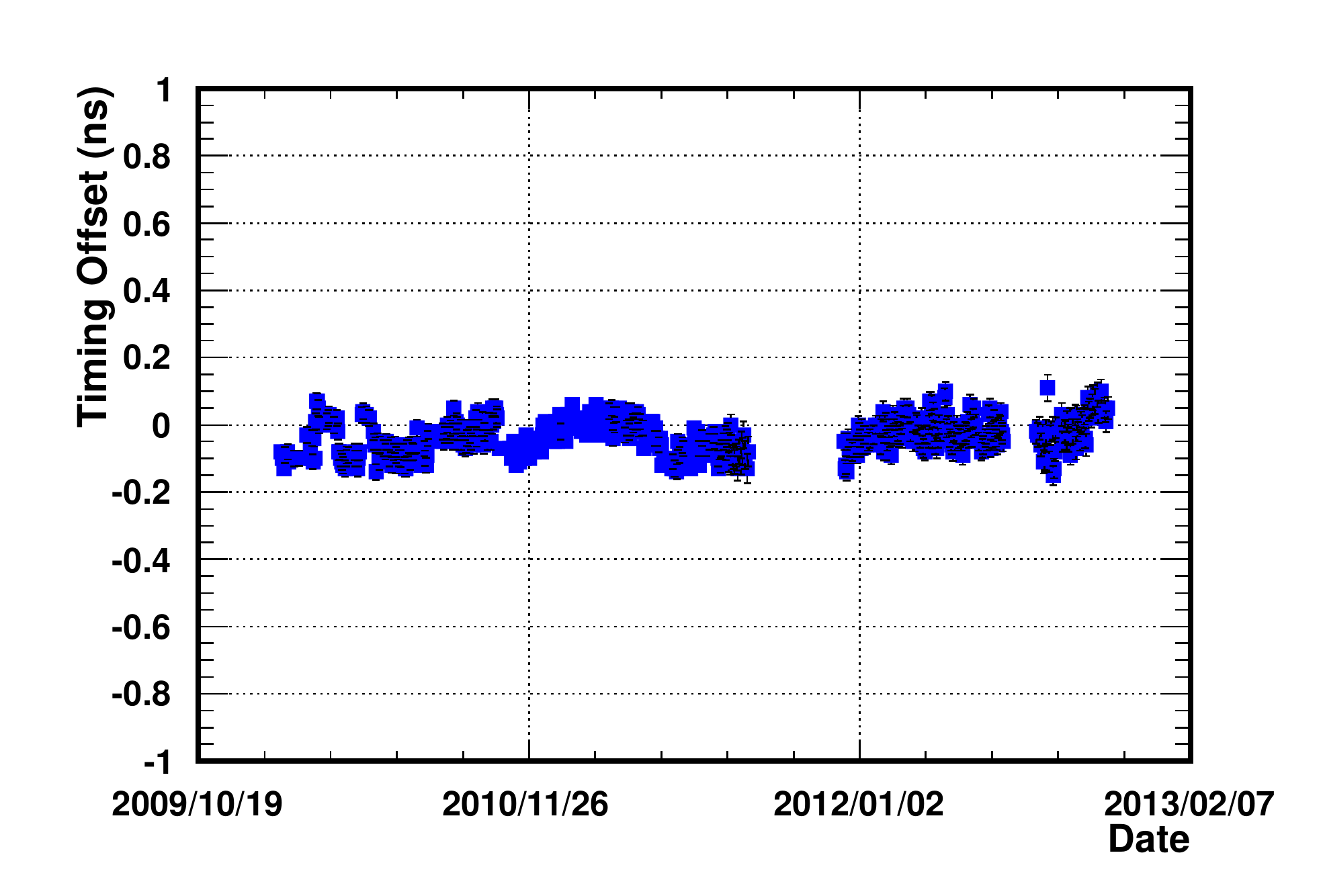}
    \caption{Timing-offset values with respect to pre-defined TQ-map for a typical
      readout channel as a function of time. The error bar on each
      point indicates statistical uncertainty.
      \label{fig:tq_toff_history}}
  \end{center}
\end{figure}

Figure~\ref{fig:tq_toff_history} shows timing-offset values as a function of time
for a typical readout channel; here, the timing offset is defined as the difference
between the measured time response from real-time monitoring and that from pre-defined TQ-map.
As shown in the figure, the time responses of readout channels have been stable 
within $\pm0.1$\,ns for a few years.

\subsection{Calibration for photon tracking}\label{sec:track}

In this section, we summarize results from measurements of water quality
and light reflection at the ID wall used in SK detector simulations (SK-MC).

\subsubsection{Light absorption and scattering in water}
\label{waterparam}

\begin{figure}[!hbtp]
\centering
\includegraphics[totalheight=0.35\textheight]{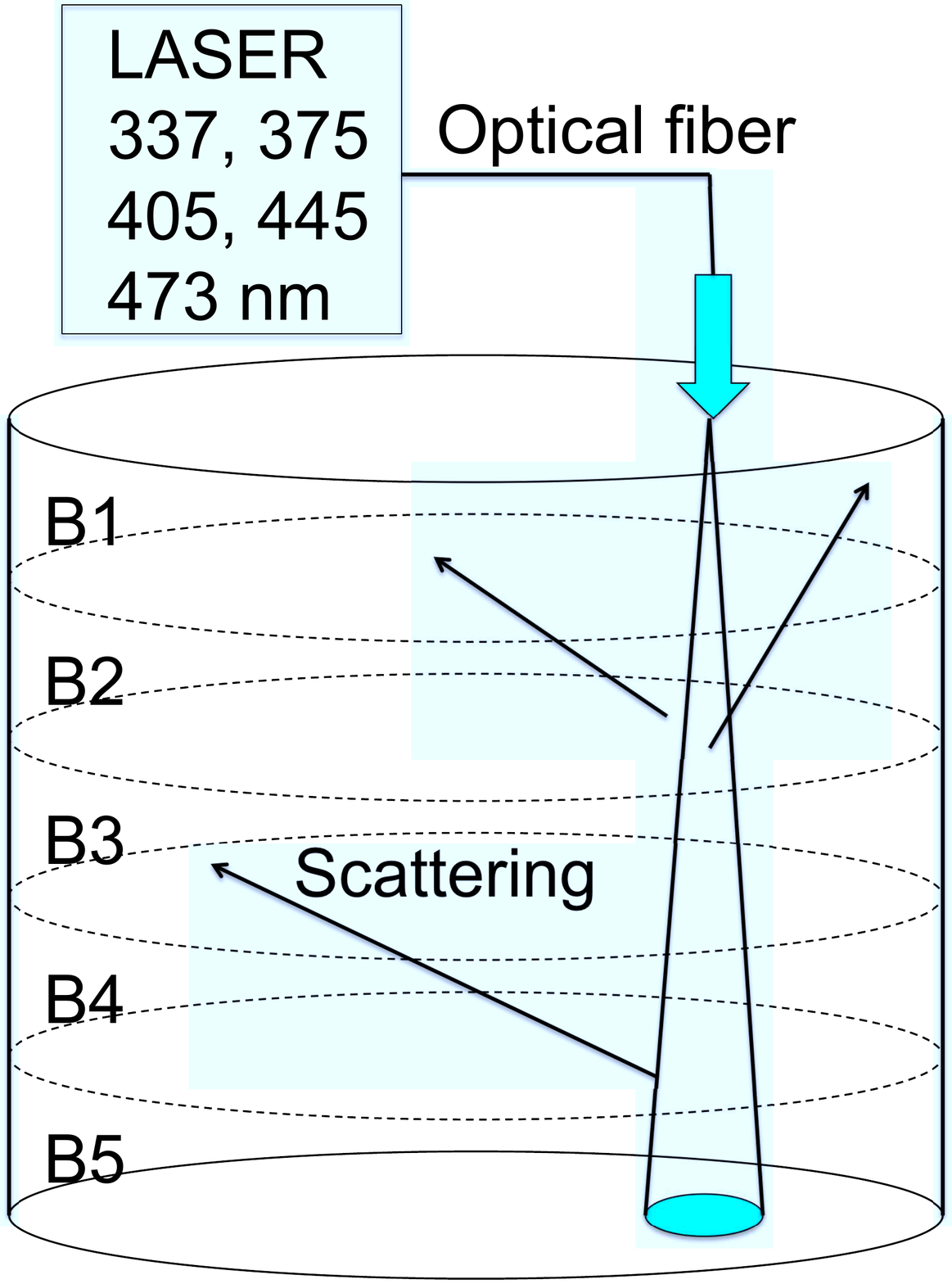}
\caption{
Real-time laser system for measurements of absorption and scattering of the Cherenkov light in water and reflectivity at the PMT surface. Analysis was performed using PMTs belonging to five divisions of the barrel region, B1 to B5, and top.
The blue shaded circle spot at the bottom region indicates the beam target
used in the TOF calculation.
}
\label{waterlasersetup}
\end{figure}

\begin{figure}[!hbtp]
\centering
\includegraphics[totalheight=0.35\textheight]{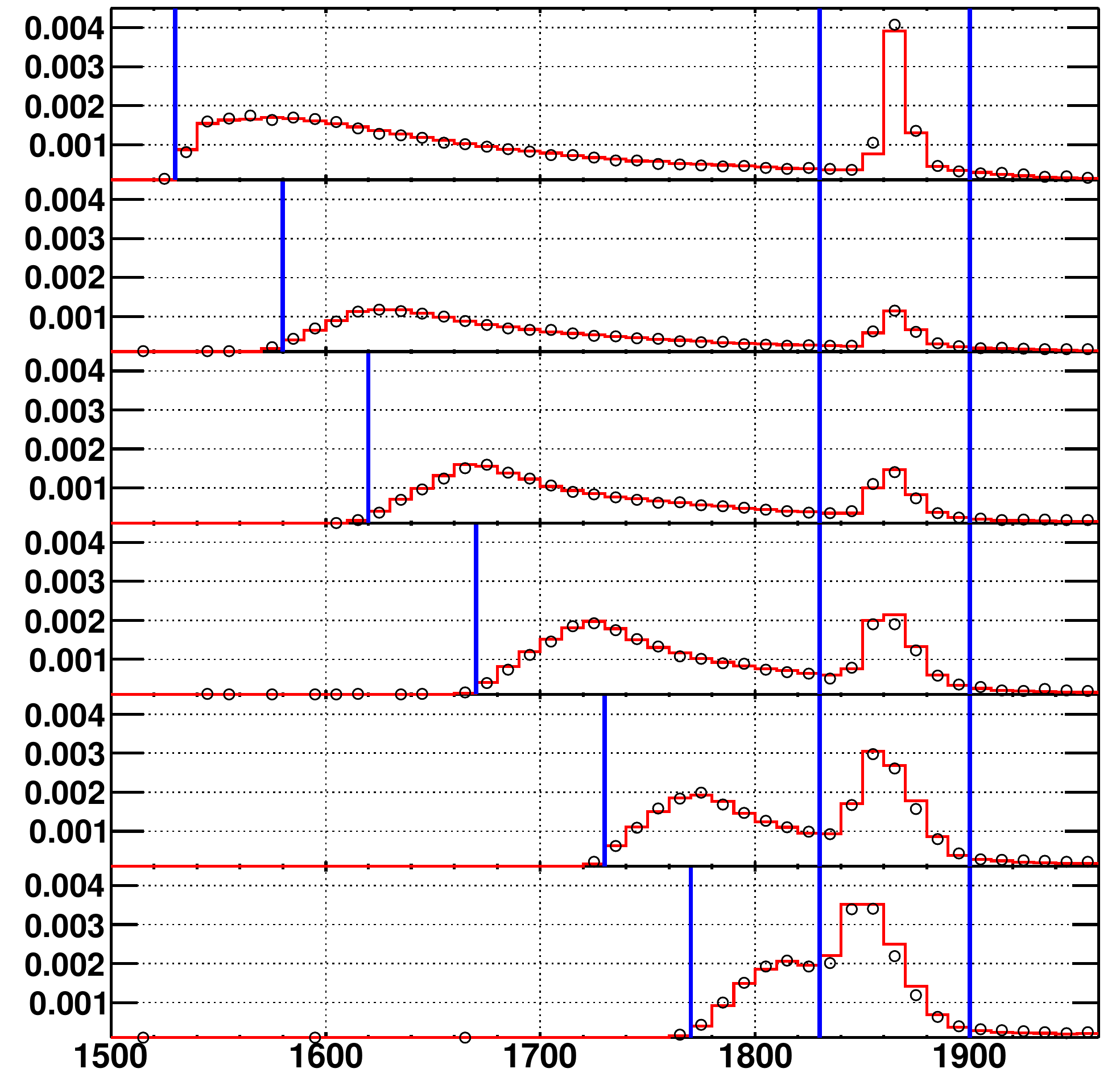}
\caption{
Typical TOF-subtracted (see text) timing distributions for the vertical down-going laser beam with the wavelength at 405\,nm between data (black circle) and MC (red histogram, the best tune) normalized by observed total photoelectrons.
The top plot is for the PMTs at the SK tank top wall.
The second to bottom plots correspond to the five barrel wall regions from top to bottom of the SK tank as indicated in Fig.~\ref{waterlasersetup}.
The time region between the left two blue solid vertical lines is used to measure the absorption and scattering described in this Section~\ref{waterparam} and the right region is used for the measurement of reflection described in Section~\ref{sec:refl}.}
\label{waterlaserdata}
\end{figure}

By injecting a collimated laser beam into the SK tank
(Fig.~\ref{waterlasersetup})
and comparing the timing and spatial distributions of the light with MC,
we can extract light absorption and scattering coefficients
as functions of wavelength.
The details of the experimental setup and data analysis are described 
in Section~8.2 \cite{Fukuda:2002uc}.
Figure~\ref{waterlaserdata} compares data and MC results for typical hit PMT
time distributions after time-of-flight (TOF based on the distance from 
the target of the laser at the bottom region to the hit PMT position).
The left region is from scattered photons,
while the right peak represents photons reflected by
the bottom PMTs and black sheets.
The total number of scattered photons and
the shape of the time distribution are used
to tune the symmetric and asymmetric scattering and the absorption parameters
for the SK-MC.
We generate various timing distributions with different MC parameters
and select the one that minimizes the $\chi^2$ value for the difference
between data and MC results.

The number of photons with a wavelength ($\lambda$)
in water decreases gradually according to:
\begin{equation}\label{Eq:I_WT}
  I(\lambda)=I_0(\lambda) e^{-\frac{l}{L(\lambda)}},
\end{equation}
where $l$ is the travel length of the light,
$I_0(\lambda)$ is the initial intensity,
$I(\lambda)$ is the intensity at $l$,
and $L(\lambda)$ is the total attenuation length
caused by scattering and absorption, which we call water transparency.
$L(\lambda)$ is defined in the SK-MC as follows:
\begin{equation}\label{Eq:WT_sym_asy_abs}
 L(\lambda) = \frac{1}{\alpha_{abs}(\lambda) + \alpha_{sym}(\lambda) +\alpha_{asy}(\lambda)},
\end{equation}
where $\alpha_{abs}(\lambda)$, $\alpha_{sym}(\lambda)$,
and $\alpha_{asy}(\lambda)$ are described below.
These are tuning parameters used in the SK-MC,
they are SK-based empirical functions
and do not exactly represent real physical properties.

\begin{figure}[!hbtp]
\centering
\includegraphics[width=8cm]{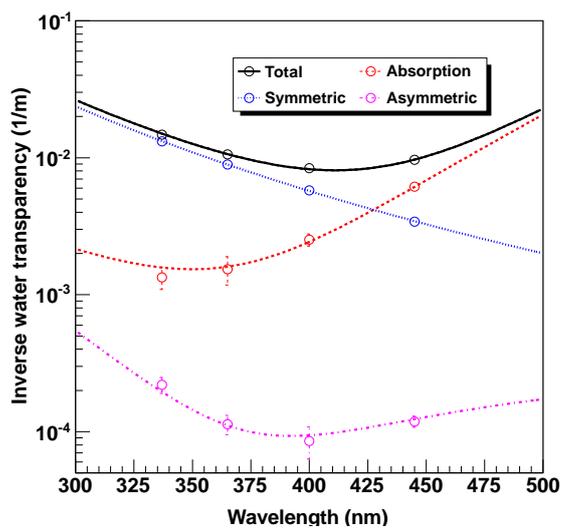}
\caption{Typical fitted water coefficient functions used in the SK-MC.
The points are the data obtained in April 2009.
The each lines through the points for absorption, symmetric scattering and asymmetric scattering show the fitted functions while the top line shows the total of all fitted functions added together.
}
\label{sk4_wtr_fit}
\end{figure}

The absorption amplitude $\alpha_{abs}$ (m$^{-1}$) 
as a function of wavelength $\lambda$ (nm)
is empirically determined 
using the laser beam data in
\begin{equation}
\alpha_{abs}(\lambda) = P_0\times\frac{P_1}{\lambda^4}+C,
\label{eqn:absform1}
\end{equation}
where the second term $C$ is the amplitude based on
the experimental data for $\lambda\ge$ 464\,nm ~\cite{PopeFry:1997ao,abstemp},
while the following formula
\begin{equation}
 C = P_0 \times P_2 \times (\lambda/500)^{P_3}
\label{eqn:absform2}
\end{equation}
is used for $\lambda\le$ 464\,nm.
Values for parameters $P_0 - P_3$
are obtained from a fit to the data.
The results of the fit of Eq.~(\ref{eqn:absform1}) and
(\label{eqn:absform2})
to the data of April 2009 are shown in Fig.~\ref{sk4_wtr_fit}.

The angular distribution of scattered light is divided into two components,
``symmetric'' (described by $1+\cos^2\theta$ where $\theta$ is
the scattered photon angle) and ``asymmetric''
for which the scattering probability increases linearly from
$\cos\theta =$ 0 to 1, and no scattering occurs for $\cos\theta<0$). 
The ``symmetric'' scattering consists of Rayleigh and symmetric Mie scattering,
while the ``asymmetric'' scattering consists of forward Mie scattering.
The Mie scattering typically has a forward-peaked distribution
for particles whose size is greater than wavelength of the light.
The ``asymmetric'' scattered angle function is a simple approximation
because details of the particle sizes are unknown.
The ``symmetric'' $\alpha_{sym}$ and ``asymmetric'' $\alpha_{asym}$
scattering amplitudes are empirically determined from
\begin{equation}
\alpha_{sym}(\lambda) = \frac{P_4}{\lambda^4}\times\left(1.0+\frac{P_5}{\lambda^2}\right),
\label{eqn:symscat}
\end{equation}
\begin{equation}
\alpha_{asym}(\lambda) = P_6\times\left(1.0+\frac{P_7}{\lambda^4}\times(\lambda-P_8)^2\right),
\label{eqn:asymscat}
\end{equation}
where $P_4 - P_8$ are fitting parameters. 
Figure~\ref{sk4_wtr_fit} shows a typical result of the scattering fit 
to Eq.~(\ref{eqn:symscat}) and (\ref{eqn:asymscat}).
%

Table~\ref{waterparadef} gives typical values of the fitted parameters
determined from the laser beam data.
%
\begin{table}
\begin{center}
\begin{tabular}{|cccc|cc|ccc|}
\hline
$P_0$ & $P_1$ & $P_2$ & $P_3$ & $P_4$ & $P_5$ & $P_6$ & $P_7$ & $P_8$ \\
\hline
0.624 & 2.96$\times$10$^7$ & 3.24$\times$10$^{-2}$ & 10.9 & 8.51$\times$10$^7$ & 1.14$\times$10$^5$ & 1.00$\times$10$^{-4}$ & 4.62$\times$10$^6$ & 392 \\
\hline
\end{tabular}
\end{center}
\caption
{\protect\small
Summary of the typical fitted water quality parameters
in Eq.~(\ref{eqn:absform1}), (\label{eqn:absform2}), (\ref{eqn:symscat}),
and (\ref{eqn:asymscat}) obtained from April 2009 data.
}
\label{waterparadef}
\end{table}
Using the values in Table~\ref{waterparadef},
the values for 1/$\alpha_{abs}$, 1/$\alpha_{sym}$, and 1/$\alpha_{asym}$ are
calculated to be 4.02$\times$10$^2$~m,
1.76$\times$10$^2$~m, and 9.89$\times$10$^3$~m, 
respectively. These lead to $L \sim$ 120~m at $\lambda$ = 400\,nm.

\begin{figure}[!hbtp]
\centering
\includegraphics[totalheight=0.35\textheight]{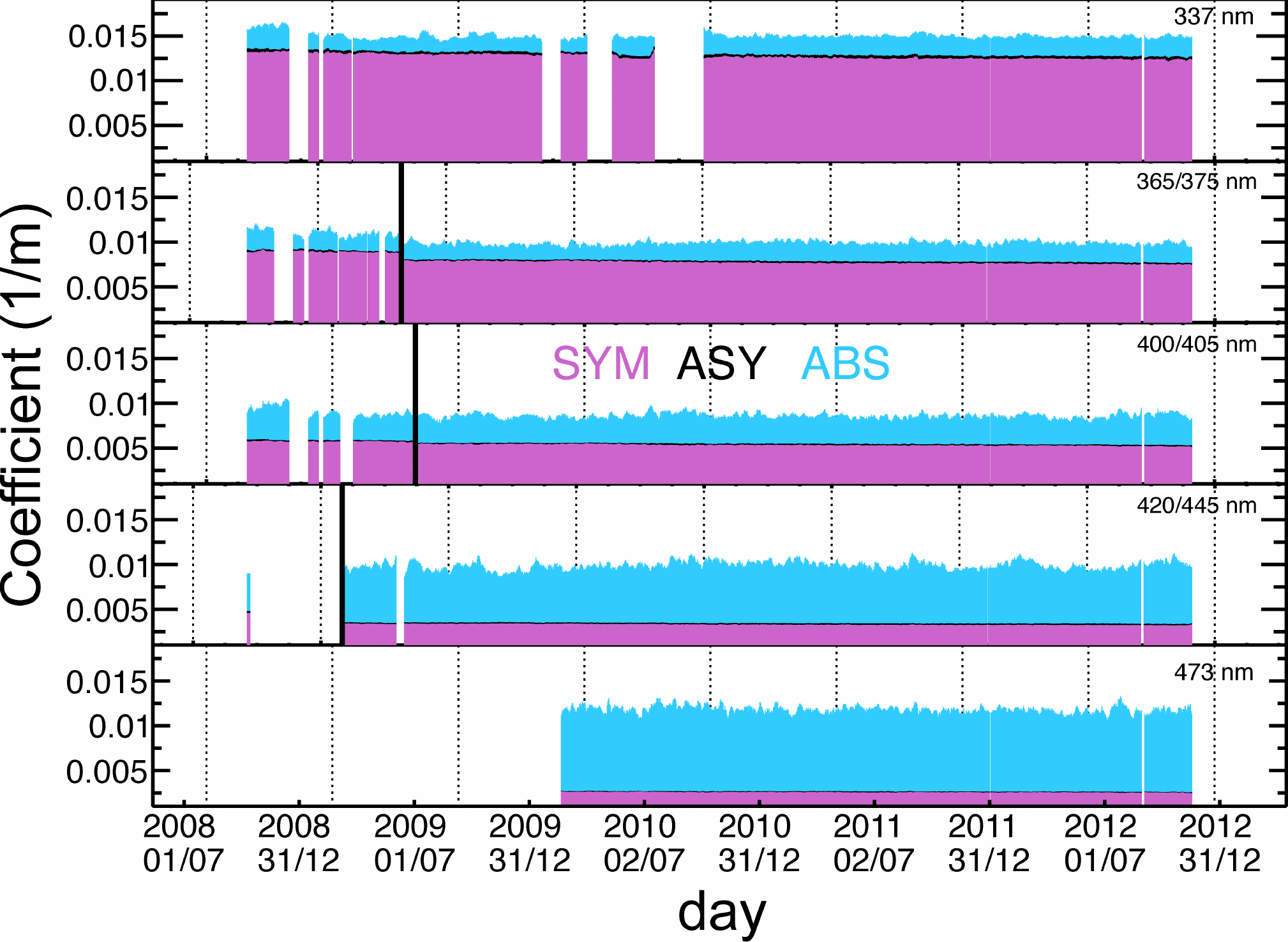}
\caption{Time variation of the water parameters in SK-IV from October 2008 to November 2012. The vertical axis is the inverse water transparency [$m^{-1}$]. The absorption, asymmetric scattering, and symmetric scattering for each wavelength are shown in blue (top), black (middle), and purple (bottom), respectively.
Note that the asymmetric scattering (black) is smaller than others
by one order of magnitude.
The wavelength was changed in 2008 
from 365\,nm to 375\,nm,
from 400\,nm to 405\,nm,
and from 420\,nm to 445\,nm
as indicated by the black vertical bars.
}
\label{sk4_wtr_timevar}
\end{figure}

 Obtaining real-time laser-beam data continues and
time variation of each coefficient are monitored at several wavelengths.
Figure~\ref{sk4_wtr_timevar} shows time variations
in absorption and scattering in SK-IV.
Each parameter was determined 
by applying the same analysis ($\chi^2$ method) on the real data
from each laser run.
%
%
The symmetric scattering coefficient is relatively stable
(the RMS/mean was about 3\% August 2008 - November 2012).
Both the absorption and asymmetric scattering coefficients have relatively
larger time dependence (about 20\%-40\% and 20\%-60\% in the same time period,
respectively).
Since asymmetric scattering is one order of magnitude smaller than absorption,
the time variations in water transparency are mainly caused by absorption.
%
%
%

\begin{figure}[htbp]
\begin{center}
\includegraphics[totalheight=0.3\textheight]{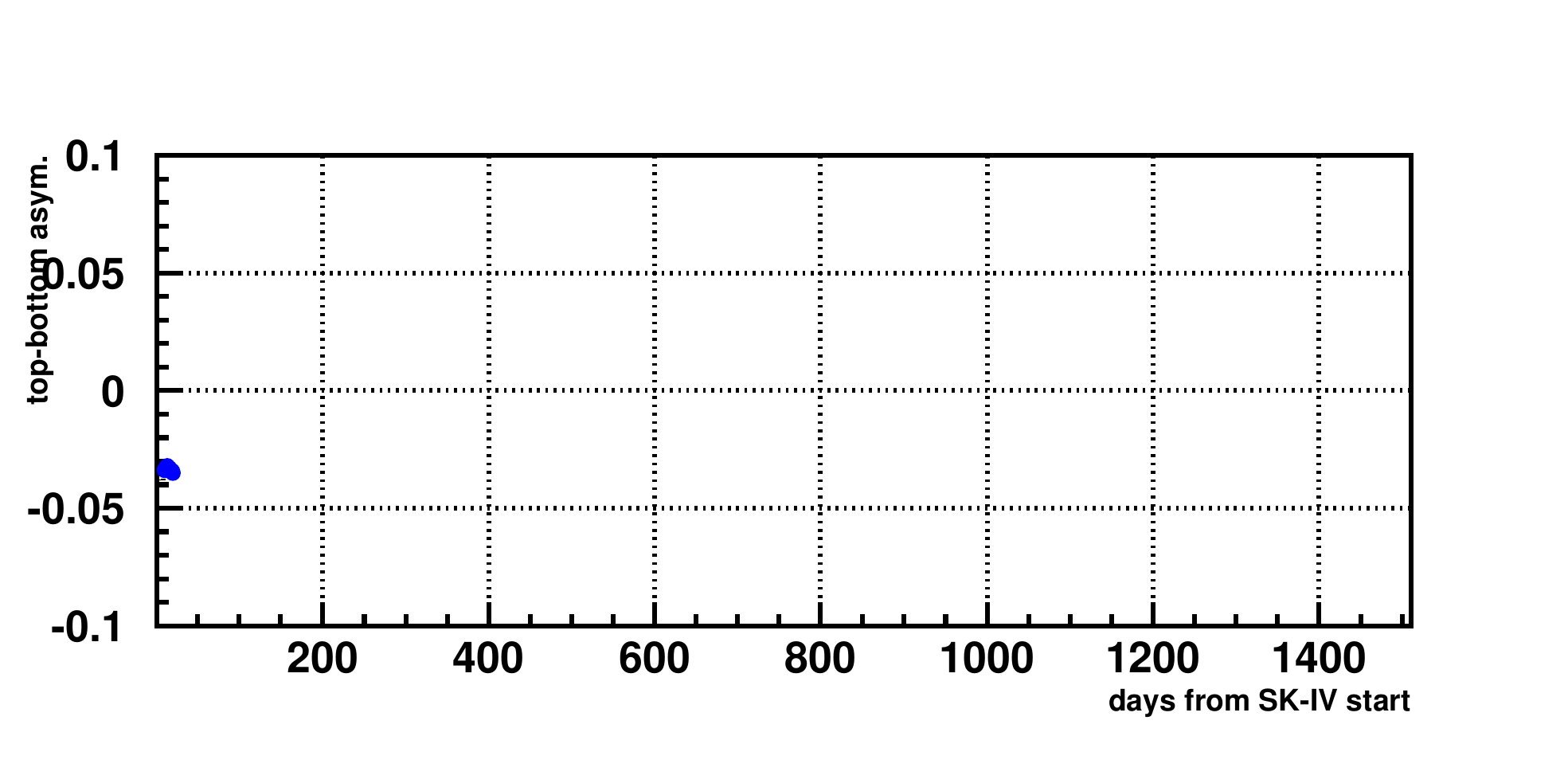}
\caption{
Top-bottom hit probability asymmetry as a function of time for SK-IV.
The vertical axis gives the value obtained using the averaged hit probability
in top, bottom and barrel PMTs defined as Eq.~(\ref{eq:tba}).
The red points show the nickel data and the blue thick curve shows the real-time Xe data described in Section~\ref{sec:hv}.
}
\label{fig:tba}
\end{center}
\end{figure}

As described in Section~\ref{water-circ}, water quality has position
dependence.
The vertical dependence is determined from the nickel data
obtained every month
and monitored with the real-time Xe system as described in
Section~\ref{sec:hv}.
A similar $z$-dependence in the water quality
is seen in other control sample data,
such as decay electrons from cosmic-ray stopping muons.
Figure~\ref{fig:tba} shows the top-bottom asymmetry
of the hit probability normalized by the average of all the PMTs
as a function of time.
The top-bottom asymmetry ($\alpha_{tba}$) is defined by
\begin{equation}
\alpha_{tba} = (\langle N_{top}\rangle-\langle N_{bottom}\rangle)/\langle N_{barrel}\rangle
\label{eq:tba}
\end{equation}
where $\langle N_{top} \rangle , \langle N_{bottom} \rangle$ and $ \langle N_{barrel} \rangle $ are the averages of the
hit probabilities of top, bottom and barrel of SK, respectively.
The average of the hit probability on the top PMTs is about 5\% smaller than
that on the bottom PMTs, as discussed in Section~\ref{water-circ}.
The nickel and Xe measurements 
are in accordance over the entire SK-IV time period.
%
Since the time variation is mainly caused by absorption, as described above,
the vertical position and time dependence of the water quality is introduced
in SK-MC by multiplying $\alpha_{abs}$ by a factor $A(z,t)$ 
in Eq.~(\ref{Eq:WT_sym_asy_abs}): 

%
\begin{eqnarray}
A(z,t) &=& 1 + z \cdot \beta(t) \quad  \mbox{for $z \geq -11$m} \nonumber\\
       &=& 1 - 11 \cdot \beta(t) \quad \mbox{for $z \leq -11$m}
\end{eqnarray}
As shown in Fig.~\ref{fig:ztemp}),
the value of $A(z,t)$ is assumed to be constant below $z=-11$~m
where water convection from the pure water inlet 
at the tank bottom is relatively stable and effective,
while we assume that absorption changes linearly 
for $z > -11$~m.
To determine the slope ($\beta$), 
the top-bottom asymmetry of the hit probability ($\alpha_{tba}$)
in the nickel data was compared with various MC samples
with different values of $\beta$ in the nickel data analysis.
We found the following relationship between $\beta$ and $\alpha_{tba}(\%)$;
\begin{equation}
\beta (1/m) = (-0.163 \times (\alpha_{tba})^2 - 3.676 \times \alpha_{tba})\times 10^{-3}
\end{equation}
For example, in April 2009 $\alpha_{tba}$ and $\beta$ 
are -4.91\% and 0.01, respectively.
%
%
%
%

\subsubsection{Light reflection at PMT and black sheet}
\label{sec:refl}
%
%
Light reflection at the PMT surface is tuned
using the same laser data 
used to determine the water quality parameters,
as described in Section~\ref{waterparam}.
Typical time distributions for data and MC after simulation tuning are shown
in Fig.~\ref{waterlaserdata}.
Four layers of material (refractive index) 
from the surface to the inside of the PMT are taken into account;
water (1.33), glass (1.472+3670/$\lambda^2$, where $\lambda$ is the
wavelength in nm),
bialkali ($n_{real} + i\cdot n_{img}$)~\cite{Motta2005217},
and vacuum (1.0). Here, $n_{real}$ and $n_{img}$ are
the real and imaginary parts of the complex refractive index,
and an appropriate thickness of the photocathode was based on
information from the manufacturer~\cite{Hamamatsu}.
The best fit values from the tuning were:  $n_{img}$ 1.667
and $n_{real}$ 2.31, 2.69, 3.06, and 3.24 at $\lambda$ = 337, 365, 400 and 420\,nm,
respectively.
Note that these fitted values include the effects
from several materials inside of the PMT
as well as the photocathode.

\begin{figure}[hbtp]
\begin{center}
 \includegraphics[totalheight=0.6\textheight,angle=-90]{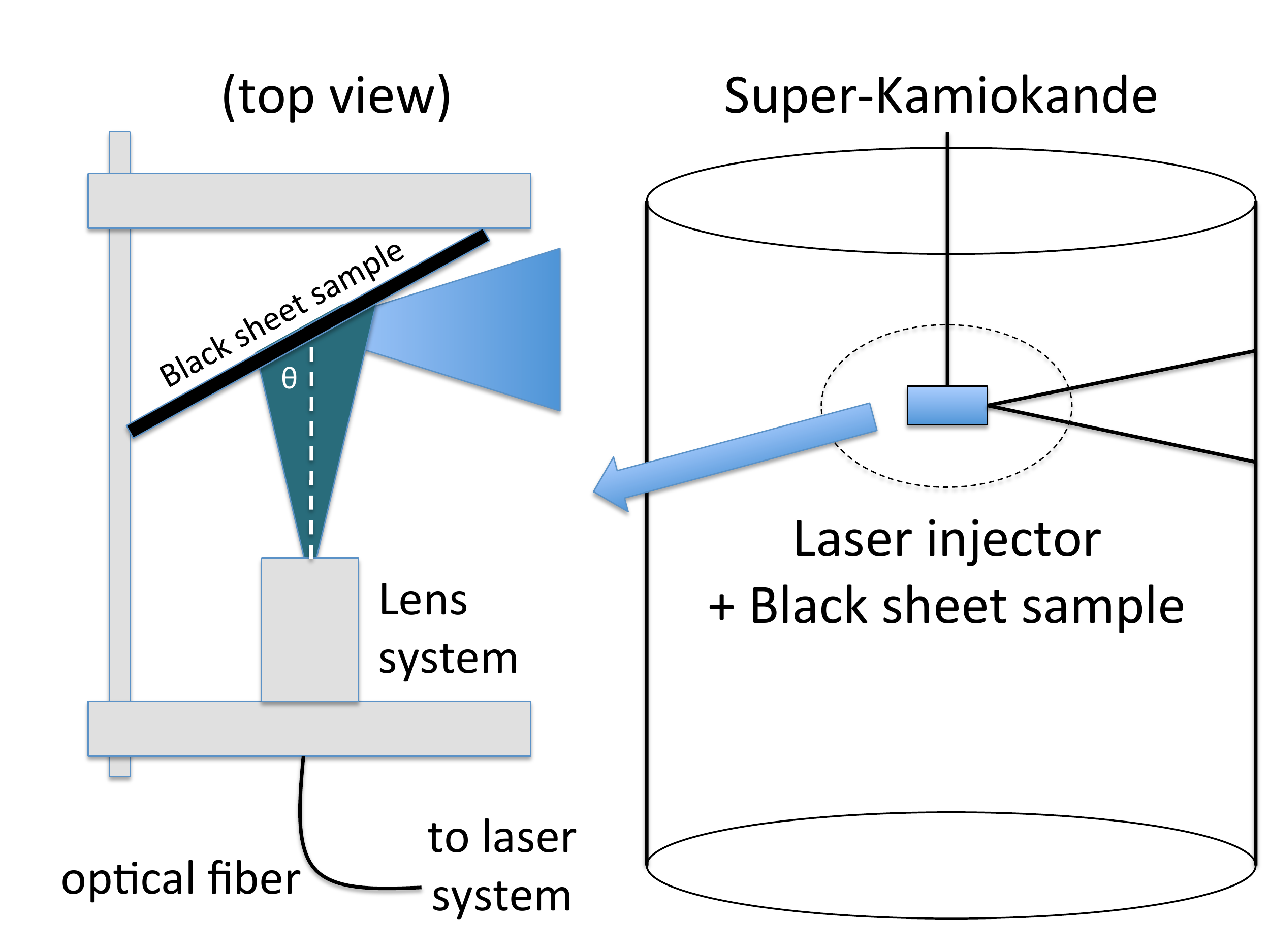}
 \caption{Schematic view of a laser light injector for the reflectivity measurement of a black sheet. The left figure shows the top view. The injector was inserted to the center part of the SK tank and the reflected light was measured by the SK ID-PMTs. For the normalization of reflectivity, data were taken without the black sheet.}
 \label{Fig:mov_li_bs}
\end{center}
\end{figure}

\begin{figure}[hbtp]
\centering
 \includegraphics[totalheight=0.35\textheight]{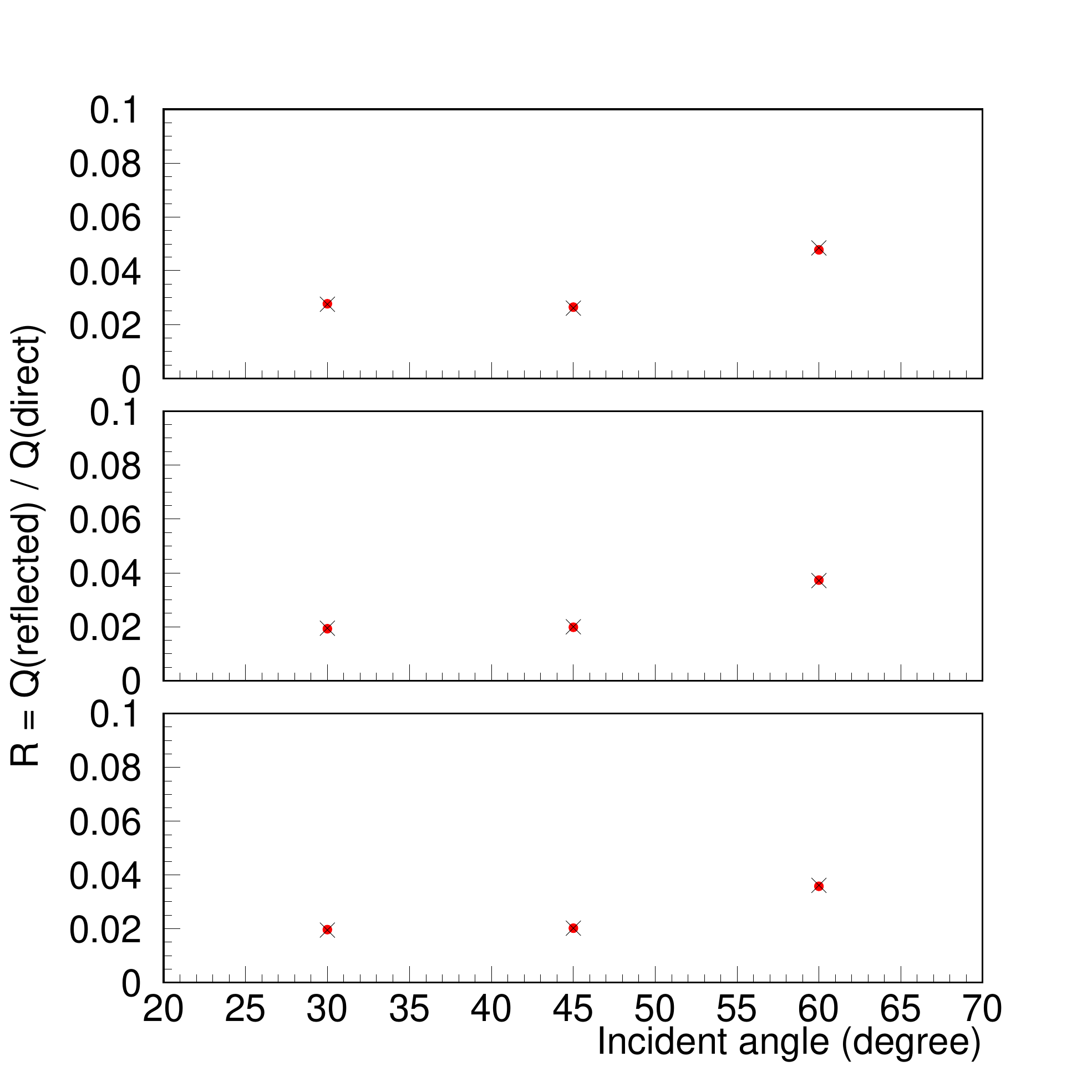}
 \includegraphics[totalheight=0.35\textheight]{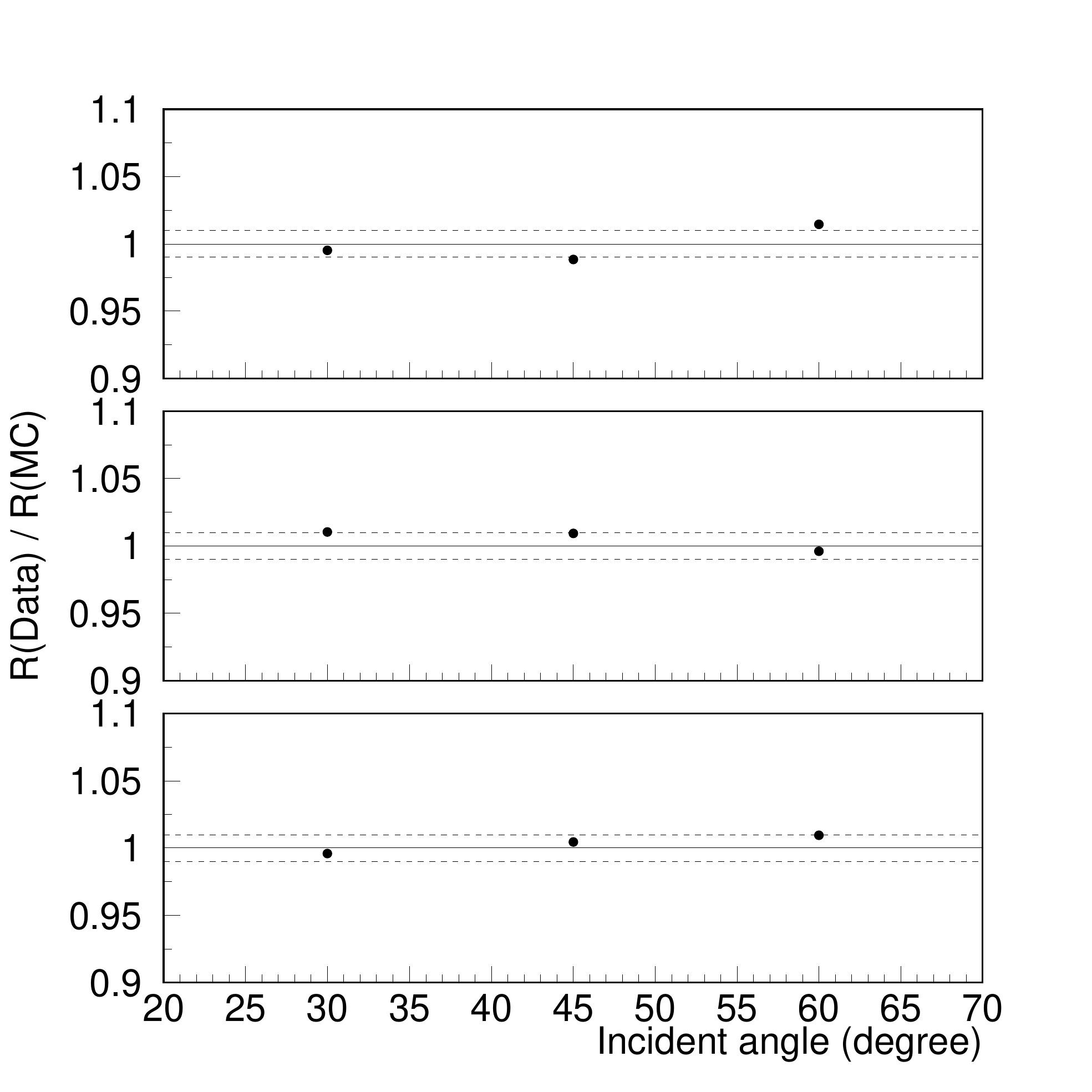}
 \caption{
The result of black sheet reflectivity tuning.
The top figure is for the wavelength at 337~nm, 
the middle is for 400~nm and the bottom is for 420~nm.
The left figure shows the charge ratio (see text)
of data (black crosses) and MC (red circles)
as a function of incident angle.
The right figure shows the ratio between data and MC
for the the left plots.
}
 \label{Fig:refres}
\end{figure}

%
Cherenkov photons are reflected or absorbed on the black sheet.
The reflectivity of the black sheet used in SK-MC is measured
by a light injector set in the SK tank.
The experimental setup is shown in Fig.~\ref{Fig:mov_li_bs}.
The reflected charge ($Q_{scattered}$) was measured
at three incident angles (30$^\circ$, 45$^\circ$, and 60$^\circ$)
with three wavelengths (337\,nm, 400\,nm, and 420\,nm).
For reference, the direct charges ($Q_{direct}$) 
without the black sheet were also measured.
The total reflectivity is tuned
using the ratio $R = Q_{scattered} / Q_{direct}$.
The tuning results are shown in Fig.~\ref{Fig:refres}.
The adjustment resulted in agreement between data and MC at better than the $\pm$1\% level
at each wavelength and position.

%
%

%
%


%
%
\section{Outer detector calibration}
\label{outer}

\subsection{Introduction}\label{od_introduction}
\begin{figure}
\begin{center}
 \includegraphics[width=10cm,clip]{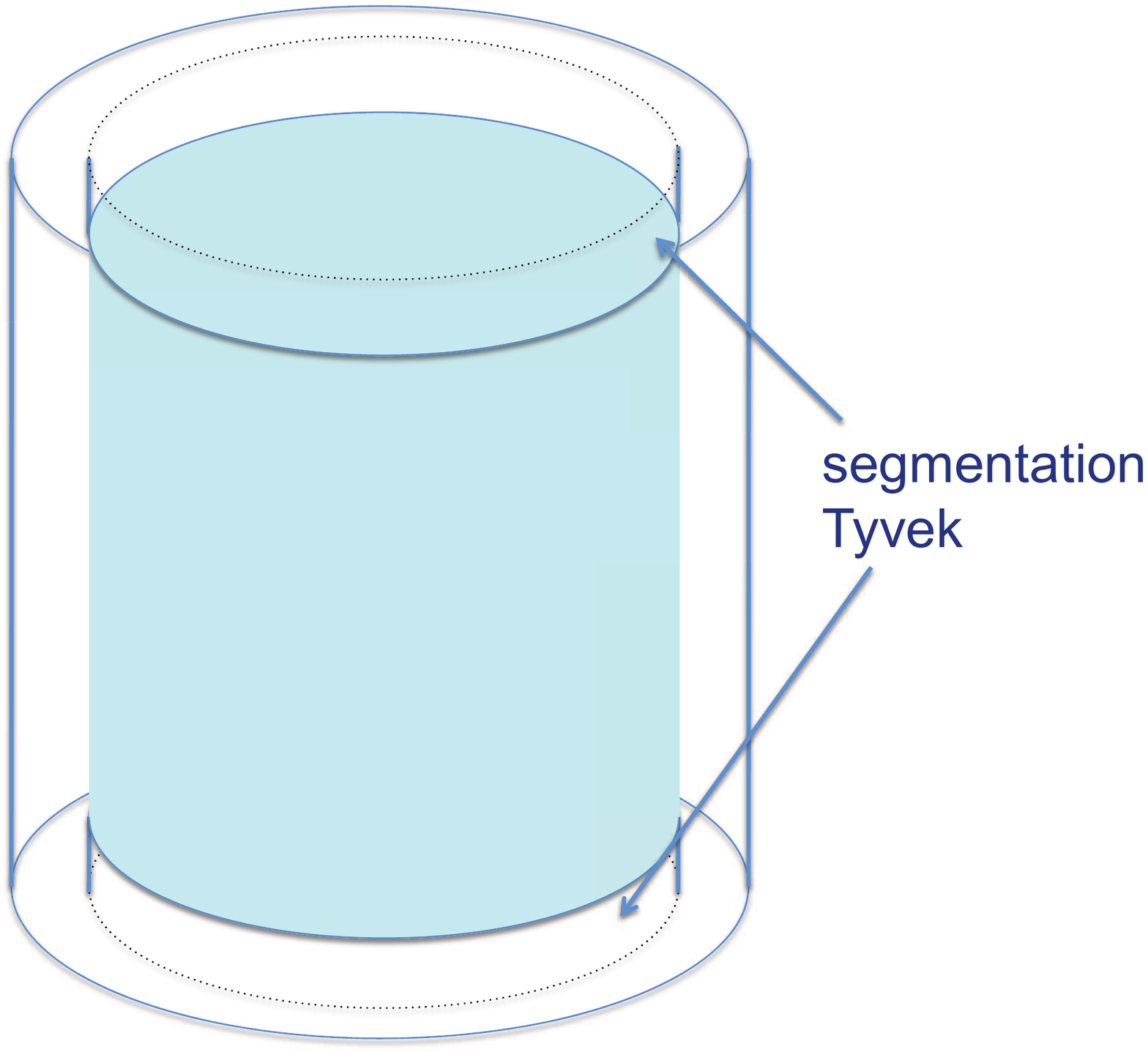}
 \caption
 {\protect \small  Schematic showing the OD segmentation Tyvek barrier to optically separate the endcap and barrel regions of the OD.
}
 \label{fig:odseg}
\end{center}
\end{figure}

The SK-I OD is described in~\cite{Fukuda:2002uc}. 
The OD was completely rebuilt to its original configuration
for SK-II. For SK-III, dead phototubes were replaced
and vertical sheets of Tyvek were installed to optically separate the barrel
and endcap regions as shown in Fig.~\ref{fig:odseg}.
The optical segmentation was implemented to enhance rejection of
background ``corner-clipping'' cosmic-ray muons from neutrino interactions
with an exiting lepton, e.g. partially-contained (PC) events used in atmospheric neutrino analyses~\cite{Ashie:2004mr,Ashie:2005}.

 For SK-IV, the electronics were completely
replaced~\cite{qbee} to use the same system as the ID (the previous custom electronics are described in~\cite{Fukuda:2002uc}).
Because the new electronics were more sensitive to cross-talk
between the ID and the OD,  the OD tube gains were reduced 
for SK-IV and the calibrations were redone.
The OD calibrations
used for SK-IV physics data analyses are summarized here.

The primary role of the OD is to identify incoming cosmic rays and atmospheric neutrino
interactions with particles leaving the ID. It also serves as a
passive buffer against low-energy particles (gamma rays and fast neutrons) from the surrounding rock.
The OD has been used for rough
reconstruction of very high-energy events (above TeV) that saturate the
ID~\cite{Swanson:2006gm}.
%
The OD is also important to separate PC
stopping and through-going events used for atmospheric neutrino
oscillation analyses~\cite{Ashie:2004mr,Ashie:2005}.
It is also used for selecting T2K beam neutrino events with light in the OD~\cite{Abe:2011ks}.
Requirements for OD calibration are not as
stringent as for the ID.
OD charge reconstruction accuracy of about 10\%-20\% and
timing to 5-10\,ns are generally sufficient for SK physics needs.
%
The calibration and simulation parameter tuning for the OD-PMTs are done differently than for the ID. 
OD time and charge calibrations are done using cosmic-ray and dark
rate data together with the nitrogen-dye laser 
that is used for the ID timing calibrations
described in Section~\ref{timing_calib}.

\subsection{OD-PMTs}\label{od_pmts}

There are 1885 8-inch PMTs installed in the OD, they are mounted on the outside of the steel structure that divides the inner and outer detectors.
Each is installed with a wavelength-shifting plate~\cite{Fukuda:2002uc}.
For SK-I, the PMTs were Hamamatsu R1408 recycled from the IMB detector.
For SK-II, some new Hamamatsu R5912 PMTs were installed.
The new tubes were installed primarily on the bottom part of the detector,
while the 
older ones were installed towards the upper part of the detector where the water pressure is lower.
For the SK-III upgrade, dead OD tubes were replaced.
For SK-III and IV there are 1293 new and 591 old PMTs in the OD.

\begin{figure}[!htbp]
\begin{centering}
\includegraphics[height=2.4in]{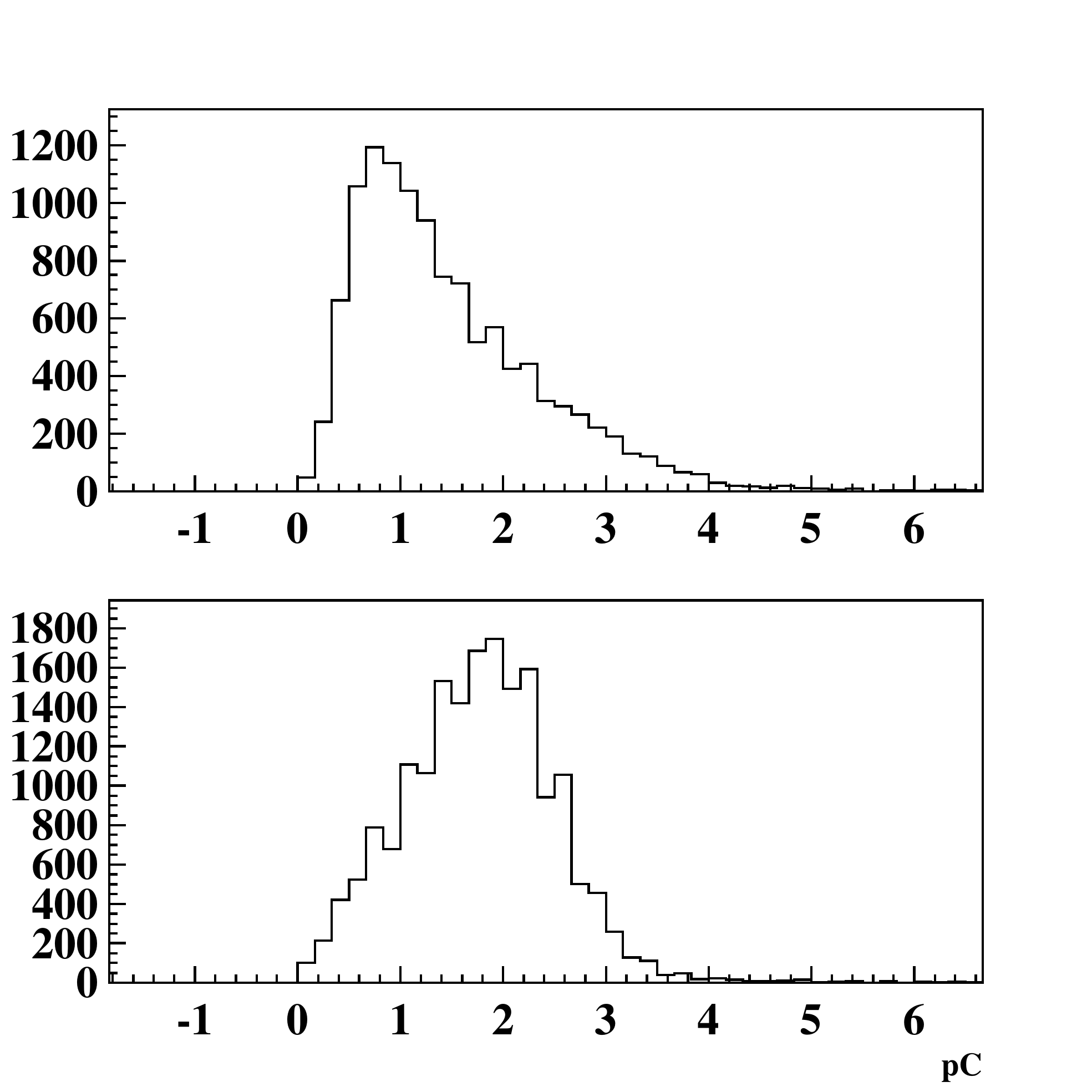}
\caption  {OD-PMT single-pe distributions, in units of pC.  Top: an example old (IMB) tube, showing roughly exponential charge distribution. Bottom: an example new tube, showing a clear single-pe peak.}
\label{fig:odspe}
\end{centering}
\end{figure}

For SK-IV, 
high-voltages applied to the OD-PMTs
were reduced in order to reduce cross-talk with the ID, to which the new electronics are sensitive.
The applied high-voltages of the new OD-PMTs were lowered by 400~V 
and those of the old OD-PMTs were lowered by 200~V.
This change reduced the gains to about 1/5 and 1/3 of the original gains
for the new and old tubes, respectively.
New and old tubes have different single-pe characteristics,
with new tubes showing generally clear
single-pe peaks and old tubes having approximately exponential single-pe charge distributions.
Figure~\ref{fig:odspe} shows typical single-pe distributions for
new and old OD-PMTs.
At the same time as the gain change, the QBEE charge threshold was changed
from about -25~mV to about -1.4~mV, in order to maintain efficiency for
single-pe detection.
Using the lower-intensity laser described in Section~\ref{od_calibration_hardware},
we experimentally confirmed that we do not lose 
single-pe hit efficiency for SK-IV
as compared to SK-III.

\subsection{OD calibration hardware}\label{od_calibration_hardware}

\begin{figure}
\begin{center}
 \includegraphics[width=10cm,clip]{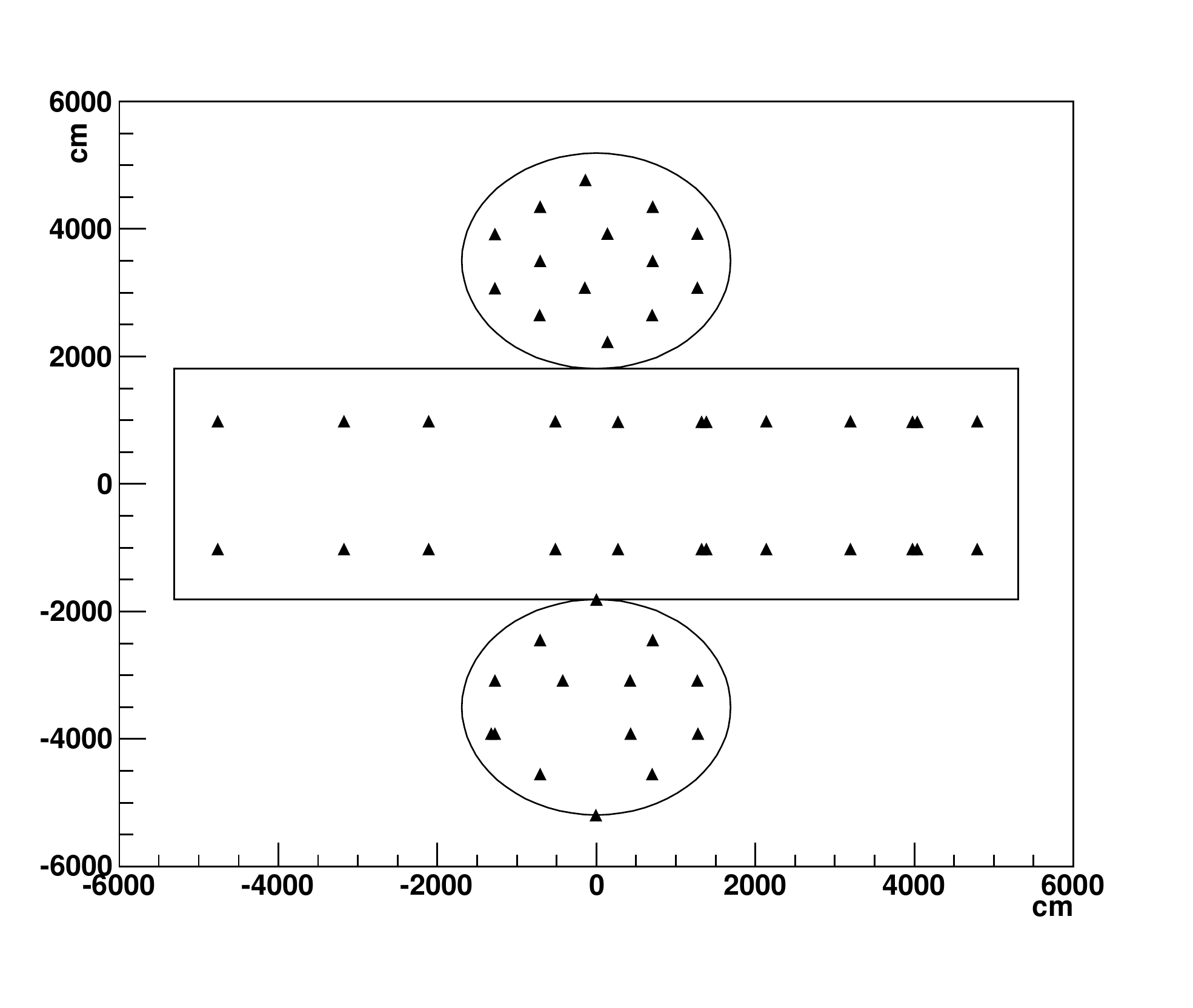}
 \caption
 {\protect \small  OD fiber end positions in ``unrolled'' view of the cylindrical tank.  Note that some fibers occupy the same position and are overlapped in the plot.
}
 \label{fiber_positions}
\end{center}
\end{figure}

The laser output is sent to 52 optical fibers that are routed to the OD.
The fibers are of UV-VIS grade
fused silica: each has a 200 $\mu$m diameter core, a 10 $\mu$m fused
silica cladding and a 10 $\mu$m protective layer.  The fibers are fed
through a hole near the middle of the top of the tank, routed to
locations around the OD, and have their ends mounted to the outer walls
to illuminate the OD-PMTs.  The OD fiber ends are mounted on white
Tyvek-wrapped backboards clamped to structural bolts on the SK water tank.
The fiber coordinates are shown in Fig.~\ref{fiber_positions}.
A total of 14 fibers are installed on the top of the tank, 12 are installed
on the walls at about $z=980$~cm in tank coordinates, 
12 are installed on the walls at about $z=-1017$~cm, and 14
are installed on the bottom.
The fibers are 72~m long except for those going to
the bottom of the tank, that are 110~m long.  Light from the fiber
ends is diffused by a mixture of optical cement and titanium dioxide
applied to the ends.


\subsection{OD charge calibration}\label{od_charge_calibration}

\subsubsection{OD linear charge response}


To determine the charge in pC corresponding to a single-pe in SK-IV,
we use a noninvasive ``dark rate method'' for charge calibration based on OD
hits outside the trigger time window.  Hits preceding
the trigger time have high probability of being single-pe hits.
Single-hit charge distributions during a several-$\mu$s time window
for each tube are accumulated for times that precede normal data triggers
by about 1 $\mu$s; the mean value is taken as pC
per photoelectron. 

To confirm the charge response per photoelectron at low light levels, where the response
of the OD-PMTs and electronics is  linear to a good approximation, the laser was flashed
at very low light levels: the occupancy $H$ (the fraction of laser pulses
resulting in a hit) is related to the mean measured photoelectron value 
according to $\langle pe \rangle = - \ln (1-H)/H$
\cite{imbcalib}.
For data with $H<0.18$, such that $\langle pe \rangle< 1.1$, the mean charge value for the hit distribution was chosen
to be the single-pe level.
The results from the laser and dark rate methods
were found to be in agreement within 10\%.  As the dark rate method for OD charge calibration
causes no detector downtime and allows for continuous monitoring, it is currently used as the primary method.
For SK-IV, typical OD pC per photoelectron (pC-per-pe) values range from one to six.
The pC-per-pe values measured this way are used to determine photoelectrons for each PMT hit in the data.

\begin{figure}[!htbp]
\begin{centering}
\includegraphics[height=3.5in]{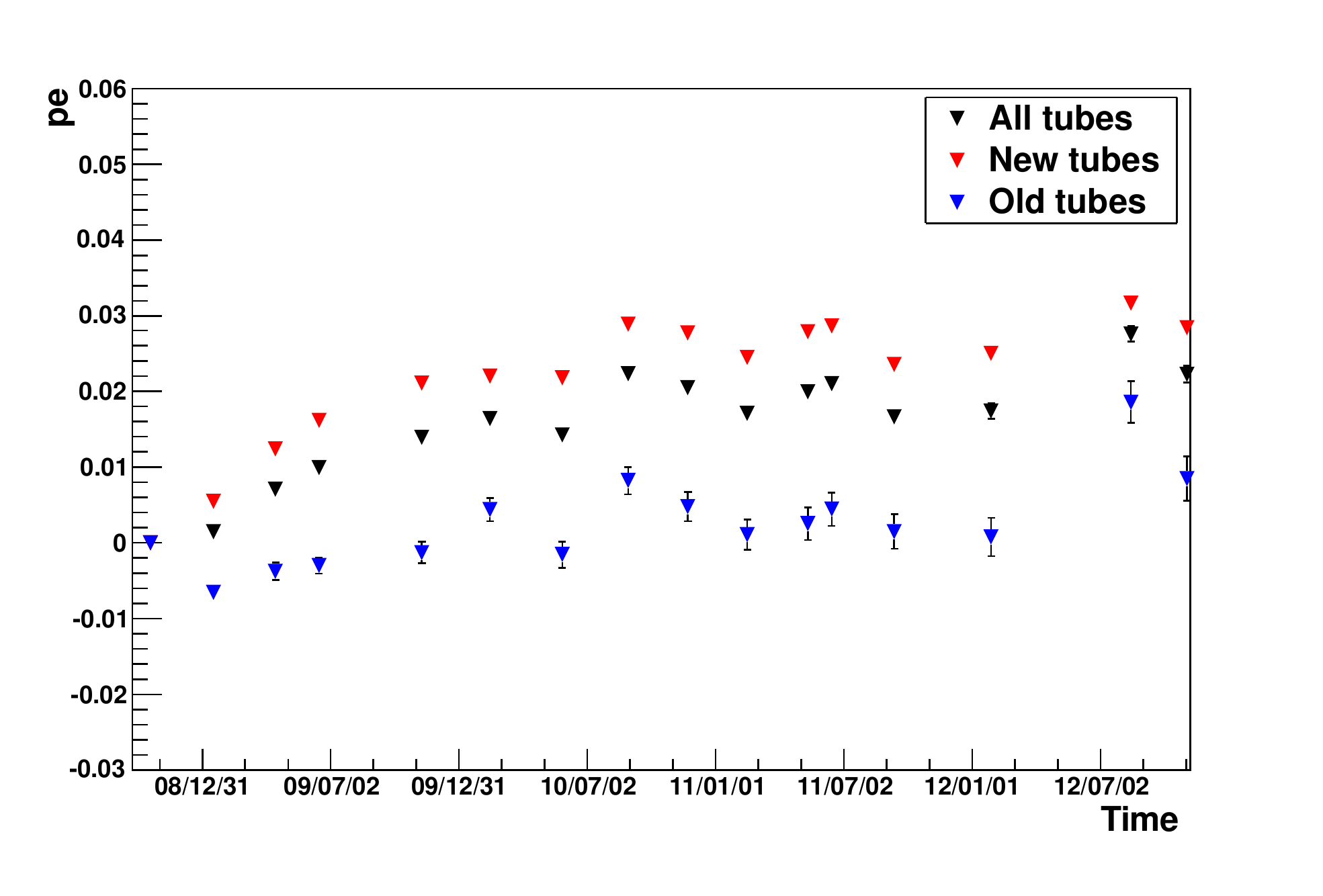}
\caption  {Average OD-PMT gain drift as a function of time, given as a fraction of a photoelectron.  The plot shows the difference in pC-per-pe, averaged over PMTs,  from the value for reference run near the beginning of SK-IV.   The plot shows new and old tubes separately. The error bars represent the error on the mean.}
\label{fig:sk4gains}
\end{centering}
\end{figure}

Average stability of the OD charge calibration in SK-IV is shown in Fig.~\ref{fig:sk4gains}.
The values of pC-per-pe are typically stable 
within at most 5\% for a given tube over a one year period.
There is a slow upward average drift as a function of time, which corresponds
 $\sim$1\% difference on average per year.  The figure also shows an increasing error on the mean gain, which implies that the PMT-to-PMT spread of gains also increases slightly with time, especially for old OD tubes.
The upward drift is more pronounced for the new (R5912) PMTs than for the recycled IMB tubes.
To account for the drift, new pC-per-pe values are determined on an approximately yearly timescale.

For SK-IV, only the linear charge conversion is applied to the data, i.e. the number of  photoelectrons is determined for a given hit using the pC-per-pe calibration constant according to pe=pC/pC-per-pe.
For OD hits in SK-IV, the MC single-pe charges are chosen separately from charge distributions for old and new OD tubes, with shapes based on 
the average old and new tube single-pe charge distributions, respectively.
In contrast to the ID case,  quantum efficiencies (or light collection efficiencies) are not measured directly, but rather are treated as parameters to be tuned in SK-MC.  Similarly, although charge nonlinearity properties of OD tubes at high charge were determined using the laser system (see below) and are simulated in SK-MC, nonlinearity corrections are not applied directly to the data.

\subsubsection{OD-PMT nonlinearity}

\begin{figure}
\begin{center}
  \includegraphics[width=10cm,clip]{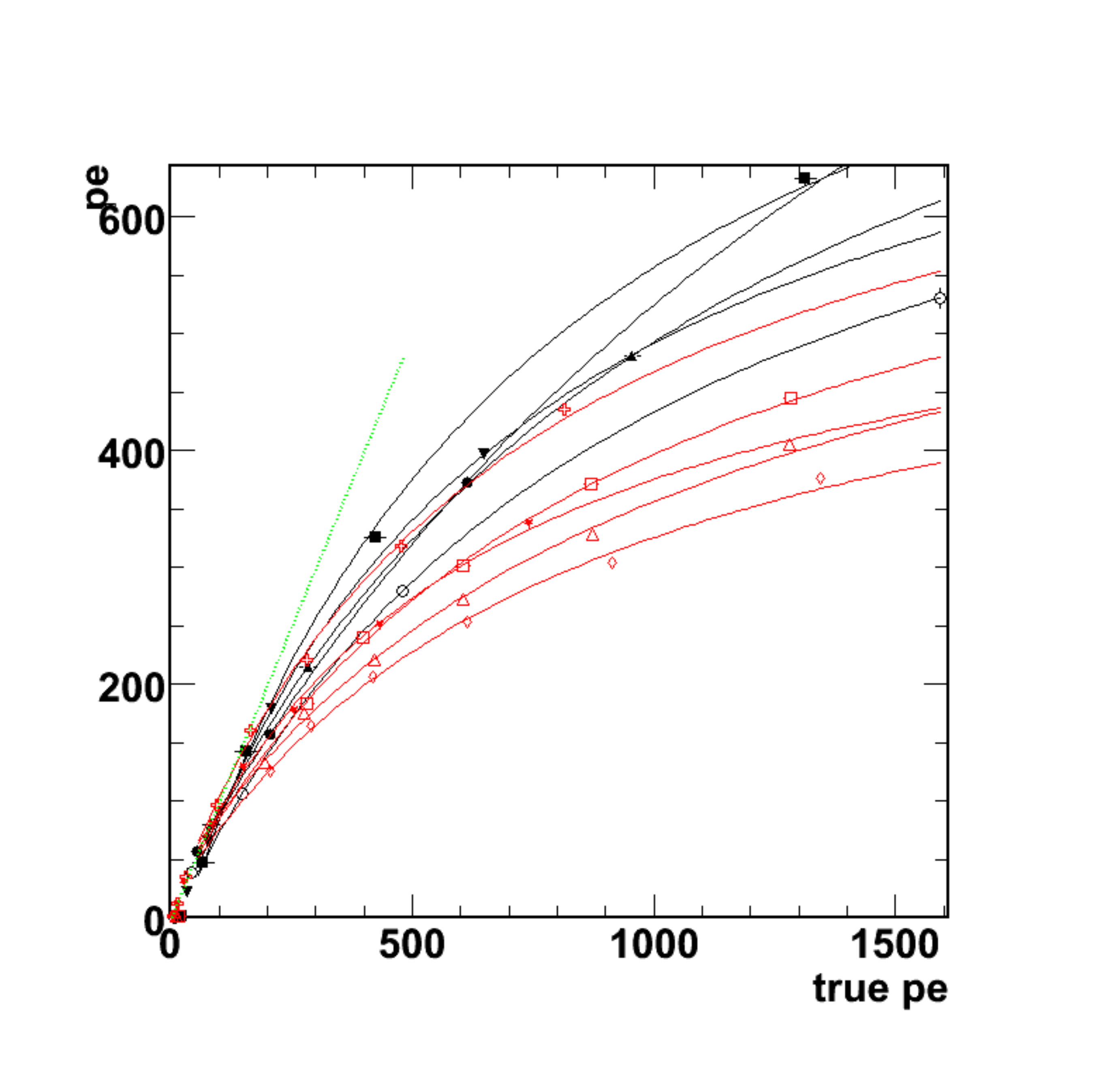}
  \caption
  {\protect \small 
Correlation between input and measured charges in pe, for 
pe=pC/pC-per-pe,
 for typical old (red) and new (black) PMTs
from laser data taken in March 2009 for SK-IV.  Each line corresponds to a particular PMT.
The result of the fit explained in the text is overlaid for each PMT.  The green line is for $\rm{pe=true~pe}$.}
 \label{od_saturation_curve}
\end{center}
\end{figure}

The response of some typical OD-PMTs to a range of light levels injected with the OD laser system is seen in Fig.~\ref{od_saturation_curve}.

Above a few hundred photoelectrons, a saturation effect appears in the measured charge.
These data are fitted to
\begin{equation}
Q_{meas} = Q_{true} / (1 + k \cdot Q_{true}),
\label{eqn:od_saturation_curve}
\end{equation}
for each PMT and implemented in the SK-IV OD detector simulation code.
In Eq.~(\ref{eqn:od_saturation_curve}), $Q_{meas}$, $Q_{true}$ represent
measured and true charges, and $k$ is the fitting parameter.
Laser light levels in this system are controlled by an attenuator.
The single-pe level was determined at low occupancy so that at low light levels,
the attenuator power law could be calibrated assuming a linear response of the PMTs.
Using the resulting power law, $Q_{true}$ was inferred.
Because some regions of the OD are less well illuminated by the laser system,
a value of $k$ was estimated for each OD-PMT with cosmic-ray muons
using the maximum charge per event ($\sim 1/k$).
Values of $k$ fall within 0.001 - 0.003.

\subsection{OD timing calibration}\label{od_timing_calibration}


Since OD timing information is not currently used for any event selection
or reconstruction in physics data analysis, except for selection of OD hits within time windows around the trigger time, the
goal of the OD timing calibration is to confirm that
both the global time offset between ID and OD and the
relative timing offsets for each OD-PMT are sufficiently small,
say within several ns (see~Section~\ref{od_introduction}).
In OD standard calibrations for the time-walk effect for OD-PMTs are not included.

For the relative timing offset,
differences in cable lengths are considered.
About 87\% of the cables (1647 among 1885) are 70~m long,
and the remaining have lengths between 71~m and 78~m.

The global time offset was determined with the laser system,
and checked with cosmic ray muons.

In October 2008, laser events were taken
by flashing the laser in the ID center and the OD top.
For this study, the same optical fiber and diffuser ball were used, so as
not to introduce any unknown systematic differences between different installations in the ID and OD.
In this way,
the global OD/ID timing offset was determined to be within several ns.
The relative offsets of individual PMTs were also confirmed to be within several ns.

\begin{figure}
\begin{center}
  \includegraphics[width=10cm,clip]{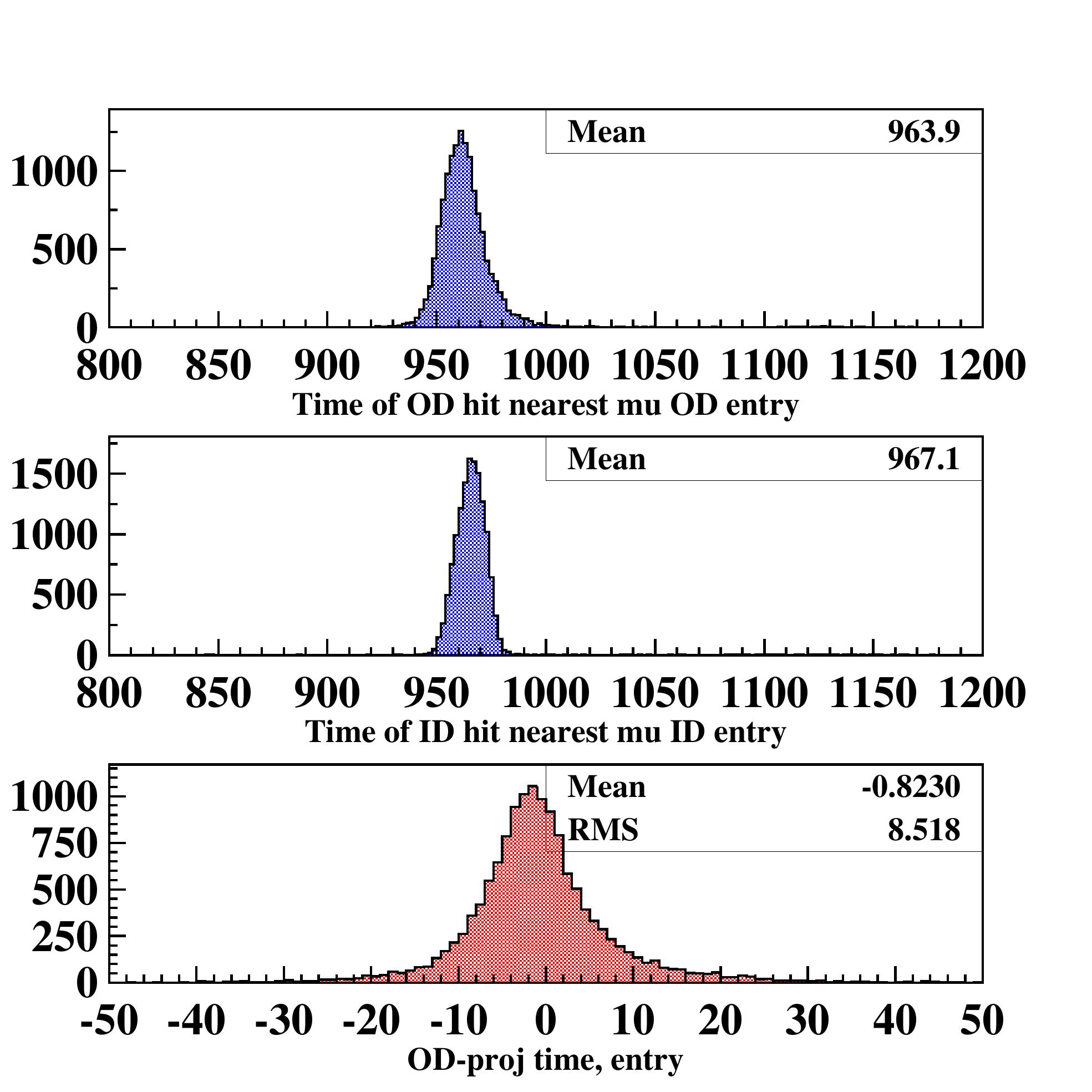}
  \caption
  {\protect \small Top: Distribution of times (ns) for the nearest hit OD-PMT to the fit tracks of downward cosmic ray muons (arbitrary zero time value).  Center: Time distribution for nearest hit ID-PMTs.  Bottom: Distribution of difference between ID and OD time after correction for time of flight of the muon.
}
 \label{fig:odtiming}
\end{center}
\end{figure}

Also in October 2008, the global time offset was independently confirmed with cosmic-ray muon data .
For each event, for muon tracks fit using ID information, 
differences were determined between recorded times for the ID and OD-PMTs nearest to the muon track.
Using the global offset parameter determined with laser data,
the global timing offset was found to be within a few ns,
and the OD resolution for determining the time the muon passed through the OD
was found to be within about 10\,ns (Fig.~\ref{fig:odtiming}).
As a stability check, the time offset was reconfirmed in February 2010.

\subsection{Optical properties of the OD}\label{od_optical_properties}

In contrast to the ID case, the optical properties of OD materials are not measured directly, but are treated as parameters to be tuned in SK-MC.   They are given nominal values and then adjusted slightly to best match quantities related to number of hits and charge for cosmic ray muons.
In the MC simulation, the reflectivity of Tyvek is modeled as a combination of Gaussian specular reflection and Lambert diffuse reflection~\cite{filevich}, with relative contributions varying as a function of angle.  To obtain a starting point for this tuning, relative amounts of Gaussian and Lambertian reflection were measured ex-situ~\cite{alvaro}, and then adjusted as part of OD simulation tuning.
Other tuned parameters in the simulation include relative reflectivity of Tyvek on each OD surface and transmissivity of Tyvek for the segmentation barriers. OD light collection efficiency is also tuned, as a separate average parameter for each of the three optically separated segments of the OD: top, bottom and barrel.
This quantity takes into account both quantum efficiency and collection of photons.

\section{Summary}
\label{summary}
Super-Kamiokande is a multi-purpose detector for the 
measurements of solar, atmospheric, and astrophysical neutrinos
as well as to search for nucleon decay. It has also been used
as far detector for long-baseline experiments using
accelerator-produced neutrino beams.
It has achieved world-leading scientific results,
such as the first confirmed discovery of neutrino oscillations.
To obtain such results, it is crucial to have precise detector
calibrations over a very wide range, from very weak one-photoelectron light levels
to quite strong, multi-hundred photoelectron light levels;
these levels correspond to physics events ranging from
a few MeV to above one TeV.
Our calibration work has been along two main lines.
One is the determination of conversion factors from raw counts by the electronics
to charges and times observed by PMTs. The other is modeling of optical properties of the water and detector.

For charge determinations, we employ a new method
that allows independent determination of PMT gain and quantum efficiency.
Our goal is to understand these quantities at the level of 1\%,
and this goal has been achieved.
We also observed some unexpected PMT features, such as a time dependence in quantum efficiency.

The time response of the readout channels (PMTs and readout electronics)
have been calibrated for different charge ranges,  covering the
entire dynamic range of charge, from $10^{-1}$ to $10^{3}$~photoelectrons.
The time resolution achieved by SK is, e.g. 2.1~ns at $\sim1$~photoelectron and
0.5~ns $\sim100$~photoelectrons.

Optical properties of the detector were measured using several light sources.
In MC simulations, we consider results of measurements of
absorption and scattering probability in SK water,
including its position dependence and reflections on PMT surfaces
and the black sheet.

Calibrations were also performed to determine charge and timing
response of the OD,  as well as its optical properties.
These parameters are now understood well enough to support physics analyses.

Through this calibration work, we have developed several new tools.
The established calibration methods described in this paper will also be useful
for future water Cherenkov detectors.

\section*{Acknowledgments}
We gratefully acknowledge the cooperation of the
Kamioka Mining and Smelting Company.
The Super-Kamiokande experiment has been built and operated
from funding by the Japanese Ministry of Education,
Culture, Sports, Science and Technology, the United
States Department of Energy, and the U.S. National
Science Foundation. Some of us have been funded by
the Korean Research Foundation (BK21),
the National Research Foundation of Korea (NRF-2009-C00046),
the State Committee for Scientific
Research in Poland (Grant No. 1757/B/H03/2008/35),
the Japan Society for the Promotion of Science, the
National Natural Science Foundation of China under Grant
No. 10875062, and the Spanish Ministry of Economy and
Competitiveness (Grant No. FPA2009-13697-C04-02).






\bibliographystyle{elsarticle-num}




\end{document}

%% file: authors.tex
\author[ICRR]{K.~Abe}
\author[ICRR,IPMU]{Y.~Hayato}
\author[ICRR]{T.~Iida\fnref{QU}} 
\author[ICRR]{K.~Iyogi} 
\author[ICRR]{J.~Kameda}
\author[ICRR,IPMU]{Y.~Kishimoto}
\author[ICRR]{Y.~Koshio\corref{cor}\fnref{OKA}}
\ead{koshio@fphy.hep.okayama-u.ac.jp}
\author[ICRR]{Ll.~Marti\fnref{IPMU2}}
\author[ICRR]{M.~Miura} 
\author[ICRR,IPMU]{S.~Moriyama} 
% ...

\address[ICRR]{Kamioka Observatory, Institute for Cosmic Ray Research, University of Tokyo, Kamioka, Gifu 506-1205, Japan}
% ...
\address[IPMU]{Kavli Institute for the Physics and Mathematics of the Universe (WPI), Todai Institutes for Advanced Study, University of Tokyo, Kashiwa, Chiba 277-8583, Japan }
% ...

\fntext[QU]{Present address: Research Center for Nuclear Physics, Osaka University, 10-1 Mihogaoka, Ibaraki, Osaka, 567-0047, Japan}
\fntext[OKA]{Present address: Department of Physics, Okayama University, Okayama, Okayama 700-8530, Japan}
\fntext[IPMU2]{Present address: Kamioka Satellite, Kavli Institute for the Physics and Mathematics of the Universe (WPI), Todai Institutes for Advanced Study, University of Tokyo, Kamioka, Gifu 506-1205, Japan}
\fntext[IPMU3]{Present address: Kavli Institute for the Physics and Mathematics of the Universe (WPI), Todai Institutes for Advanced Study, University of Tokyo, Kashiwa, Chiba 277-8583, Japan }
\fntext[OSAKA2]{Present address: Department of Earth and Space Science, Osaka University, Toyonaka, Osaka, 560-0043, Japan}
\fntext[OCU]{Present address: Department of Physics, Osaka City University, Sugimoto 3-3-138, Sumiyoshi-ku, Osaka, 558-8585, Japan}
\fntext[GW]{Present address: GIST College, Gwangju Institute of Science and Technology, Gwangju 500-712, Korea}
\fntext[TOHOKU]{Present address: Research Center for Neutrino Science, Tohoku University, Sendai 980-8578, Japan}
\fntext[KOITO]{Present address: Shizuoka Plant, Koito manufacturing CO., LTD., 500, Kitawaki, Shimizu-ku, Shizuoka 424-8764, Japan}
\fntext[KAMI]{Present address: Kamioka Observatory, Institute for Cosmic Ray Research, University of Tokyo, Kamioka, Gifu 506-1205, Japan}
\fntext[CERN]{Present address: CERN PH-Division, Bat 32, R-A-18, CH-1211, Geneva, Switzerland}
\fntext[DEC]{Deceased.}
\cortext[cor]{Corresponding author. Tel.: +81 86 251 7817; fax: +81 86 251 7830.}
